%% file: blasfeo_blas_api.tex
\documentclass[a4paper]{article}

\usepackage[margin=3.0cm]{geometry}
\usepackage{amsmath}
\usepackage{amssymb}
\usepackage{xfrac}
\usepackage{graphics}
\usepackage{subfig}
\usepackage{color}
\usepackage{comment}
\graphicspath{{./}}

\newif\ifcommentandrea
\commentandreatrue
\usepackage{calc}

\newif\ifcommenttom
\commenttomtrue
\usepackage{calc}

\newif\ifcommentdimitris
\commentdimitristrue

\usepackage{soulutf8}

\newif\iftodo
\todotrue
\usepackage{calc}

\author{Gianluca Frison$^1$, Tommaso Sartor$^1$, Andrea Zanelli$^1$, Moritz Diehl$^{1,2}$ \\ \it \small University of Freiburg, $^1$ Department of Microsystems Engineering (IMTEK), $^2$ Department of Mathematics \\ \rm \small email: name.surname@imtek.uni-freiburg.de}
\title{The BLAS API of BLASFEO: \\ optimizing performance for small matrices}

\begin{document}

\makeatletter
\renewcommand*\env@matrix[1][*\c@MaxMatrixCols c]{%
  \hskip -\arraycolsep
  \let\@ifnextchar\new@ifnextchar
  \array{#1}}
\makeatother

\maketitle

\thanks{\footnotesize This research was supported by the German Federal Ministry for Economic Affairs and Energy (BMWi) via eco4wind (0324125B) and DyConPV (0324166B), and by DFG via Research Unit FOR 2401.}

\begin{abstract}

BLASFEO is a dense linear algebra library providing high-performance implementations of BLAS- and LAPACK-like routines for use in embedded optimization and other applications targeting relatively small matrices.
BLASFEO defines an API which uses a packed matrix format as its native format.
This format is analogous to the internal memory buffers of optimized BLAS, but it is exposed to the user and it removes the packing cost from the routine call.
For matrices fitting in cache, BLASFEO outperforms optimized BLAS implementations, both open-source and proprietary.
This paper investigates the addition of a standard BLAS API to the BLASFEO framework, and proposes an implementation switching between two or more algorithms optimized for different matrix sizes.
Thanks to the modular assembly framework in BLASFEO, tailored linear algebra kernels with mixed column- and panel-major arguments are easily developed.
This BLAS API has lower performance than the BLASFEO API, but it nonetheless outperforms optimized BLAS and especially LAPACK libraries for matrices fitting in cache.
Therefore, it can boost a wide range of applications, where standard BLAS and LAPACK libraries are employed and the matrix size is moderate.
In particular, this paper investigates the benefits in scientific programming languages such as Octave, SciPy and Julia.

%BLASFEO is a dense linear algebra library providing high-performance implementations of BLAS- and LAPACK-like routines for use in embedded optimization, and is therefore targeting relatively small matrices.
%One of the key features of BLASFEO is the use of a packed matrix format, named panel-major, as the native format in its API.
%For matrices fitting in cache, this matrix format is analogous to the packed sub-matrices in the memory buffers of optimized BLAS libraries, but with the key advantage of removing the packing cost from the routine call.
%Thanks to that, for small matrices BLASFEO outperforms optimized BLAS implementations, both open-source and proprietary, on a wide range of computer architectures.
%This paper investigates the addition of a standard BLAS API to the BLASFEO framework, and proposes an implementation switching between two or more algorithms, which are optimized for different matrix sizes.
%By leveraging the modular nature of the assembly kernels framework in BLASFEO, tailored linear algebra kernels with mixed column- and panel-major arguments are easily developed.
%This BLAS API has generally lower performance than the BLASFEO API, but it nonetheless outperforms optimized BLAS and especially LAPACK libraries for matrices fitting in cache.
%Therefore, it can be used to boost a wide range of applications, where standard BLAS and LAPACK libraries are employed and the matrix size is moderate.
%In particular, this paper investigates the benefits in scientific programming languages such as Octave, Python SciPy and Julia.

\end{abstract}

\input{blasfeo_body}

\end{document}

%% file: blasfeo_body.tex
%%%%%%%%%%%%%%%%%%%%%%%%%%%%%%%%
\section{Introduction} \label{sec:intro}
%%%%%%%%%%%%%%%%%%%%%%%%%%%%%%%%

This paper describes the implementation of a standard BLAS application programming interface (API) in the BLASFEO framework~\cite{Frison2018}, and investigates its use in scientific programming languages.

BLASFEO (Basic Linear Algebra Subroutines For Embedded Optimization) is an open-source dense linear algebra (DLA) library~\cite{BLASFEO2016} aimed at providing high-performance implementations of BLAS- and LAPACK-like routines for use in embedded optimization and small-scale high-performance computing in general~\cite{Ferreau2017, Frison2015a, Jerez2014}.
One of the most remarkable features of BLASFEO is the definition of a novel interface for DLA routines, where high-performance implementations of the computing routines operate on a performance-optimized matrix format (denoted panel-major, and resembling the packed format of the internal memory buffers used in many optimized BLAS implementations~\cite{Goto2008, Zee2016}), and packing and unpacking routines are exposed to the user.
From now on in this paper, this interface is indicated as the BLASFEO API.
The BLASFEO API allows one to significantly reduce the overhead of DLA routines, especially for small sized matrices, in which case the quadratic cost of packing and unpacking data is not negligible with respect to the cubic cost of the DLA operation itself.
However, the BLASFEO API is not compatible with the standard DLA interfaces.

BLAS~\cite{Lawson1979} and LAPACK~\cite{Anderson1999} are the de facto standard DLA interfaces, and have been so for several decades.
They are extremely widespread and form the basis of the major part of numerical and scientific software, and in this they represent a very successful example of combining high-performance and portability.
Several highly optimized implementations exist, both open-source (e.g. ATLAS~\cite{Whaley1999}, OpenBLAS~\cite{OpenBLAS2011}, BLIS~\cite{Zee2015}) and proprietary (e.g. MKL~\cite{MKL}).
From now on in this paper, the BLAS and LAPACK standard interface is indicated as the BLAS API.

The aim of this research work is to investigate the addition of a BLAS API to the BLASFEO framework, alongside the original BLASFEO API.
By analogy with an optimization problem, it can be said that the objective is unchanged: attaining the highest performance for the routines of a DLA library, in the case of matrices fitting in cache.
However, there are two more constraints, one hard and one soft: the strict compliance with the BLAS API, and the desire of reusing as much code as possible between the BLAS API and the BLASFEO API.
As it happens in an optimization problem, the introduction of additional constraints can lower the optimal value attained by the objective.

The numerical experiments in Section~\ref{sec:perf} show that, compared to the BLASFEO API, the performance loss for the BLAS API of BLASFEO ranges between negligible up to 20\% depending on the routine, matrix size and architecture, with an average of about 10-15\%.
For the matrix sizes of interest, this still puts the BLAS API of BLASFEO solidly ahead of (the single-threaded version of) other optimized BLAS API implementations (both open-source and proprietary), with only a handful of cases where it is roughly on par with them.
The addition of multi-threaded capabilities to the BLASFEO framework is outside the scope of the current paper\footnote{Note that one strategy to get multi-threaded linear algebra routines for relatively large matrices is to use libraries such as PLASMA~\cite{Buttari2009} or libflame~\cite{Zee2009}, which employ a task scheduler with sequential BLAS and LAPACK for tasks on modest sized tiles. Thanks to its good performance for small matrices, BLASFEO is an excellent choice to provide the sequential BLAS and LAPACK routines.}. %, but it is definitely of interest for future investigations.
%Furthermore, in the case of small matrices, the multi-threaded version of other BLAS API implementations performs worse than the single-threaded counterparts.
%Therefore, only the single-threaded version of DLA libraries is considered in this paper.
%Furthermore, it should be noted that 
The comparisons in this paper are limited to standard BLAS API library implementations of DLA routines.
Therefore alternative approaches for small-scale DLA routines such as code generation \cite{Houska2011, Heinecke2016, Spampinato2018}, C++ templates \cite{blaze,eigen-hp} or specialized compilers \cite{Spampinato2016} are not considered.
Likewise, approaches based on batched BLAS~\cite{Masliah2016} are also not considered.

The constraint about strict compliance with the BLAS API has important consequences on the algorithmic choices, as discussed in Section~\ref{sec:alg} taking {\tt dgemm} as an example.
In particular, the two extremes of working natively in column-major on all input and output matrices, or alternatively packing/unpacking them all in BLASFEO native panel-major format, both have major performance flaws.
Two better algorithmic variants are identified, which only pack some matrices (panels from the left factor $A$, and additionally the right factor $B$ for one variant) while operating natively in column-major for all other matrices.
The need to perform packing `on-line', within each routine call, is a major change compared to the original BLASFEO API implementation scheme.
However, if carefully performed and implemented as discussed in Section~\ref{sec:alg}, this packing does not introduce excessive overhead, and it even represent the tool now used to complete the BLASFEO API itself with the routine variants disregarded in the original implementation scheme in~\cite{Frison2018} (such as the `TN' and `TT' versions of {\tt dgemm}, where the left factor matrix $A$ is transposed).
A further consequence of the compliance with the BLAS API is the need to stick to a destructive routine interface, whereas the BLASFEO API adds an additional matrix argument reserved for the output and therefore it is not destructive.
In many applications, a non-destructive interface can avoid the need to perform additional matrix copies, which in the case of small matrices are comparably costly.

The desire of reusing as much code as possible affects the details of implementation, as discussed in Section~\ref{sec:impl}.
In particular, in the BLASFEO framework linear algebra kernels are defined as the innermost loop coded as a stand-alone routine, and are implemented in a modular fashion by exploiting the flexibility of the assembly language.
These kernels are split into smaller sub-routines which perform elementary operations such as loading or storing of a matrix to/from registers, multiplication of two panels, or factorization of a register-resident sub-matrix.
The sub-routines make use of a custom function calling convention, which allows one to pass data in floating point (FP) registers and reduces the calling overhead.
This framework naturally allows the implementation of linear algebra kernels operating on matrix arguments with different storage formats (in this case, column-major and panel-major), by simply `gluing together' sub-routines operating on different storage formats.
A high-degree of code reuse is attained, since the actual operations are implemented in the shared sub-routines, while the tailored kernels are simply a sequence of calls to the sub-routines themselves.
More details are found in Section~\ref{sec:impl:kernel}.

One of the most particular features of the BLASFEO kernels with respect to other optimized BLAS implementations like OpenBLAS or BLIS is the fact that the BLASFEO kernels operate on the left and right factors packed into panels of identical height $p_s$.
In case of non-square kernels, this implies that multiple panels are swept at once from the left or the right factor, as opposed to a single higher panel in other optimized BLAS implementations.
In the BLASFEO API, this is a requirement, since the same panel-major matrix must be freely employed as left or right factor (and as output matrix too).
In the BLAS API, this feature is maintained, mainly to maximize the code shared with the BLASFEO API.
Even if the sweeping of multiple shorter panels is possibly slightly sub-optimal compared to the sweeping of a single higher one (due to cache associativity), in practice the BLASFEO API shows excellent performance on all tested architectures~\cite{Frison2018}.
Conversely, in the BLAS API this feature turns out to be an intrinsic advantage in the implementation of routines with symmetric nature (like e.g. {\tt dsyrk} or {\tt dpotrf}), since the mathematical symmetry of the operation can be carried over to the working memory buffers, which are shared between left and right factors, reducing the amount of packing.
More details are found in Section~\ref{sec:impl:blas}.

Other design choices compatible with the BLAS API compliance constraint are carried on unchanged from the BLASFEO API.
No cache blocking is performed, and level 3 routines are implemented as nested triple loops, whose innermost loop is coded as a stand-alone routine, the kernel.
The modular nature of the assembly kernels allows one to easily code kernels tailored to the different DLA routines and fully exploiting vectorization also in the corner cases.
LAPACK routines like factorizations are implemented similarly to the level 3 BLAS routines (i.e., as nested triple loops whose innermost loop is a tailored kernel), and not on top of them.
As in the case of the BLASFEO API, it can be said that the block size in blocked LAPACK routines shrinks to the kernel size, while unblocked routines coincide with the tailored kernels and operate on fixed-size register-resident sub-matrices.

Since BLAS and LAPACK are at the foundation of modern scientific programming languages, Section~\ref{sec:appl} of this paper evaluates the use of the BLAS API of BLASFEO in Octave, Python SciPy and Julia, and investigates to what extent a DLA library optimized for small matrices can speedup computations in high-level languages.
These languages are chosen because of their widespread use and their open-source nature.
As such, by default, they are distributed with open-source BLAS and LAPACK implementations such as OpenBLAS.
Being open-source software too, BLASFEO can contribute to improve the performance of such languages possibly more than proprietary BLAS and LAPACK implementations do, while keeping a fully open-source software stack.

%%%%%%%%%%%%%%%%%%%%%%%%%%%%%%%%
\section{Comparison of {\tt dgemm} algorithms for small matrices} \label{sec:alg}
%%%%%%%%%%%%%%%%%%%%%%%%%%%%%%%%

This section proposes and compares several algorithms to implement the BLAS routine {\tt dgemm} in the BLASFEO framework, with focus on optimizing performance for small matrices.
Similar findings apply in the implementation of other BLAS and LAPACK routines. %versions of the {\tt dgemm} routine.

%%%%%%%%
\subsection{Notation} \label{sec:alg:notation}
%%%%%%%%

The general matrix-matrix multiplication routine {\tt dgemm} is certainly the most famous and important BLAS routine, and therefore it is used here as an example.
%For readability, it is assumed that it implements the operation $C \Leftarrow A \cdot B + C$, where transposition options
In order to keep the notation more readable, the {\tt dgemm} routine is here assumed to implement the operation $C \Leftarrow A \cdot B + C$.
Therefore a call in the C language is assumed to be in the form (where the function arguments of the transposition options, the scalar factors $\alpha$ and $\beta$ and the leading size of the matrices are omitted)
\begin{verbatim}
void dgemm_(int *m,
            int *n,
            int *k,
            double *A,
            double *B,
            double *C);
\end{verbatim}
where $m\times n$ is the size of the result matrix and $k$ is the length of each dot product, $A$ and $B$ denote the left and right matrix factors, and $C$ denotes the result matrix.
In the standard BLAS API, all matrices $A$, $B$ and $C$ are stored in column-major format.

It is well known that in double precision all level 3 BLAS and many LAPACK routines such as factorizations can be implemented using {\tt dgemm}~\cite{Kaagstroem1998,Goto2008a}.
In the BLASFEO framework, this is exploited in the design of kernels for level 3 BLAS and LAPACK routines, which all share the common structure (where for compactness the function arguments of the scalar factors $\alpha$ and $\beta$ and the size of the buffers are omitted)
\begin{verbatim}
void kernel_dgemm(int k_max,
                  double *A_k,
                  double *B_k,
                  double *C_k,
                  double *D_k);
\end{verbatim}
with the only exception of {\tt dtrsm} kernels, which take as an extra argument the triangular matrix of the system to be solved
\begin{verbatim}
void kernel_dtrsm(int k_max,
                  double *A_k,
                  double *B_k,
                  double *C_k,
                  double *D_k,
                  double *E_k);
\end{verbatim}
Here $k_{\max}$ is the length of the innermost loop, $A_k$ and $B_k$ are the sub-matrices of the left and right factors $A$ and $B$, $C_k$ and $D_k$ are the read and written sub-matrices of the result matrix $C$, and $E_k$ is the triangular sub-matrix of the system to be solved in the case of {\tt dtrsm}.
Note the subscript $_k$, which denotes that these quantities refer to the kernel sub-matrices (as opposed to the routine matrices, which are denoted without the subscript $_k$).
These kernel sub-matrices may be stored in column-major or in panel-major formats depending on the kernel implementation.

%%%%%%%%%%%%%%%%
\subsection{Algorithmic variants} \label{sec:alg:var}
%%%%%%%%%%%%%%%%

This section describes five algorithmic variants for the implementation of the standard BLAS API {\tt dgemm} routine leveraging the existing BLASFEO framework.
It is assumed that the high-performance backend of BLASFEO is employed, such that the panel-major matrix format is the memory layout of the BLASFEO matrices {\tt blasfeo\_dmat}.

%%%%
\paragraph{Algorithm `A'}
%%%%

It is straightforward to add a standard BLAS API {\tt dgemm} routine in the BLASFEO framework by only using existing BLASFEO routines for packing, unpacking and the BLASFEO API {\tt blasfeo\_dgemm\_nt} routine.
This is the algorithmic variant named `A'.

In more detail, the routine {\tt blasfeo\_memsize\_dmat} returns the amount of memory that should be provided to create a {\tt blasfeo\_dmat} matrix.
The memory for all arguments $A_b$, $B_b$, $C_b$ of the BLASFEO {\tt dgemm\_nt} routine is allocated dynamically and aligned to cache line boundaries, and the {\tt blasfeo\_dmat} matrices $A_b$, $B_b$, $C_b$ are created with the {\tt blasfeo\_create\_dmat} routine.
The whole input matrices $A$, $B$, $C$ are packed into the {\tt blasfeo\_dmat} matrices $A_b$, $B_b$, $C_b$ using the routines {\tt blasfeo\_pack\_dmat} and {\tt blasfeo\_pack\_tran\_dmat}.
At this point, the {\tt blasfeo\_dgemm\_nt} routine can be employed to compute the matrix product.
Finally, the result is stored back in the matrix $C$ using the routine {\tt blasfeo\_unpack\_dmat}.
Overall, the amount of packing is of $mk+nk+2mn$ matrix elements.

All parametric variants of {\tt dgemm} can be handled by transposing while packing the factor matrices $A$ and $B$.
The `NT' variant of the {\tt dgemm} routine is chosen as the computational kernel, as this is the best-performing variant in the BLASFEO API.

This algorithm is easy to implement without coding any new linear algebra routine or kernel, but building entirely on top of the existing BLASFEO API.
However, it unnecessarily performs packing and unpacking of $C$ (which is not needed for performance), and it does not exploit the order of linear algebra loops within the BLASFEO {\tt dgemm} routine.
Furthermore, for very small matrices the dynamic memory allocation can severely affect performance.

%%%%
\paragraph{Algorithm `B'}
%%%%

By opening the lid on the internals of the BLASFEO API {\tt dgemm} routine, it is possible to obtain better performing algorithms.
The algorithmic variant named `B' operates natively on the matrix $C$ in column major, reducing by a factor 2 (in the case of square matrices) the amount of packing, amounting to $mk+nk$ matrix elements.
This requires the implementation of new kernels, in which $A_k$ and $B_k$ are in panel-major and $C_k$ and $D_k$ are in column-major.
The factor matrices $A$ and $B$ can still be freely transposed while packed, and therefore also in this algorithmic variant only the `NT' variant of the {\tt dgemm} kernel is employed.

Additionally, the fact that the {\tt dgemm} routine is implemented as a triple-loop, with the outer loop over $m$, is exploited.
At each iteration of the loop over $m$, a sub-matrix of $A$ is determined, where the number of rows is fixed and equal to the height $m_r$ of the computational kernel, and the number of columns is equal to $k$.
The two innermost loops repeatedly stream this sub-matrix of $A$ while sweeping over the entire matrix $B$.
%The two innermost loops sweep over sub-matrices of $A$ where the number of rows is fixed and equal to the height $m_r$ of the computational kernel, and the number of columns is equal to $k$.
%These sub-matrices of $A$ are accessed only once.
Therefore, $A$ can be packed $m_r$ rows at a time in the outermost loop, in a memory buffer of size $\mathcal O(k)$.
Regarding $B$, it is packed as in algorithmic variant `A', outside the triple loop.

The main drawbacks of this algorithm are the overhead due to the packing of both $A$ and $B$, and the usage of dynamic memory allocation, which adversely affect performance for small matrices.
Dynamic memory allocation is employed because the buffer size for the $B$ matrix grows as $\mathcal O(nk)$, and in the case of square matrices it quickly gets large with the matrix size.
Therefore, its size can easily exceed the stack size, which can be rather small in the case of embedded devices.

%%%%
\paragraph{Algorithm `C'}
%%%%

The aim of algorithmic variant `C' is to further reduce the amount of packing and to avoid the dynamic memory allocation and therefore to improve performance for small matrices.
In order to do so, this variant operates natively on both matrices $C$ and $B$, each stored in column-major order.
The matrix $A$ is still packed (and possibly transposed) as in algorithmic variant `B', in a memory buffer of size $\mathcal O(k)$.
The amount of packing reduces to $mk$ matrix elements, which is independent of $n$.
Since the matrix $B$ is used natively and therefore not explicitly transposed, the two variants `NN' and `NT' of the {\tt dgemm} kernels are employed (handling the {\tt dgemm} variants `NN' and `TN', and `NT' and `TT' respectively).

Automatic memory allocation (that allocates memory on the stack) is employed for the $A_k$ buffer, of size $\mathcal O(k)$, avoiding any dynamic memory allocation.
The main drawback of this algorithmic variant is that the column-major memory layout of $B$ adversely affects cache usage, especially in the `NT' and `TT' {\tt dgemm} variants where $B$ is accessed across columns.

%%%%
\paragraph{Algorithm `Ct'}
%%%%

Algorithmic variant `Ct' is analogous to algorithmic variant `C', and loosely speaking it can be considered its transposed (hence the `t' in the name).
Compared to algorithmic variant `C', the variant `Ct' packs the $B$ matrix instead of the $A$ matrix, the order of outer loops is swapped (outermost over $n$), and finally the kernel sizes and types are transposed\footnote{For example, {\tt kernel\_dgemm\_tt\_4x12\_libc4cc} for algorithmic variant `Ct' is the equivalent of {\tt kernel\_dgemm\_nn\_12x4\_lib4ccc} for algorithmic variant `C'.}.
In particular, the matrix $A$ is used natively and therefore not explicitly transposed, and the two variants `NT' and `TT' of the {\tt dgemm} kernels are employed, handling the {\tt dgemm} variants `NN' and `NT', and `TN' and `TT', respectively.
The number of matrix elements packed by variant is $nk$, which is independent of $m$.
The $B_k$ buffer, of size $\mathcal O(k)$, is allocated on the stack using automatic memory allocation.

%%%%
\paragraph{Algorithm `D'}
%%%%

Algorithmic variant `D' natively operates on all matrices in column-major, and therefore it does not need any memory allocation nor packing.
As long as all matrix arguments are accessed natively and therefore not explicitly transposed, this variant requires a different kernel for each of the {\tt dgemm} variants `NN', `NT', `TN', `TT'.
In the column-major format, the variants `TN' and `TT' (which are the variants where the matrix $A$ has to be transposed in the multiplication) are less friendly to vectorization, since contiguous $A$ elements are used at different iterations of the loop over $k$.
Therefore, on modern architectures, native kernels for these variants perform worse, with the performance penalty increasing with the vector size.
In the case of the `TT' variant, a workaround consists in implementing the kernel based on a kernel for the `NN' variant as $C \Leftarrow \beta \cdot C + \alpha \cdot (B \cdot A)^T$.
In the case of the `TN' variant the same workaround cannot be employed, as in the transposition $(A^T \cdot B)^T = (B^T \cdot A)$ the left factor is still transposed.
Therefore, the implementation of this variant requires the explicit transposition of either matrix factor, and therefore it is better replaced by algorithmic variants `C' or `Ct', where the explicit transposition is combined with packing into a better performing matrix format.

Therefore, this algorithmic variant is only implemented in the {\tt dgemm} variants `NN', `NT' and `TT'.

%%%%%%%%%%%%%%%%
\subsection{Comparison of algorithmic variants} \label{sec:alg:comp}
%%%%%%%%%%%%%%%%

Table~\ref{tab:alg} summarizes the key features of the five algorithmic variants for the implementation of {\tt dgemm}, as described in Section~\ref{sec:alg:var}.
It reports the memory layout of $A_k$, $B_k$ and $C_k$ as employed in the computational kernel, and the number and type of computational kernels needed to compute all four {\tt dgemm} variants.
The table also reports the amount of memory used for internal buffers (and therefore excludes the memory needed to store the $A$, $B$ and $C$ arguments of {\tt dgemm}) and the type of memory allocation employed: automatic memory allocation on the stack, or dynamic memory allocation on the heap\footnote{Static memory allocation is not a viable option, as it would make the routines not thread-safe.}.
Finally, the number of matrix elements packed is also listed.

\begin{table}
\centering
\caption{Key features of algorithmic variants for the implementation of the BLAS routine {\tt dgemm}.
In the kernel names, the first 2 characters encode the transposition of $A$ and $B$ (`n' non-transposed, `t' transposed); the last 3 characters encode the memory layout of $A_k$, $B_k$ and $C_k$ (`c' column-major, `p' panel-major).
Note that in the BLAS API {\tt dgemm} implementation $D_k$ coincides with $C_k$.
Kernels denoted with the exponent $^*$ are not vectorization-friendly and generally perform poorly on modern architectures.}
\label{tab:alg}
\begin{tabular}{l||c|c|c|c|c|c|c}
%alg. & $A_k$ & $B_k$ & $C_k$ & mem. use & mem. alloc. & packing & kernels \\
alg. & $A_k$ & $B_k$ & $C_k$ & mem. use & alloc. & packing & kernels \\
\hline
\hline
`A' & p & p & p & $\mathcal O(mk+nk+mn)$ & heap & $mk+nk+2mn$ & nt\_ppp \\
`B' & p & p & c & $\mathcal O(nk)$ & heap & $mk+nk$ & nt\_ppc \\
`C' & p & c & c & $\mathcal O(k)$ & stack & $mk$ & \{nn, nt\}\_pcc \\
`Ct' & c & p & c & $\mathcal O(k)$ & stack & $nk$ & \{nt, tt\}\_cpc \\
`D' & c & c & c & $\mathcal O(1)$ & none & none & \{nn, nt, tn$^*$, tt\}\_ccc \\
\end{tabular}
\end{table}

%%%%%%%%%%%%%%%%
\subsubsection{Comparisons for square matrices} \label{sec:alg:comp:square}
%%%%%%%%%%%%%%%%

\begin{figure}[!t]
\centering
\subfloat[dgemm\_nn]{\includegraphics[width=0.45\linewidth]{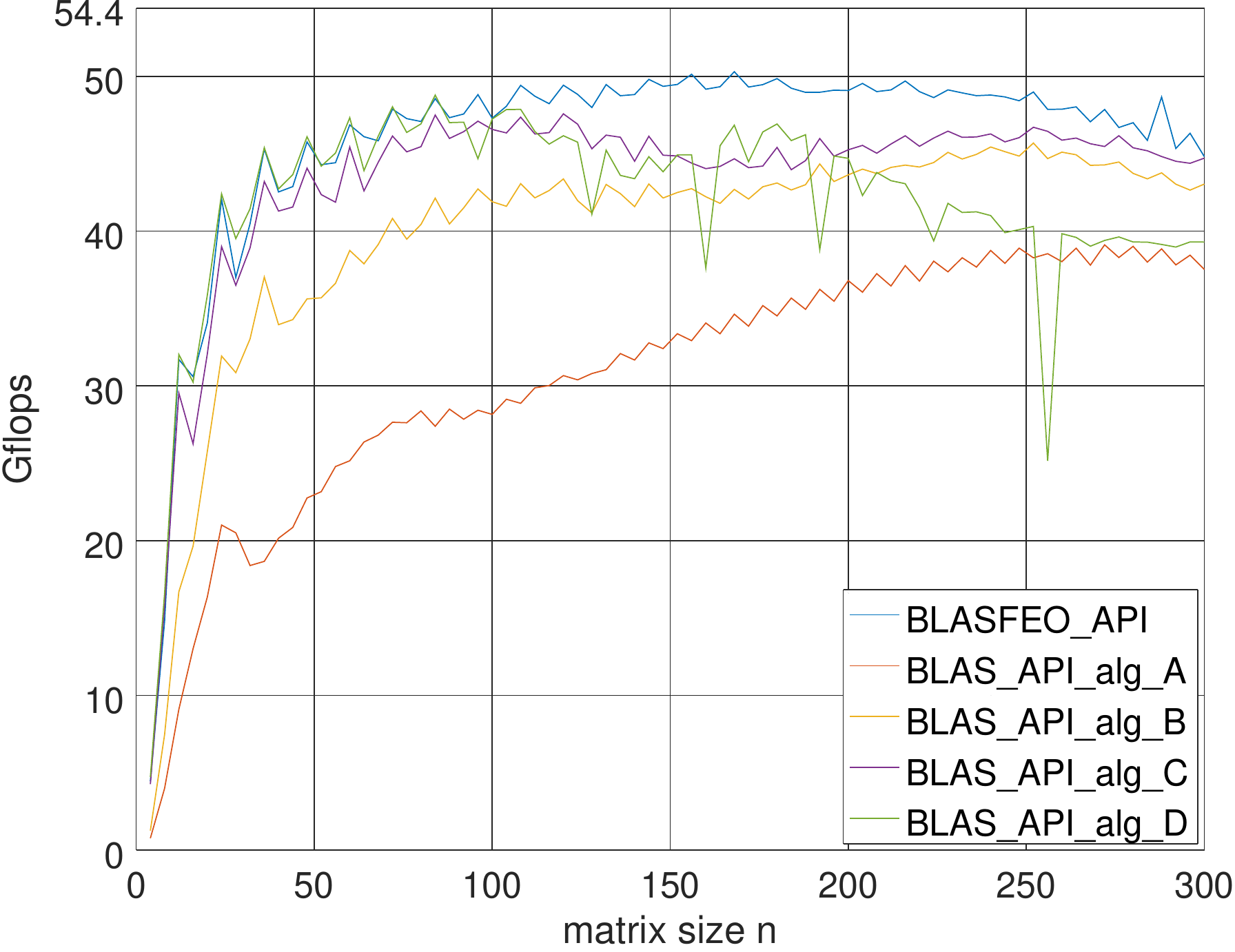} \label{fig:alg:haswell:dgemm_nn}} %\\
\subfloat[dgemm\_nt]{\includegraphics[width=0.45\linewidth]{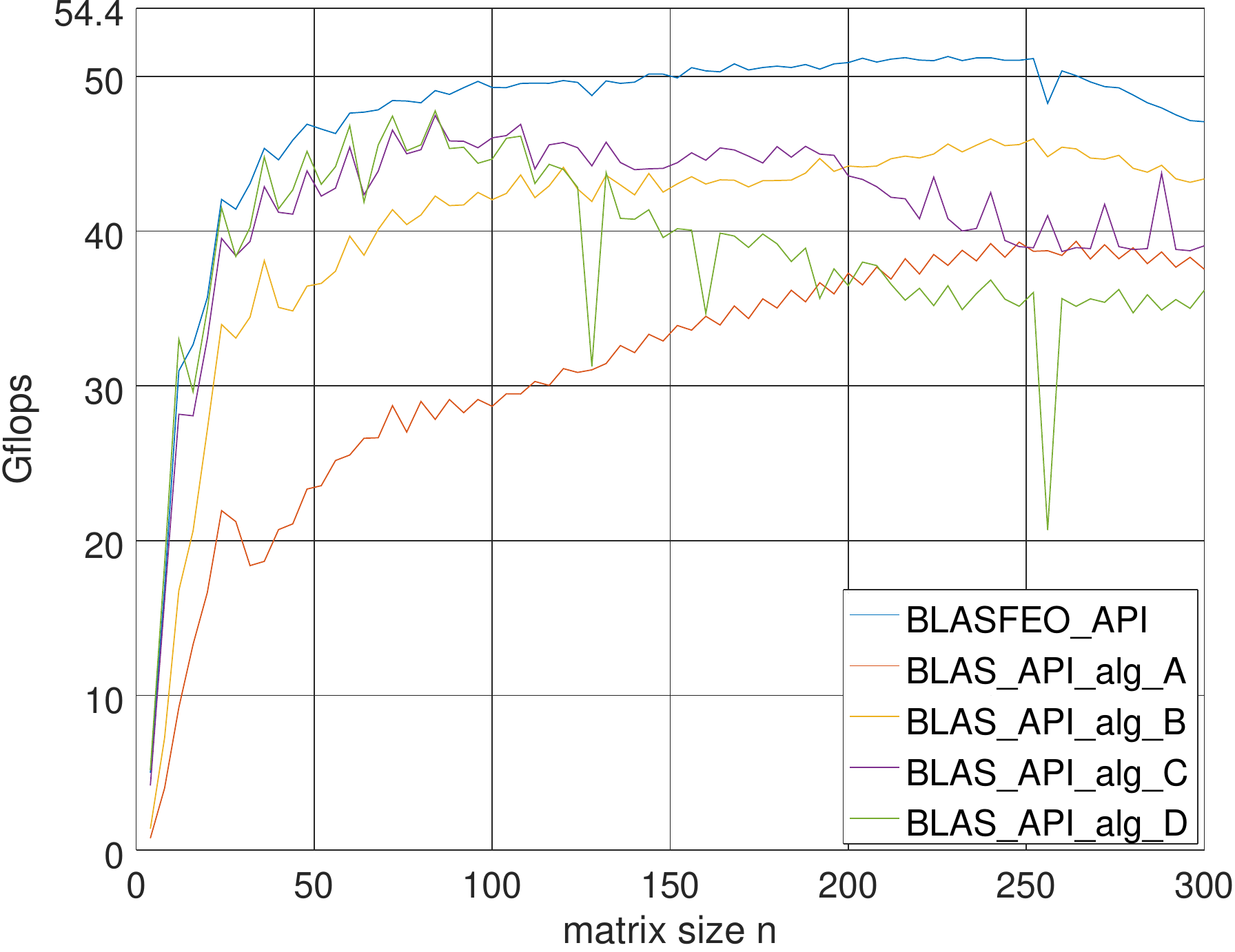} \label{fig:alg:haswell:dgemm_nt}} %\\
\caption{Comparison of algorithmic variants for the implementation of the BLAS routine {\tt dgemm} in the case of square matrices: performance on Intel Haswell Core i7 4810MQ.
Left picture: {\tt dgemm} with {\tt transa=`n'} and {\tt transb=`n'}.
Right picture: {\tt dgemm} with {\tt transa=`n'} and {\tt transb=`t'}.}
\label{fig:alg:haswell}
\end{figure}

In case of square matrices, algorithmic variants `C' and `Ct' are equivalent: therefore only variant `C' is considered here.
Figure~\ref{fig:alg:haswell} shows the performance plot for the four algorithmic variants `A', `B', `C' and `D', and for the `NN' and `NT' {\tt dgemm} variants, on an Intel Haswell Core i7 4810MQ.
The performance of the BLASFEO API routines is also added for comparison.

Algorithmic variant `A' shows a large overhead, especially for small matrices, being at least two times slower than the BLASFEO API routines for matrices smaller than 50.
Algorithmic variant `B' significantly reduces overhead, and it always performs better than variant `A' in the matrix size range of interest.
Both variants perform similarly across all four {\tt dgemm} variants, as they employ a single computational kernel and the transpositions are performed while packing at no extra cost.

Algorithmic variant `C' further reduces overhead for matrices smaller than about 200, approaching the performance of the BLASFEO API routines.
However, when the matrix $B$ is employed transposed, the performance of this variant degrades when L1 cache size is exceeded.
Finally, compared with algorithmic variant `C', variant `D' slightly reduces overhead only for very small matrices, and its performance degrades more when L1 cache size is exceeded.
Software prefetching can mitigate but not solve these performance degradations.
%Furthermore, in the case the matrix $A$ is employed transposed, the kernels {\tt tn\_ccc} and {\tt tt\_ccc} are not vectorization-friendly, and therefore perform poorly on modern architectures.

%%%%%%%%%%%%%%%%
\subsubsection{Considerations for rectangular matrices} \label{sec:alg:comp:skinny}
%%%%%%%%%%%%%%%%

In the case of rectangular matrices, the data in Table~\ref{tab:alg} leads to some additional considerations.
Section~\ref{sec:skinny} in the appendix contains some benchmarks in the case of both square and skinny matrices.

When the result matrix $C$ is rectangular, algorithmic variants `C' and `Ct' are no longer equivalent.
They perform the same number of flops and they have the same memory usage, but the amount of packing is different.
In particular, algorithmic variant `C' requires packing the $mk$ elements of the matrix $A$ (amount independent of $n$), and therefore it is advantageous when $C$ is a fat matrix with $m<n$.
Conversely algorithmic variant `Ct' requires packing the $nk$ elements of the matrix $B$ (amount independent of $m$), making it the best choice when $C$ is a thin matrix with $m>n$.

When $m$ and $n$ are both very small, for any value of $k$ the cost of packing any of the matrices $A$ or $B$ cannot be well amortized over the number of flops, namely $2mnk$.
Therefore, in this case algorithmic variant `D' (if available for the {\tt dgemm} variant) is likely the best choice.

More generally, the case of square matrices maximises the ratio between number of flops and number of matrix elements, and thus minimizes the impact of packing on the computation time.
Therefore, in the case of rectangular matrices, algorithmic variants involving less packing become more advantageous compared to the findings in Section~\ref{sec:alg:comp:square}.

%%%%%%%%%%%%%%%%
\subsection{Proposed approach} \label{sec:alg:prop}
%%%%%%%%%%%%%%%%

The implementation of a generic DLA routine in the BLAS API of BLASFEO makes use of algorithmic variant `C' for moderately sized matrices, and algorithmic variant `B' for larger matrices.
Other algorithmic variants can possibly supplement these variants in special cases.
As an example, in the case of non-symmetric routines like {\tt dgemm}, algorithmic variant `Ct' can replace algorithmic variant `C' in the case of thin result matrices.
Algorithmic variant `D' (if a high-performance implementation is available, like for the `NN', `NT' and `TT' variants, but not for the `TN' variant of {\tt dgemm}) can be employed in the case of very small matrices.

The optimal switching point differs for different linear algebra routines and architectures.
In the examples in Figure~\ref{fig:alg:haswell}, the switching point between algorithmic variants `B' and `C' is for $m=n=k\approx400$ for the `NN' {\tt dgemm} variant\footnote{Switching point not visible in Figure~\ref{fig:alg:haswell:dgemm_nn}.} and for $m=n=k\approx200$ for the `NT' {\tt dgemm} variant.
The switching point can be manually limited to a maximum value using a compilation flag to ensure that the memory employed by the algorithmic variant `C' does not exceed the stack size on the specific OS and architecture; algorithmic variant `B' employs dynamic memory allocation and therefore safely works for much larger matrix sizes.

In case of the {\tt dgemm} routine, the proposed generic approach requires the implementation of three kernel types: {\tt nt\_ppc}\footnote{The kernel naming convention is explained in the caption of Table~\ref{tab:alg}.} for algorithmic variant `B' and {\tt nn\_pcc} and {\tt nt\_pcc} for algorithmic variant `C'.
These add to the existing kernel types {\tt nn\_ppp} and {\tt nt\_ppp} used in the BLASFEO API.
Regarding the supplementary variants, algorithmic variant `Ct' requires the implementation of the kernels {\tt nt\_cpc} and {\tt tt\_cpc}, which are internally based on the kernels for algorithmic variant `C' thanks to the matrix multiplication transposition rule (so no intrinsically new kernel type is required).
Algorithmic variant `D' requires the implementation of the kernels {\tt nn\_ccc}, {\tt nt\_ccc} and {\tt tt\_ccc}, out of which the first two are intrinsically new kernels, while the {\tt tt\_ccc} kernel is internally based on the {\tt nn\_ccc} kernel, again thanks to the matrix multiplication transposition rule.
Section~\ref{sec:impl:kernel} explains how all these kernels are coded in assembly in a modular fashion.

%%%%%%%%%%%%%%%%%%%%%%%%%%%%%%%%
\section{Details of implementation} \label{sec:impl}
%%%%%%%%%%%%%%%%%%%%%%%%%%%%%%%%

The BLAS API of BLASFEO has the same aim as the BLASFEO API: providing a linear algebra library performance-optimized for matrices fitting in cache.
Therefore, at the high level many implementation choices are unchanged: lack of cache blocking, ordering of loops, tailoring of custom assembly kernels for each linear algebra routine.

As discussed in Section~\ref{sec:alg:var}, the main difference is the packing of data: in the BLASFEO API this is split from the linear algebra routines, while in the BLAS API this has to happen within the linear algebra routines.

%%%%%%%%%%%%%%%%
\subsection{Implementation of BLAS and LAPACK routines} \label{sec:impl:blas}
%%%%%%%%%%%%%%%%

As a general rule, in the BLAS API of BLASFEO all level 3 BLAS and LAPACK routines are implemented using the algorithmic variants `C' (for small matrices) and `B' (for larger matrices) described in Section~\ref{sec:alg:var}.
The exact switching point depends on the specific routine and computer architecture, but typically in the case of square matrices it is for matrix sizes in the range 64 to 400.

This section describes the details of some implementation choices for selected BLAS and LAPACK routines in the BLAS API of BLASFEO.

%%%%%%%%
\subsubsection{{\tt dgemm}} \label{sec:impl:blas:gemm}
%%%%%%%%

The implementation of {\tt dgemm} was discussed in details in Section~\ref{sec:alg}.

%%%%%%%%
\subsubsection{{\tt dsyrk}} \label{sec:impl:blas:syrk}
%%%%%%%%

In the case of {\tt dsyrk}, the $A$ and $B$ matrices of {\tt dgemm} coincide.
However generally, in the {\tt dsyrk} implementation in optimized BLAS libraries, the memory buffers for $A_k$ and for $B_k$ cannot coincide, since in the case of non-square kernels they are packed with different panel sizes.

Conversely, in the BLASFEO framework, the panel sizes for $A_k$ and for $B_k$ are equal even for non-square kernels, since this is a requirement to make the panel-major the default matrix format for all BLASFEO API routines.
Therefore, in the implementation of {\tt dsyrk}, the same memory buffer is employed for $A_k$ and $B_k$.
For the `B' algorithmic variant, the $A$ matrix is packed at once into $B_k$, and then $A_k$ simply points to a panel in $B_k$, without re-packing it.
This reduces the amount of necessary packing by a factor 2.
For the `C' algorithmic variant, the kernels used to compute the diagonal blocks have both $A_k$ and $B_k$ (which coincide) available in panel-major, and therefore use the faster {\tt ppc} kernel variants.
This gives the BLASFEO framework an inherent advantage over other optimized implementations of {\tt dsyrk}.

%%%%%%%%
\subsubsection{{\tt dtrsm}} \label{sec:impl:blas:trsm}
%%%%%%%%

In the BLASFEO framework, the {\tt dtrsm} routines are implemented using custom kernels, which consist of the {\tt dgemm} loop followed by a custom and fully-unrolled sub-routine performing the solution of a fixed-size register-resident triangular system.
In the implementation of these kernels, there is a fundamental difference between left and right variants of the {\tt dtrsm} routine.
Before proceeding, it is useful to note that in BLASFEO optimal kernels are generally square or tall, meaning that $m_r\geq n_r$, where $m_r \times n_r$ is the kernel size (i.e. the size of the result sub-matrix computed by the kernel).

Left variants of the {\tt dtrsm} kernel require one to solve a few ($n_r$) large ($m_r \times m_r$) triangular systems of equations, each computing an independent $m_r \times 1$ sub-column of $D$.
The implementation of this operation is not vector friendly, as the vectorization happens in the computation of a single sub-column of $D$.
The triangular matrix is the left factor and therefore accessed with vector loads, requiring masking at the edges and performing useless flops.
%Furthermore, solution requires a large $O(m_r \times m_r)$ number of dependent operations to compute one sub-column of $C$, and the number of sub-columns, $n_r$, is small and therefore it gives little possibility to extract instruction-level parallelism. ?????
Furthermore, the solution of larger $m_r \times m_r$ triangular systems requires bigger and more complex custom routines, and it implies that less flops are processed by the efficient {\tt dgemm} loop and more by these custom routines.

Right variants of the {\tt dtrsm} kernel require one to solve many ($m_r$) small ($n_r \times n_r$) triangular systems of equations, each computing an independent $1 \times n_r$ sub-row of $D$.
The implementation of this operation is vector friendly.
The computation of a single sub-row of $D$ is not vectorized, but $n_r$ of such sub-rows can fit in a vector register (where $n_r$ is the size of the vector register) and be processed in parallel using vector instructions.
The triangular matrix is at the right factor and its elements are broadcast, and therefore no useless flops or memory loads need to be performed.
Furthermore, the solution of smaller $n_r \times n_r$ triangular systems requires smaller and simpler routines, and more flops are processed by the more efficient {\tt dgemm} loop.

Given the above considerations, which get increasingly important as the SIMD width increases, in BLASFEO only kernels tailored to the right variants of {\tt dtrsm} are implemented.
The left variants {\tt dtrsm} are implemented using transposition and the kernels for the right {\tt dtrsm} variant.
E.g. the triangular system solution $A^{-1} B$ (with $A$ a triangular matrix) is implemented as $(B^T A^{-T})^T$, with the transposition of $B$ performed at no extra cost, since it has to be packed anyway being now the left factor.

%%%%%%%%
\subsubsection{{\tt dtrmm}} \label{sec:impl:blas:trmm}
%%%%%%%%

In the implementation of {\tt dtrmm}, considerations analogous to {\tt dtrsm} apply.

%%%%%%%%
\subsubsection{{\tt dpotrf}} \label{sec:impl:blas:potrf}
%%%%%%%%

In the implementation of {\tt dpotrf}, considerations analogous to {\tt dsyrk} apply.
In fact, also in this case the symmetry of the operation can be exploited to reduce the amount of required data packing.
This gives to the BLASFEO framework an inherent advantage over other optimized implementations of {\tt dpotrf}.

%%%%%%%%
\subsubsection{{\tt dgetrf}} \label{sec:impl:blas:getrf}
%%%%%%%%

The implementation of the {\tt dgetrf} routine is rather different compared to other routines in BLASFEO.
At the high level, it is computed by column-blocks, as opposed to row-blocks of the other routines.
Internally, however, the column-blocks stored in column-major are packed and transposed into row-blocks store in panel-major, which is the format used for computing.
This enables using the more efficient {\tt dgemm\_nt} kernel, and to perform triangular system solutions with the triangular matrix as the right factor.
Furthermore, in the case of tall kernels this enables reusing the same L1-resident tall factor across calls to the {\tt dtrsm} and {\tt dgemm} kernels operating at the same iteration of the outer loop.

%%%%%%%%%%%%%%%%
\subsection{Custom function calling convention: modular assembly kernels} \label{sec:impl:kernel}
%%%%%%%%%%%%%%%%

The linear algebra implementation as proposed in Sections~\ref{sec:alg:prop} and \ref{sec:impl:blas} requires the implementation of a large number of custom kernels, which generally share a large fraction of the code.
In BLASFEO, these kernels are written in assembly (with the exception of the generic target, coded in C).
The flexibility of the assembly language is exploited to code the kernels in a modular fashion enhancing code reuse, and yet attaining high performance.

%%%%%%%%
\subsubsection{Modular kernel design} \label{sec:impl:kernel:mod}
%%%%%%%%

In the BLASFEO framework, a kernel is a function with global scope, implementing the innermost loop of the linear algebra routines, and operating on a sub-matrix of the result matrix $D_k$.
The size $m_r\times n_r$ of the kernel is the size of the sub-matrix of the result matrix $D_k$ computed by the kernel itself.

A kernel algorithm is split into more elementary operations, each of which is coded in a separate function (named inner function) with module scope.
Therefore, the kernels themselves do not contain code for any numerical operation, and they are simply implemented as a sequence of calls to inner functions.
Typically, the same inner function is called by many different kernels, ensuring a high degree of code reuse.
For example, the BLASFEO API kernel for the `NT' version of {\tt dgemm}, and the BLAS API {\tt ppc} kernel for the `L' version of {\tt dpotrf} are decomposed into inner functions as
\begin{verbatim}
GLOB_FUN_START(kernel_dgemm_nt_4x4_lib4444)
PROLOGUE

ZERO_ACC

// set some GP registers ...
CALL(inner_kernel_dgemm_nt_4x4_lib44)

// set some GP registers ...
CALL(inner_scale_ab_4x4_lib4)

// set some GP registers ...
CALL(inner_store_4x4_lib4)

EPILOGUE
RETURN
GLOB_FUN_END(kernel_dgemm_nt_4x4_lib4444)

GLOB_FUN_START(kernel_dpotrf_nt_l_4x4_lib44cc)
PROLOGUE

ZERO_ACC

// set some GP registers ...
CALL(inner_kernel_dgemm_nt_4x4_lib44)

// set some GP registers ...
CALL(inner_scale_m11_4x4_libc)

// set some GP registers ...
CALL(inner_edge_dpotrf_4x4_libc)

// set some GP registers ...
CALL(inner_store_l_4x4_libc)

EPILOGUE
RETURN
GLOB_FUN_END(kernel_dpotrf_nt_l_4x4_lib44cc)
\end{verbatim}
Here, each `4' after `lib' means that the corresponding matrix argument is in panel-major layout with panel size $p_s=4$, while each `c' after `lib' means that the corresponding matrix argument is in column-major layout.
For example, {\tt inner\_kernel\_dgemm\_nt\_4x4\_lib44} takes for its two matrix arguments $A_k$ and $B_k$ two panels of size $p_s=4$, while {\tt inner\_store\_l\_4x4\_libc} stores the lower triangular of a $4\times 4$ sub-matrix in column-major format.
Note that the two functions share the same innermost loop, coded in the inner function {\tt inner\_kernel\_dgemm\_nt\_4x4\_lib44}.
%`PROLOGUE' and `EPILOGUE' perform the operations needed by the standard function calling convention, like storing and loading callee-saved registers.

Generally, in the standard function calling conventions it is not possible to pass data in FP vector registers.
Therefore, if the proposed approach is coded using C code, similarly to the $\nu$-BLACs in {\tt  \textsc{LGen}}~\cite{Spampinato2016} and {\tt  \textsc{SLinGen}}~\cite{Spampinato2018}, the inner functions need to be inlined, otherwise the overhead would be unacceptably large.
Even in the case where all inner functions are inlined, the compilers seem to struggle to find a globally optimal register allocation strategy across the numerous inlined functions.
In any case, inlining makes only the source code modular, but not the object code, whose size grows considerably.

The approach proposed in BLASFEO employs assembly code instead.
The inner functions have module scope, and therefore they do not need to follow the standard function calling convention.
In particular, some FP registers are designated to pass numerical data between several inner functions in a consistent way.
Similarly, some general purpose (GP) registers are designated to pass arguments (like loop sizes and pointers) from the kernels to the inner functions.
These registers are different from the registers used to pass arguments in the standard function calling convention (which remain unmodified), ensuring that inner functions can be called in any order.

In this way, both modularity and performance are attained.
Overhead is kept at a minimum, as it only consists of setting some GP registers and performing some unconditional jumps for the inner functions calls.
Optionally, BLASFEO also gives the possibility to treat the entire inner functions as macros and inline them, removing these unconditional jumps at the expense of an increase in object code size.

In BLASFEO, the {\tt dgemm} inner kernels are the backbone of the kernels for all level 3 BLAS and LAPACK routines.
For performance reasons, these inner kernels are only implemented for the `NN' and `NT' {\tt dgemm} variants, where the left factor is non-transposed.
Conversely, and remembering that matrix elements are stored contiguously on the same column in both the considered column-major and panel-major matrix formats, when the left factor is transposed then contiguous matrix elements on the same column are employed at different iterations of the inner loop over $k$.
The vectorization of such `TN' and `TT' {\tt dgemm} inner kernels is intrinsically less efficient since it is dot-product-based: it uses register space less effectively (it requires an entire vector register for each kernel matrix element, to hold a partial dot-product in each vector element); it requires a final reduction step (to sum up the partial dot-products across vector elements); it needs larger values of $k$ to perform well (and the value increases with the vector width); it has a smaller degree of reuse of factor matrix elements in registers (since the kernel size is smaller, and this is especially important since BLASFEO does not employ cache blocking).
As a replacement for the functionality of such inner kernels, BLASFEO employs only the more efficient `NN' and `NT' {\tt dgemm} inner kernels together with the explicit transposition of the left $A_k$ or right $B_k$ factors, or with the application of the matrix multiplication transposition rule $(A\cdot B)^T = (B^T \cdot A^T)$ to the register-resident sub-matrix of the result matrix computed by the kernel.

%%%%%%%%
\subsubsection{Custom function calling convention for inner functions} \label{sec:impl:kernel:call}
%%%%%%%%

This section describes the custom calling convention used for the inner functions in BLASFEO for the currently supported targets.
For compactness reasons, the usage of FP registers is only described for real double precision kernels.

Note that in the BLASFEO kernels integer numbers are passed by value and FP numbers by reference.
Therefore, BLASFEO kernels only have arguments of integer type (integer numbers or pointers), and do not employ the FP part of the standard function calling convention.

%%%%
\paragraph{x86\_64}
%%%%

In {\tt x86\_64} architectures, the standard function calling convention is different in Linux and macOS (where the first 6 integer function arguments are passed in the GP registers {\tt rdi, rsi, rdx, rcx, r8, r9}, and the remaining on the stack), and Windows (where the first 4 integer function arguments are passed in the GP registers {\tt rdx, rcx, r8, r9}, and the remaining on the stack).

The {\tt x86\_64} architecture has 16 64-bit GP registers.
In BLASFEO, the registers {\tt rdi, rsi, rdx, rcx, r8, r9} are reserved for the arguments of the kernel function, and never modified; the register {\tt esp} is reserved for the stack pointer.
The registers {\tt r10, r11, r12, r13, r14, r15, rax, rbx, rbp} are available for the BLASFEO custom function calling convention: they are used, in this order, to pass integer and pointer arguments to the inner functions, and for internal computations within the inner functions.
As an example, the inner function {\tt inner\_kernel\_dgemm\_nt\_4x4\_lib44} takes the arguments
\begin{verbatim}
// r10d <= int k_max
// r11  <= double *A_k
// r12  <= double *B_k
\end{verbatim}

The 64-bit {\tt x86\_64} architecture with SSE and AVX ISAs (Instruction Set Architecture) has 16 FP registers.
The number of FP registers used as accumulator registers depends on the kernel size and on the instruction set.
In double precision, in the case of the AVX 256-bit registers, the accumulation registers are the registers {\tt ymm0-ymm3} for the $4\times 4$ kernels, the registers {\tt ymm0-ymm7} for the $8\times 4$ kernels and the registers {\tt ymm0-ymm11} for the $12\times 4$ kernels.
In the case of the SSE 128-bit registers, they are the registers {\tt xmm0-xmm7} for the $4\times 4$ kernels.
The remaining FP registers can be freely used for internal computations in the inner functions.

%%%%
\paragraph{x86}
%%%%

Currently, for the 32-bit {\tt x86} architecture BLASFEO has only been ported to Linux with {\tt gcc} and {\tt clang} compilers.
In this case, the standard function calling convention is to pass all function arguments on the stack.

The {\tt x86} architecture has 8 32-bit GP registers.
The register {\tt esp} is reserved for the stack pointer.
The remaining 7 registers {\tt eax, ebx, ecx, edx, esi, edi, ebp} are used in the BLASFEO custom function calling convention. %, in this order, to pass integer arguments to the inner functions, and for internal computations within the inner functions.
%As an example, the inner function {\tt inner\_kernel\_dgemm\_nt\_4x4\_lib44} takes the arguments
%\begin{verbatim}
%// eax  <= int k_max
%// ebx  <= double *A_k
%// ecx  <= double *B_k
%\end{verbatim}

The {\tt x86} architecture with SSE and AVX ISAs has 8 FP registers.
At most 4 can be used as accumulator registers: the 128-bit registers {\tt xmm0-xmm3} for the $4\times 2$ kernels targeting SSE ISAs and the 256-bit registers {\tt ymm0-ymm3} for the $4\times 4$ kernels targeting the AVX ISAs.
%In the case of the SSE 128-bit registers, they are the registers {\tt xmm0-xmm3} for the $4\times 2$ kernels.
%In the case of the AVX 256-bit registers, they are the registers {\tt ymm0-ymm3} for the $4\times 4$ kernels.
%The remaining FP registers can be freely used for internal computations in the inner functions.

%%%%
\paragraph{ARMv8A}
%%%%

Currently, for the 64-bit {\tt ARMv8A} architecture BLASFEO has only been ported to Linux with {\tt gcc} and {\tt clang} compilers.
The standard function calling convention is to pass the first 8 integer function arguments in the GP registers {\tt x0-x7} (which in the case of 32-bit arguments are named {\tt w0-w7}) and the remaining on the stack.

The {\tt ARMv8A} architecture has 30 GP registers ({\tt x0-x29}), one procedure link register ({\tt x30}), and one zero register ({\tt x31}).
In BLASFEO, the registers {\tt x0-x7} are reserved for the arguments of the kernel function, and never modified; the registers {\tt x30} and {\tt x31} are not employed.
The remaining registers {\tt x8-x29} are used in the BLASFEO custom function calling convention. %are used, in this order, to pass integer arguments to the inner functions, and for internal computation within the inner functions.
%As an example, the inner function {\tt inner\_kernel\_dgemm\_nt\_4x4\_lib44} takes the arguments
%\begin{verbatim}
%// w8   <= int k_max
%// x9   <= double *A_k
%// x10  <= double *B_k
%\end{verbatim}

The {\tt ARMv8A} architecture with NEONv2 ISA has 32 128-bit FP registers.
The FP registers used as accumulator registers are the registers {\tt q0-q7} for the $4\times 4$ kernels, the registers {\tt q0-q15} for the $8\times 4$ kernels, and the registers {\tt q0-q23} for the $12\times 4$ kernels.
%The remaining FP registers can be freely used for internal computations in the inner functions.

%%%%
\paragraph{ARMv7A}
%%%%

Currently, for the 32-bit {\tt ARMv7A} architecture BLASFEO has only been ported to Linux with {\tt gcc} and {\tt clang} compilers.
The standard function calling convention is to pass the first 4 integer function arguments in the GP registers {\tt r0-r3} and the remaining on the stack.

The {\tt ARMv7A} architecture has 13 GP registers ({\tt r0-r12}), a stack pointer register ({\tt r13}), a link register ({\tt r14}) and a program counter register ({\tt r15}).
In BLASFEO, the registers {\tt r0-r3} are reserved for the arguments of the kernel function and never modified; the registers {\tt r13-r15} are not employed.
The registers {\tt r4-r12} are used in the BLASFEO custom function calling convention. %, in this order, to pass integer arguments to the inner functions, and for internal computation within the inner functions.
%As an example, the inner function {\tt inner\_kernel\_dgemm\_nt\_4x4\_lib44} takes the arguments
%\begin{verbatim}
%// r4   <= int k_max
%// r5   <= double *A_k
%// r6   <= double *B_k
%\end{verbatim}

The {\tt ARMv7A} architecture has two extensions for FP computations, VFP (offering single and double precision scalar FP instructions) and NEON (offering single precision SIMD FP instructions).
VFP comes in two versions, offering 32 or 16 64-bit FP registers.
In BLASFEO, only the more common version with 32 FP registers is targeted.
In double precision, the FP registers used as accumulator registers are the registers {\tt d0-d15} for the $4\times 4$ kernels.
%The remaining FP registers can be freely used for internal computations in the inner functions.

%%%%%%%%%%%%%%%%%%%%%%%%%%%%%%%%
\section{Performance evaluation} \label{sec:perf}
%%%%%%%%%%%%%%%%%%%%%%%%%%%%%%%%

In this section, the performance of the BLAS API of BLASFEO is evaluated against other optimized BLAS libraries.
Namely, we consider the open-source implementations OpenBLAS version 0.3.4-dev, BLIS version 0.5.0 (coupled with LAPACK 3.5.0 to provide the {\tt dpotrf} and {\tt dgetrf} routines), and the proprietary implementation MKL version 2019.1.144.
Additionally, the performance of the BLASFEO API is added as a reference, even if this does not implement the standard BLAS API.
The sequential version of all libraries is employed\footnote{Generally the multi-threaded version with only one thread enabled has worse performance due to the additional overhead of the parallelization framework}.
All open-source libraries are compiled with standard options.
MKL is used with the MKL\_DIRECT\_CALL\_SEQ option.
The same setup is used for all experiments in the paper.

It should be stressed that these comparisons are limited to library implementations of linear algebra routines.
Therefore they do not consider program generated, just-in-time (JIT) compiled or template-based implementations like e.g. {\tt LIBXSMM}, {\tt Eigen}, {\tt Blaze}, {\tt  \textsc{LGen}}, {\tt  \textsc{SLinGen}}.
The interested reader can find some comparisons involving also these approaches in the original BLASFEO article~\cite{Frison2018}.

%It should also be noted that the single-threaded version of all libraries is considered.
%The development and evaluation of the multi-threaded implementation in BLASFEO is outside the scope of this article.
The performance of the linear algebra routines is evaluated for matrices of size up to 300, as this is large enough for the embedded optimization applications of interest, and it is a good size to show the behavior of the routines.
In fact, since BLASFEO does not implement cache blocking, 300 as a matrix size is large enough to show some performance degradation when the data footprint exceeds the cache size.
Matrix sizes are not known at compile time and all experiments are performed with hot cache\footnote{This is generally the case in embedded optimization, where optimization problems are repeatedly solved at each sampling time by using iterative algorithms.}.

The performance is evaluated on three microarchitectures: Intel Haswell\footnote{The exact same optimized code is employed also on the client version of the more recent Intel Skylake microarchitecture, with analogous results.}, ARM Cortex A57 and ARM Cortex A53.
Intel Haswell is a high-performance microarchitecture commonly employed in laptops, workstations and servers, and high-end embedded devices.
The ARM microarchitectures are commonly employed in mobile and embedded devices, with a focus on performance as well as cost and power consumption.
In particular, the ARM Cortex A53 is a rather small and cheap core, and it supports in-order execution.

For all microarchitectures, the same six routines are evaluated.
The {\tt dgemm} routine is evaluated in the variants `NN' and `NT'\footnote{In BLASFEO, the kernels employed in these routines are the backbone for all algorithmic variant of the BLAS API, as well as the BLASFEO API.}.
The `UT' (upper-transposed) variant of the {\tt dsyrk} routine makes use of the `NN' inner kernel in the BLASFEO API, and the `NT' inner kernel in the BLAS API.
Therefore this is an example of a routine performing similarly in the two APIs, since the faster kernel in the BLAS API offsets the lower overhead of the BLASFEO API.
The `RLTU' (right-lower-transposed-unit\_diagonal) variant of the {\tt dtrsm} routine makes use of the `NT' inner kernel for both the BLAS API and the BLASFEO API.
Therefore this is an example of a routine performing better in the BLASFEO API.
Finally the Cholesky and the row-pivot LU factorizations (implemented in {\tt dpotrf} (`L' variant) and {\tt dgetrf} respectively) are two routines widely employed in the solution of linear systems of equations, and their performance is critical in many embedded optimization applications.

%%%%%%%%%%%%%%%%
\subsection{{\tt x86\_64} Intel Haswell} \label{sec:perf:haswell}
%%%%%%%%%%%%%%%%

Intel Haswell is a deeply out-of-order microarchitecture that performs aggressive hardware prefetch.
Regarding the implementation of linear algebra routines, the Haswell core can perform two 256-bit wide FP fused-multiplication-accumulate every clock cycle, giving a throughput of 16 flops per cycle in double precision.
The client version of Intel Skylake, the Haswell successor microarchitecture, does not show significant differences in this regard, and the same optimized code is employed.

It is relatively easy to write {\tt dgemm} kernels achieving high performance provided that at least 10 accumulation registers are employed.
In the implementation of BLASFEO, the panel size $p_s$ is 4 in double precision.
The optimal {\tt dgemm} kernel size is $12\times 4$.
Hardware prefetch can detect the streaming of data along panels or along columns.

The test processor is the Intel Core i7 4810MQ, a quad-core Haswell processor running at 3.4 GHz when the 256-bit execution units are employed (3.8 GHz when they are idle).
The maximum double-precision throughput per core is 54.4 Gflops.
The memory is 8 GiB of DDR3L-1600 RAM in dual-channel configuration, giving a bandwidth of 25.6 GB/s.
Each core has 32 KB 8-way set associative data and 32 KB 8-way set associative instruction L1 caches, and 256 KB 8-way set associative L2 cache.
All cores share 6 MB 12-way set associative L3 cache.
The cache line size is 64 bytes.

\begin{figure}[!t]
\centering
\subfloat[dgemm\_nn]{\includegraphics[width=0.45\linewidth]{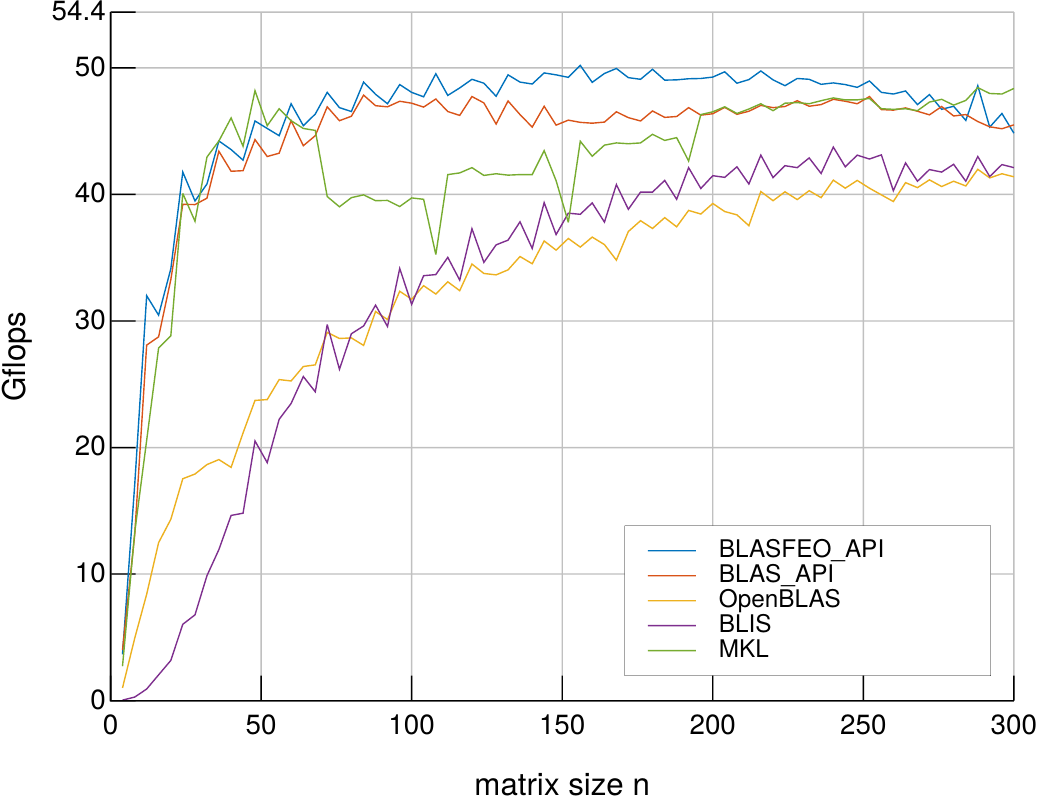} \label{fig:perf:haswell:dgemm_nn}} %\\
\subfloat[dgemm\_nt]{\includegraphics[width=0.45\linewidth]{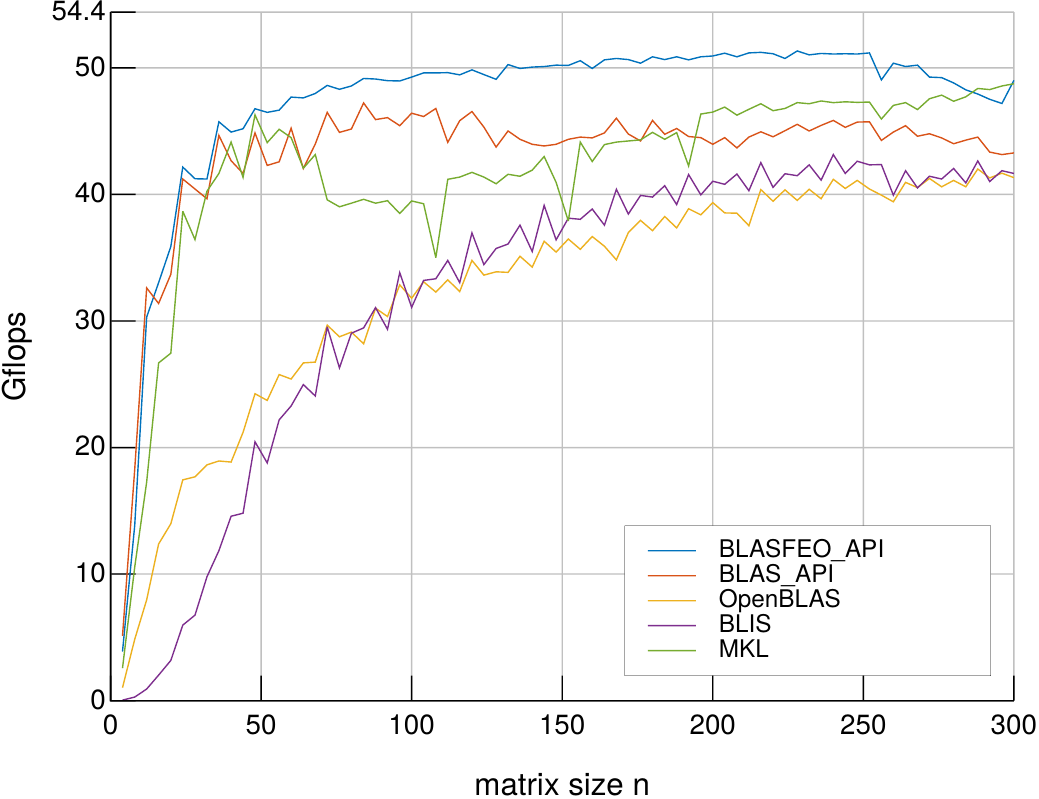} \label{fig:perf:haswell:dgemm_nt}} \\
\subfloat[dsyrk\_ut]{\includegraphics[width=0.45\linewidth]{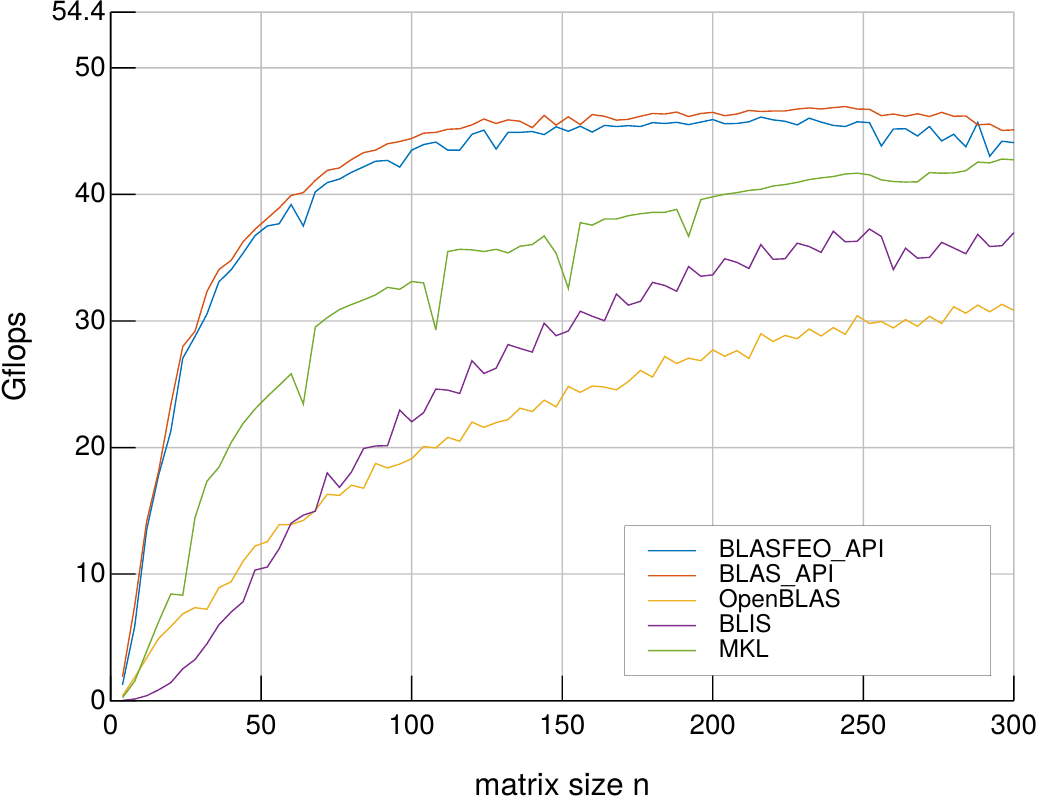} \label{fig:perf:haswell:dsyrk_ut}} %\\
\subfloat[dtrsm\_rltu]{\includegraphics[width=0.45\linewidth]{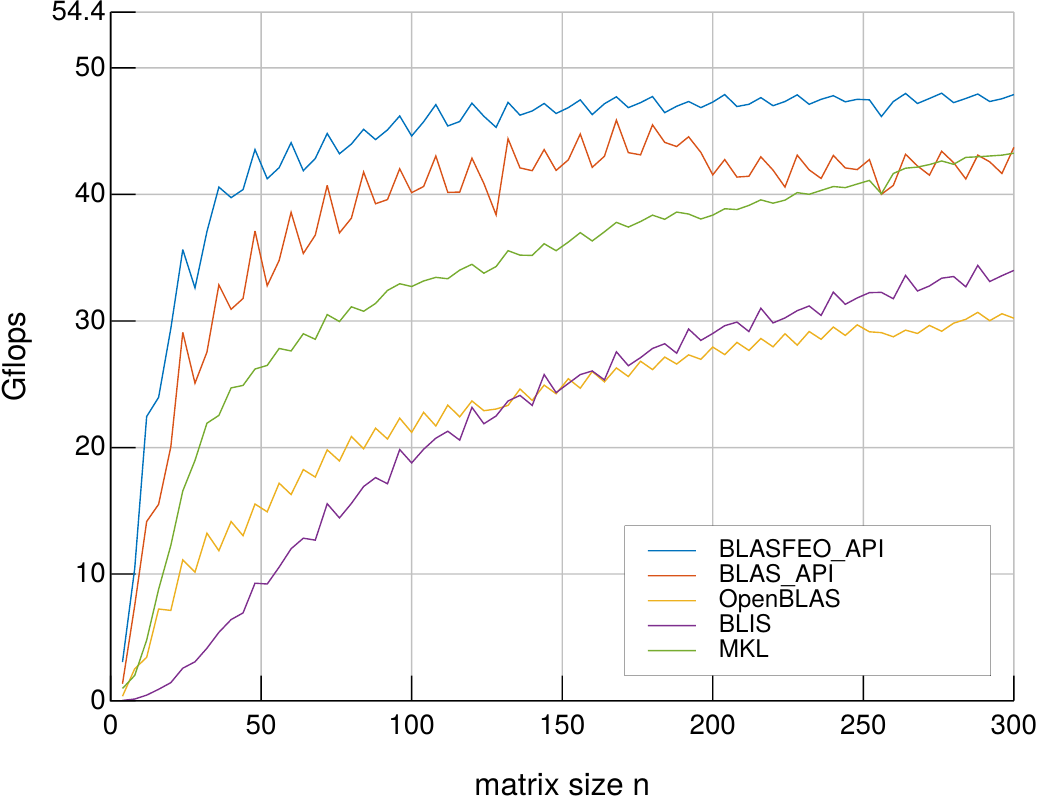} \label{fig:perf:haswell:dtrsm_rltu}} \\
\subfloat[dpotrf\_l]{\includegraphics[width=0.45\linewidth]{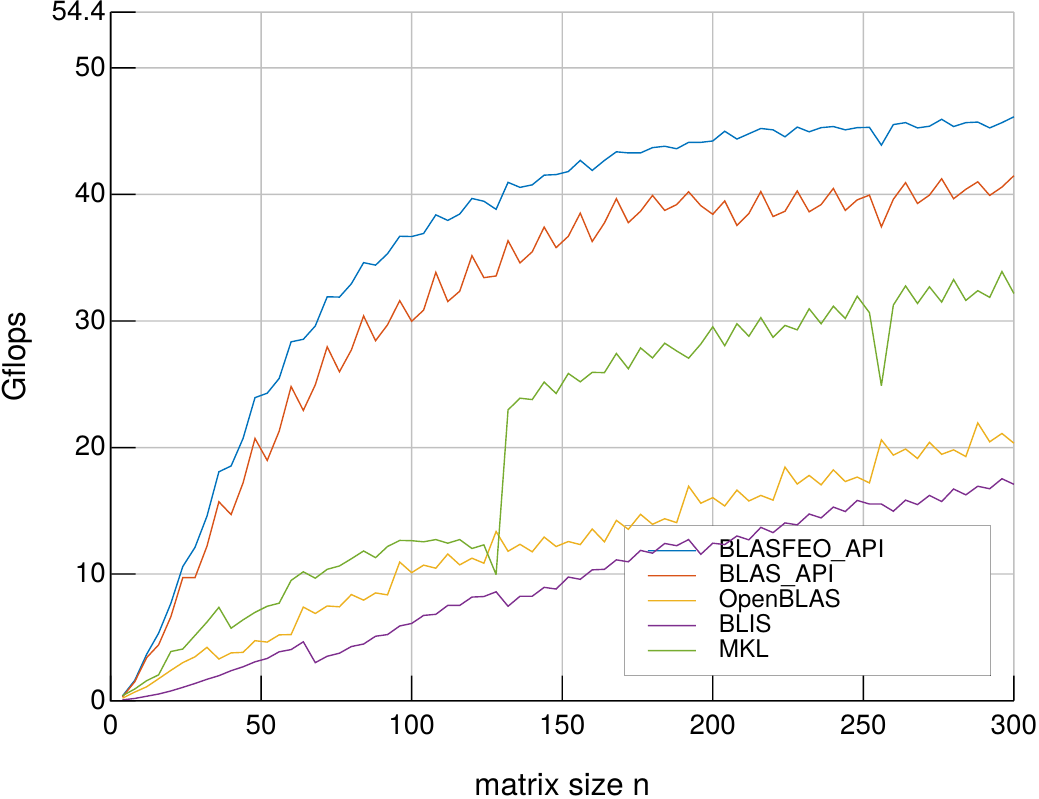} \label{fig:perf:haswell:dpotrf_l}} %\\
\subfloat[dgetrf]{\includegraphics[width=0.45\linewidth]{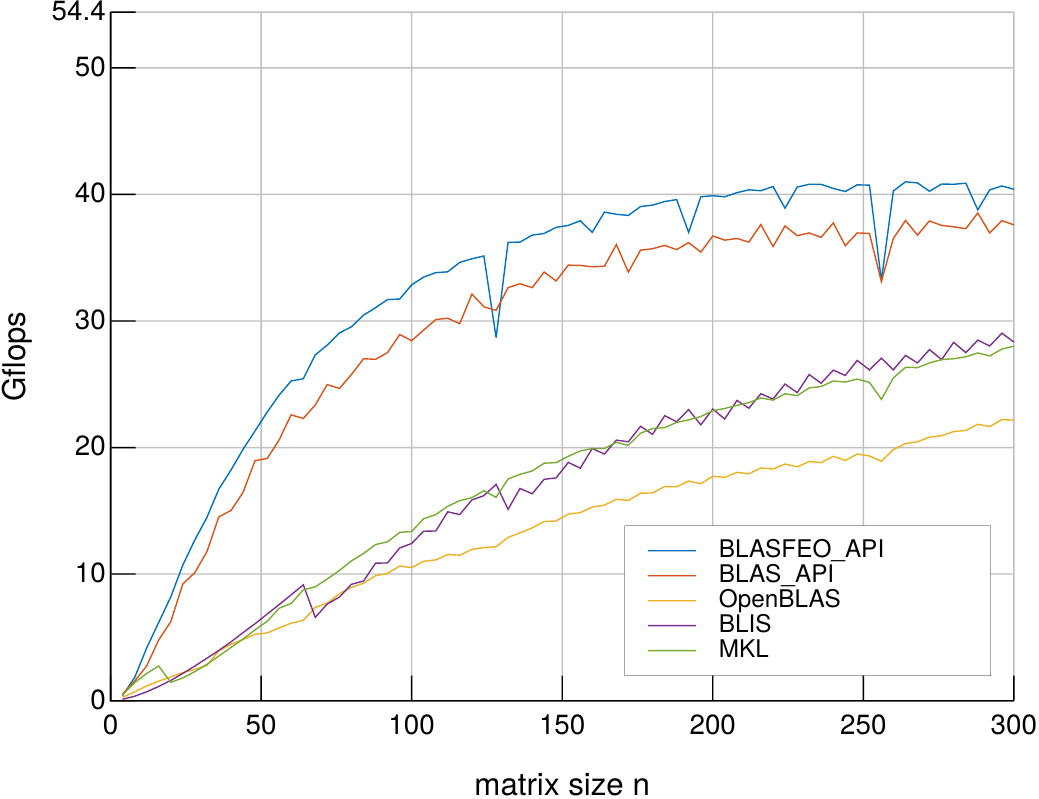} \label{fig:perf:haswell:dgetrf}} \\
\caption{Performance of some BLAS and LAPACK routines on one core of the Intel Haswell Core i7 4810MQ @ 3.4 GHz.}
\label{fig:perf:haswell}
\end{figure}

Performance plots are in Figure~\ref{fig:perf:haswell}.
For the matrix sizes of interest, the BLAS API of BLASFEO is competitive with Intel MKL in the implementation of {\tt dgemm}, and it outperforms MKL for the other level 3 BLAS and especially for LAPACK routines.
It also outperforms the other open-source implementations OpenBLAS and BLIS for all tested routines, and by a factor 2 or 3 for matrix sizes in the order of tens.
The use of the BLASFEO API gives an additional 10-15\% speedup in most routines (except {\tt dsyrk\_ut}).

The drop in performance visible especially for the BLASFEO API implementation of {\tt dgemm} is due to the fact that 4 panels (3 for $A_k$ and 1 for $B_k$ simultaneously streamed by the {\tt dgemm} kernel) of height $p_s=4$ exceed the L1 data cache size for $n\geq 256$.

%%%%%%%%%%%%%%%%
\subsection{{\tt ARMv8A} ARM Cortex A57} \label{sec:perf:cortex_a57}
%%%%%%%%%%%%%%%%

The ARM Cortex A57 is a relatively low-power microarchitecture, and it is the 64-bit successor of the ARM Cortex A15.
It is a 3-way superscalar microarchitecture with out-of-order execution.
The NEON ISA in the ARMv8A architecture supports vectorization in both single and double precision, with 4- and 2-wide vectors respectively.
Regarding the implementation of linear algebra routines, the Cortex A57 core can perform one 128-bit wide FP fused-multiplication-accumulate at every clock cycle, giving a throughput of 8 and 4 flops per cycle in single and double precision respectively.
The Cortex A72 and A75, the successor microarchitectures, do not show significant difference in this regard, and the same optimized code is employed.

In the implementation of BLASFEO, the panel size $p_s$ is 4 in double precision.
The optimal {\tt dgemm} kernel size is $8\times 4$.
Software prefetch is employed for both the left and the right factors in all matrix formats, slightly improving performance.

The test processor is the NVIDIA Tegra TX1 SoC in the Shield TV featuring 4 Cortex A57 cores, each paired with one of the 4 low-power Cortex A53 cores.
The tests are performed on a single Cortex A57 core, running at 2.15 GHz during all tests.
The maximum double-precision throughput per core is 8.6 Gflops.
The memory interface is 64-bit LPDDR4-3200 giving 25.6 GB/s of bandwidth.
The amount of memory is 3 GiB.
Each Cortex A57 core has 48 KB 3-way set associative instruction L1 cache and 32 KB 2-way set associative data L1 cache.
All Cortex A57 cores share a 2 MB 16-way set associative unified L2 cache.
The cache line size is 64 bytes.

\begin{figure}[!t]
\centering
\subfloat[dgemm\_nn]{\includegraphics[width=0.45\linewidth]{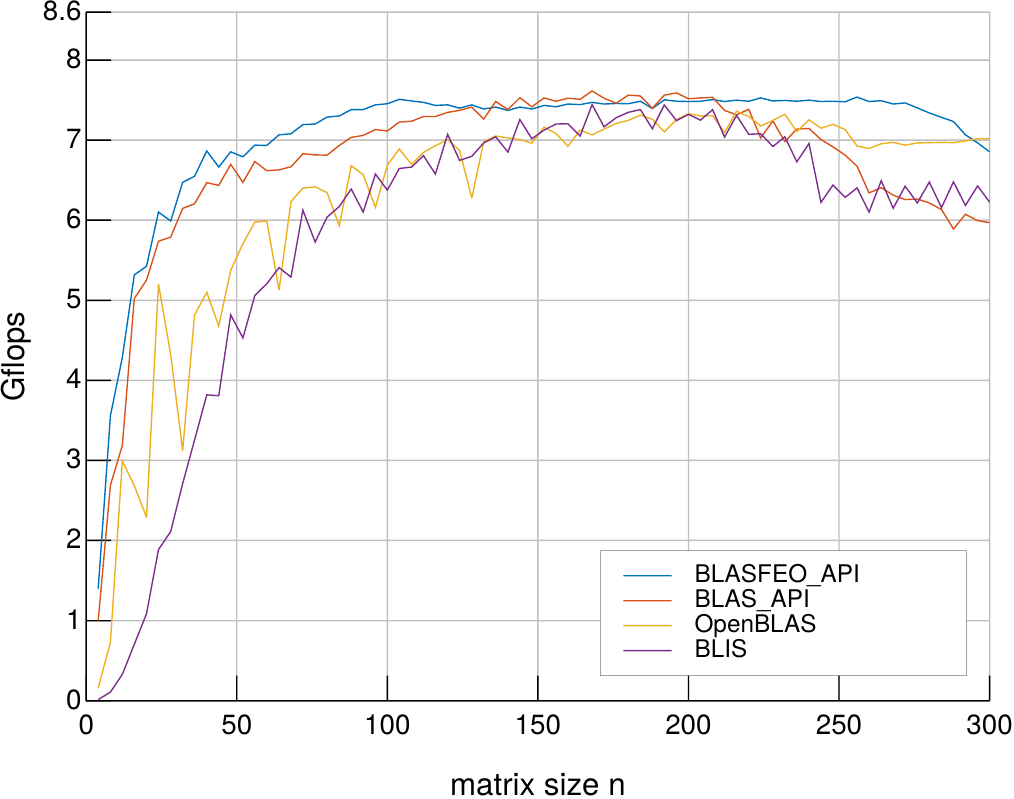} \label{fig:perf:cortexa57:dgemm_nn}} %\\
\subfloat[dgemm\_nt]{\includegraphics[width=0.45\linewidth]{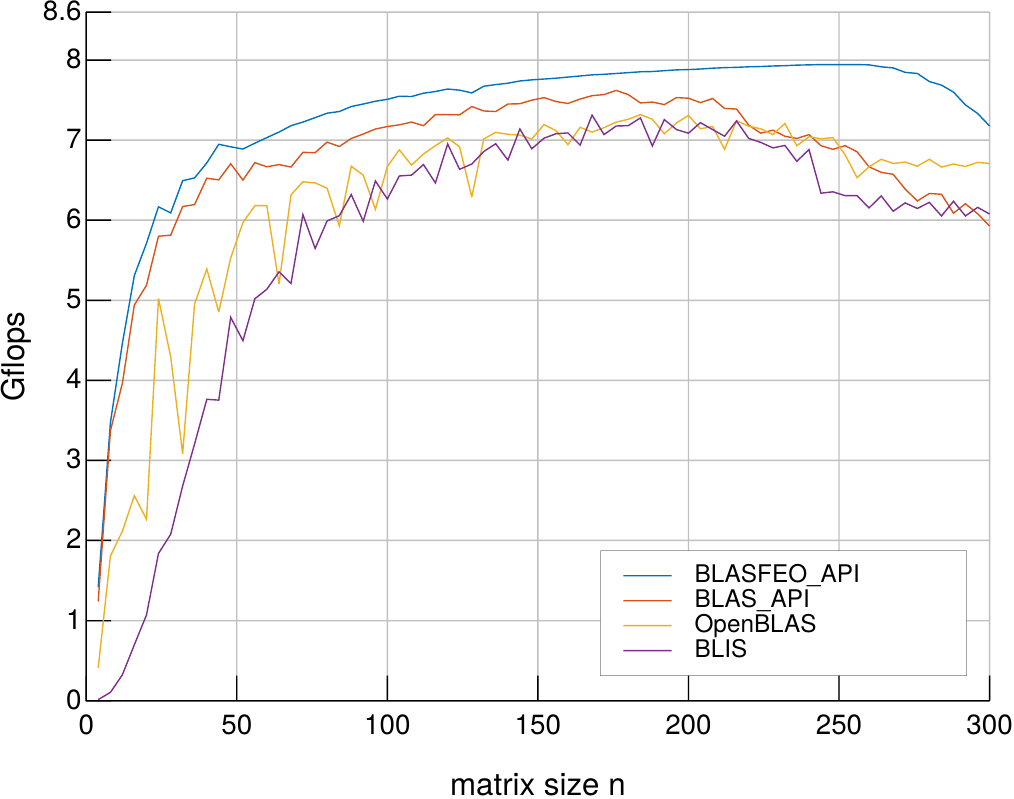} \label{fig:perf:cortexa57:dgemm_nt}} \\
\subfloat[dsyrk\_ut]{\includegraphics[width=0.45\linewidth]{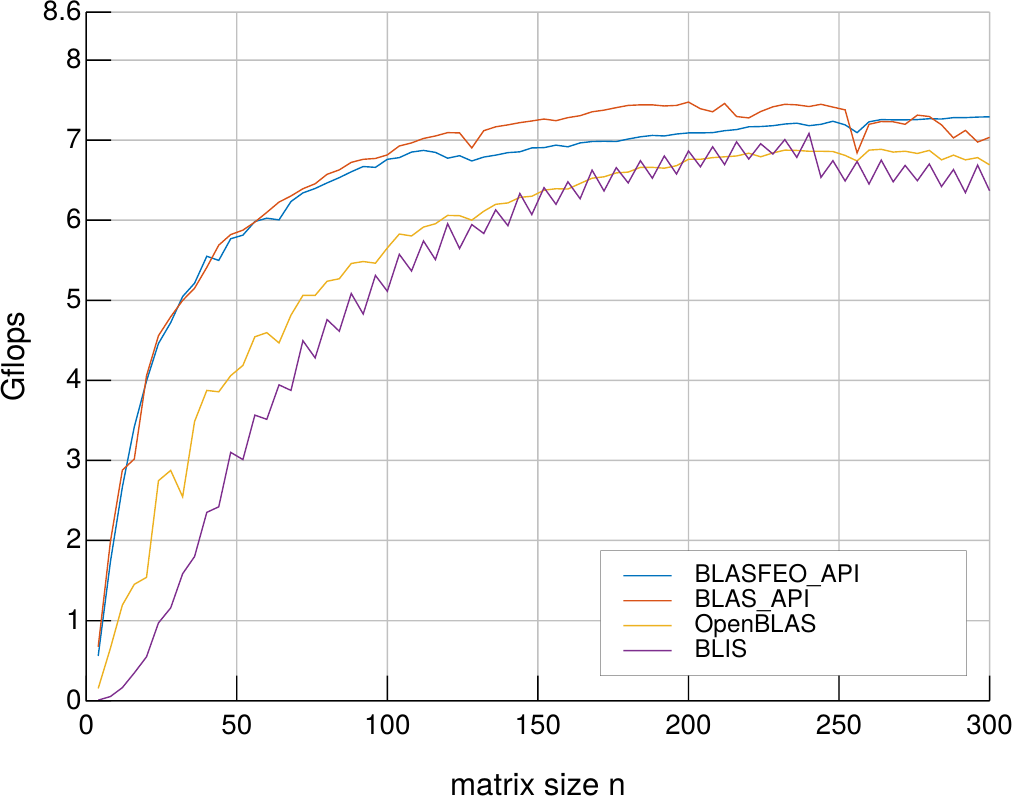} \label{fig:perf:cortexa57:dsyrk_ut}} %\\
\subfloat[dtrsm\_rltu]{\includegraphics[width=0.45\linewidth]{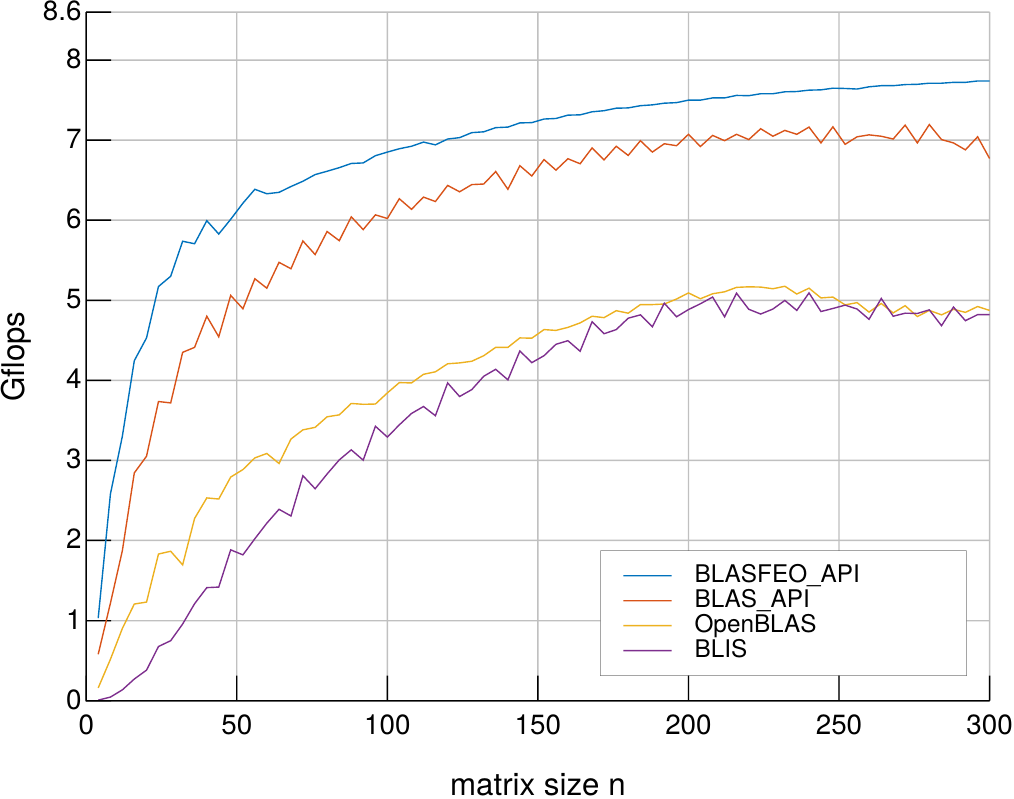} \label{fig:perf:cortexa57:dtrsm_rltu}} \\
\subfloat[dpotrf\_l]{\includegraphics[width=0.45\linewidth]{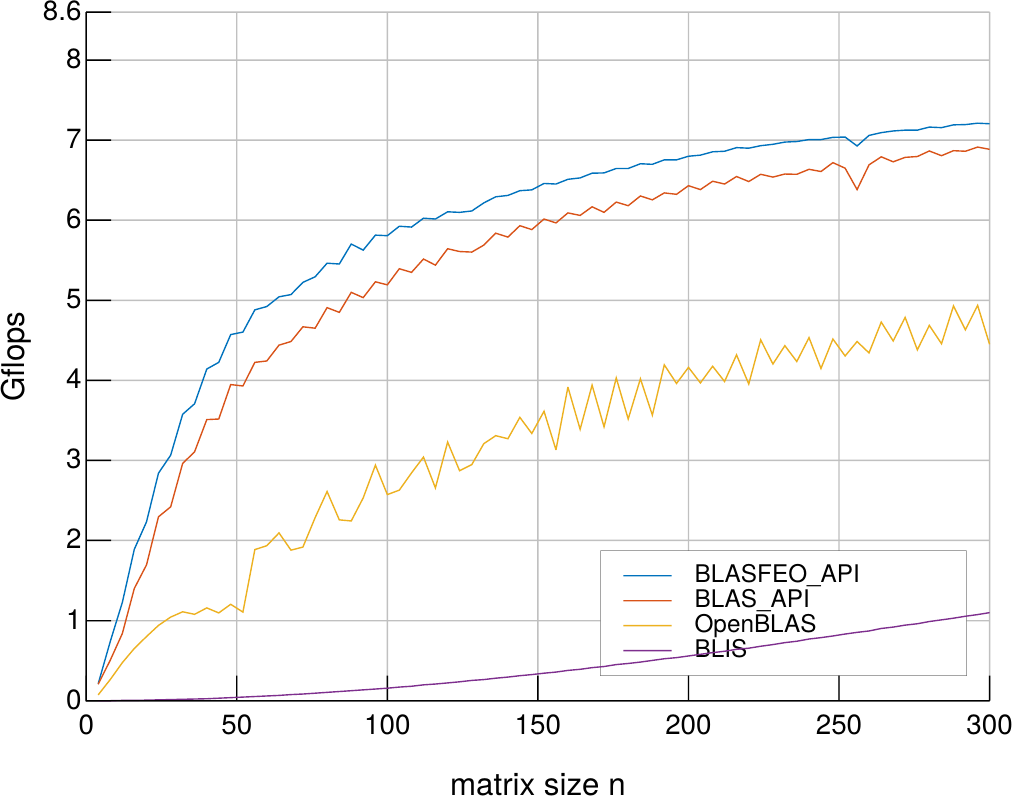} \label{fig:perf:cortexa57:dpotrf_l}} %\\
\subfloat[dgetrf]{\includegraphics[width=0.45\linewidth]{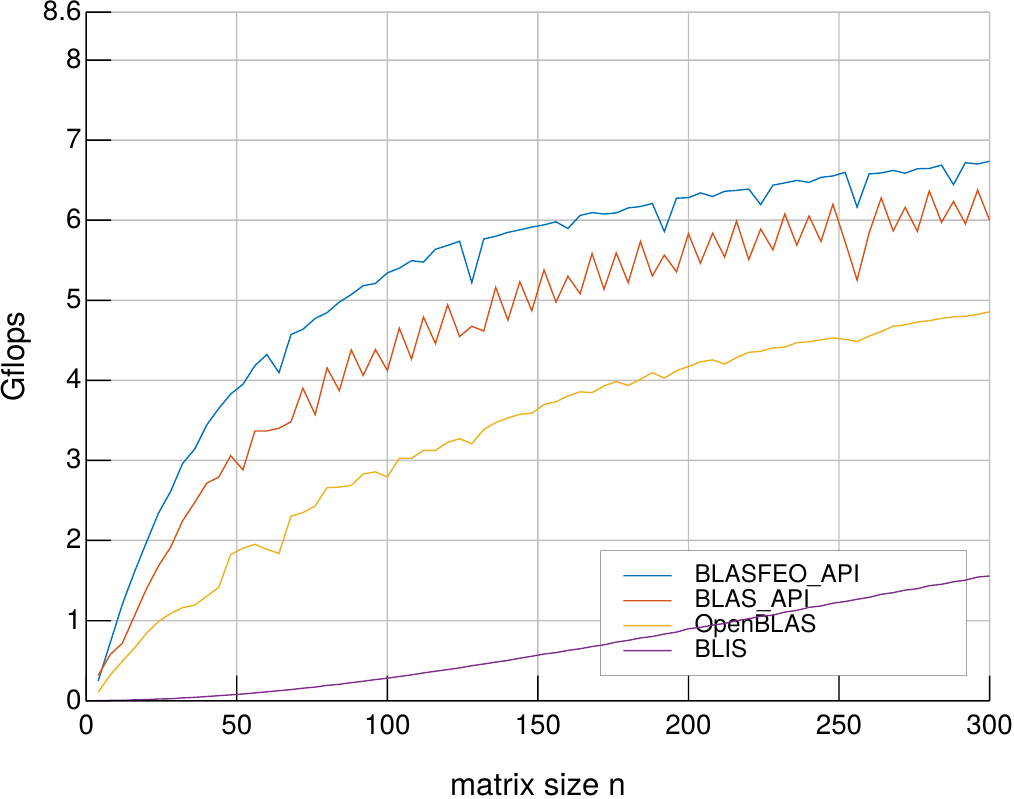} \label{fig:perf:cortexa57:dgetrf}} \\
\caption{Performance of some BLAS and LAPACK routines on one core of the ARM Cortex A57 @ 2.15 GHz.}
\label{fig:perf:cortexa57}
\end{figure}

Performance plots are in Figure~\ref{fig:perf:cortexa57}.
In the case of {\tt dgemm}, the BLAS API of BLASFEO is competitive or outperforms OpenBLAS (whose performance shows significant drops for some sizes) and BLIS (which shows considerable overhead for small matrices).
The performance advantage for the other level 3 BLAS and especially LAPACK routines is larger, and in the range of a speedup factor of 2 for matrix sizes in the order of tens.
The use of the BLASFEO API gives an additional 10-15\% speedup in most routines.

The drop in performance visible especially for the BLASFEO API implementation of {\tt dgemm} is due to the fact that 3 square matrices ($A$, $B$ and $C$) of doubles exceed the L2 cache size for $n\geq295$ (with cache associativity and share of cache between data and instructions slightly lowering this value).
The BLAS API employs more memory (in particular, the matrix $B$ is entirely packed at once into a buffer), and therefore the drop happens for smaller matrix sizes.

%%%%%%%%%%%%%%%%
\subsection{{\tt ARMv8A} ARM Cortex A53} \label{sec:perf:cortex_a53}
%%%%%%%%%%%%%%%%

The ARM Cortex A53 is a low-power core, intended to be used alone or as the little companion core of the big cores like Cortex A57 or A72.
It can be considered the 64-bit successor of the ARM Cortex A7.
It is a partially 2-way superscalar microarchitecture (only certain combinations of instructions can be dual-issued) with in-order execution.
The Cortex A53 is ISA-compatible with the Cortex A57, and therefore it also supports the NEON ISA.
Regarding the implementation of linear algebra routines, the Cortex A53 core can perform one 128-bit wide FP fused-multiplication-accumulate at every clock cycle, giving a throughput of 8 and 4 flops per cycle in single and double precision respectively (same as the Cortex A57).

In the implementation of BLASFEO, the panel size $p_s$ is 4 in double precision.
The optimal {\tt dgemm} kernel size is $12\times 4$.
The low-power features of the ARM Cortex A53 (in-order execution, partial dual issue, small number of execution ports) make it difficult to write high-performing code, and e.g. the code optimized for the Cortex A57 performs poorly.
In particular, as described in~\cite{Hasan2017} a 64-bit FP load and an FMA cannot be co-issued, and no 64-bit FP load can be performed on the 4th cycle after any FMA; 64-bit loads to GP registers can be co-issued with FMA, and loads of more than 64-bit cannot be issued in one single cycle.
Therefore, an optimal scheme alternates 3 FMAs and 1 64-bit FP load, with the remaining data loaded using 64-bit loads in GP registers and data transfers between GP and FP registers.
The maximum attainable performance with this scheme is 3 128-bit FMAs every 4 clock cycles, that is 75\% of peak FMA throughput.

The test processor is the Amlogic S905 SoC in the ODROID-C2 board, featuring 4 Cortex A53 cores.
The tests are performed on a single Cortex A53 core, running at 1.536 GHz during all tests.
The maximum double-precision throughput per core is 6.144 Gflops.
The memory interface is 32-bit DDR3-1824 giving 7.296 GB/s of bandwidth.
The amount of memory is 2 GiB.
Each Cortex A53 core has 32 KB 2-way set associative instruction L1 cache and 32 KB 4-way set associative data L1 cache.
All cores share 512 KB 16-way set associative unified L2 cache.
The cache line size is 64 bytes.

\begin{figure}[!t]
\centering
\subfloat[dgemm\_nn]{\includegraphics[width=0.45\linewidth]{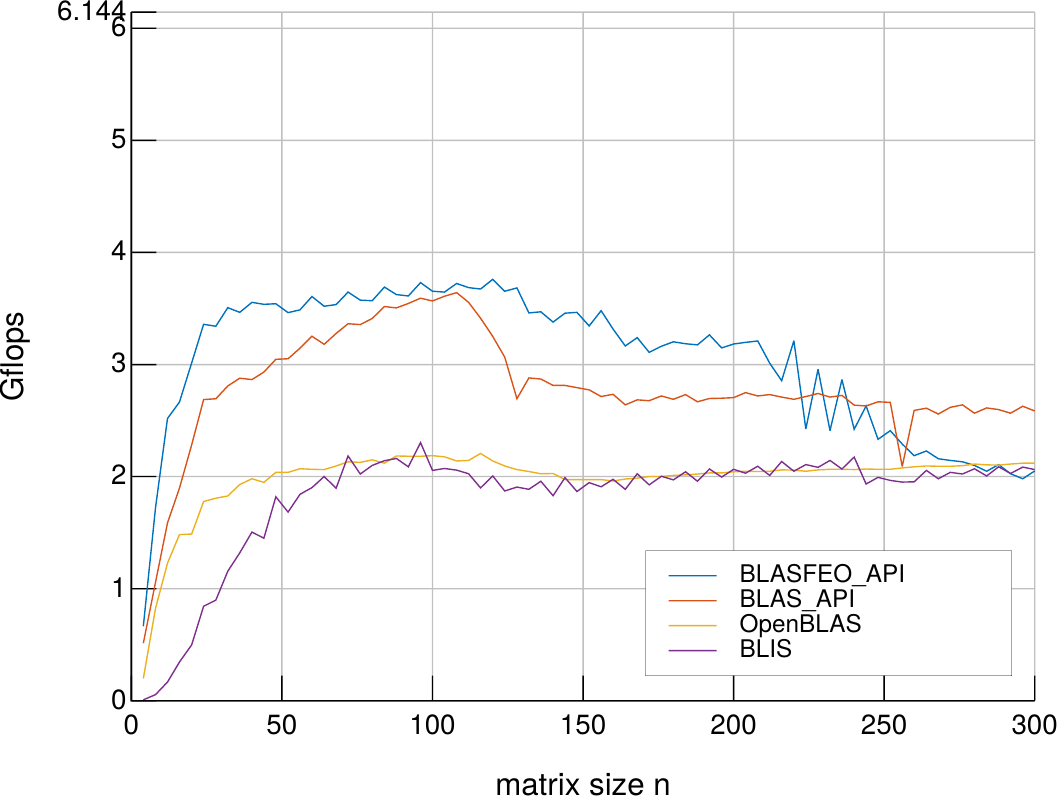} \label{fig:perf:cortexa53:dgemm_nn}} %\\
\subfloat[dgemm\_nt]{\includegraphics[width=0.45\linewidth]{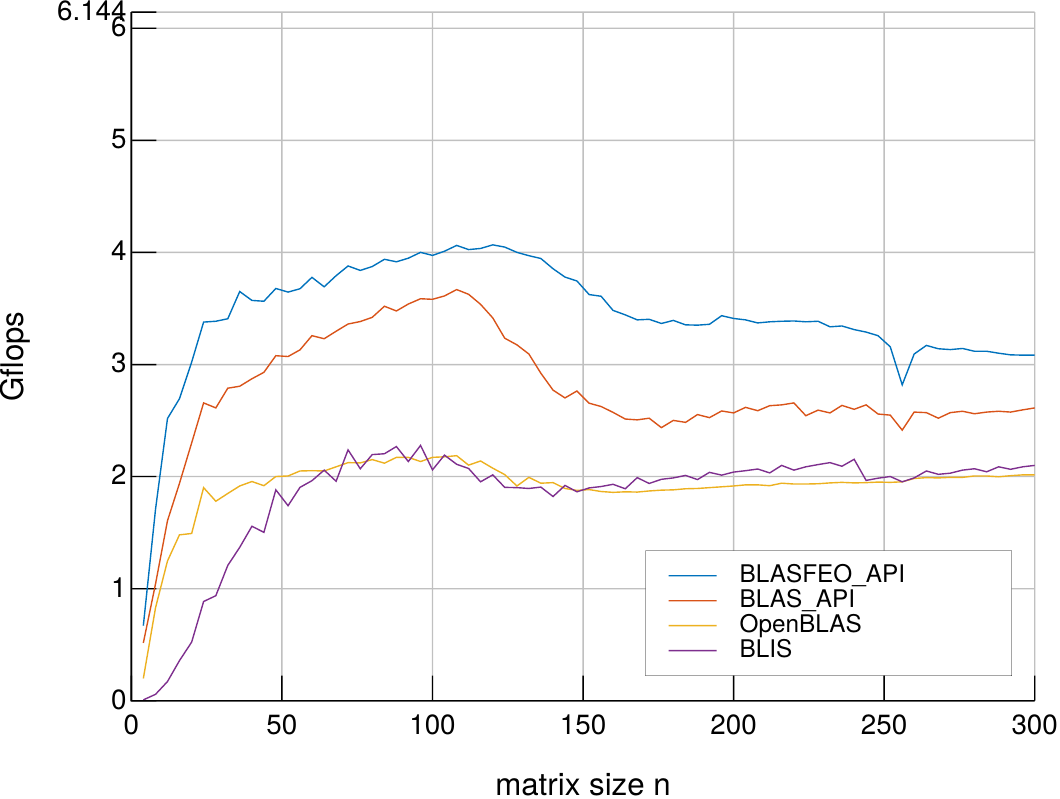} \label{fig:perf:cortexa53:dgemm_nt}} \\
\subfloat[dsyrk\_ut]{\includegraphics[width=0.45\linewidth]{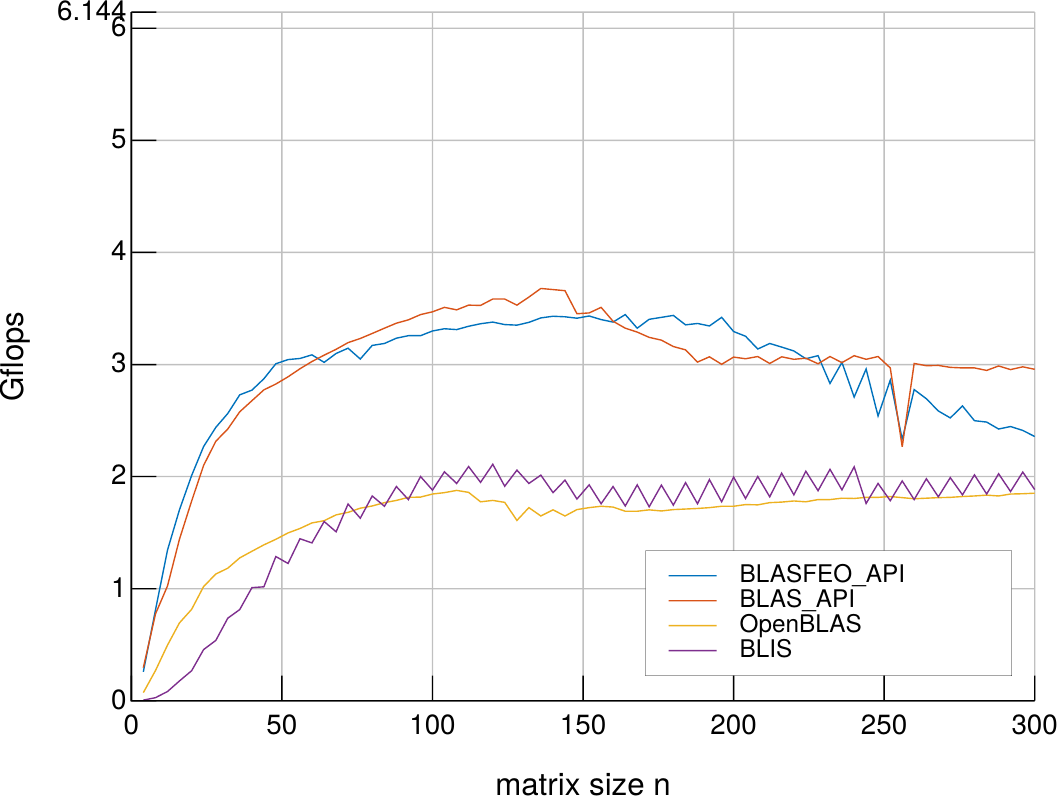} \label{fig:perf:cortexa53:dsyrk_ut}} %\\
\subfloat[dtrsm\_rltu]{\includegraphics[width=0.45\linewidth]{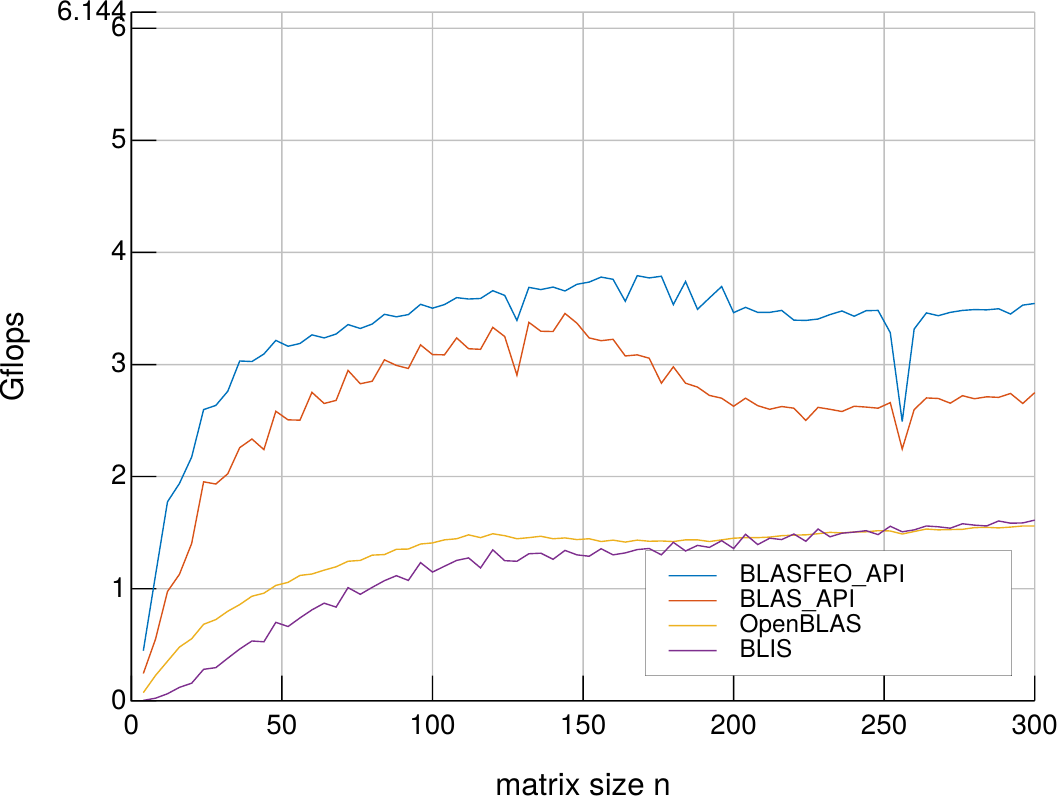} \label{fig:perf:cortexa53:dtrsm_rltu}} \\
\subfloat[dpotrf\_l]{\includegraphics[width=0.45\linewidth]{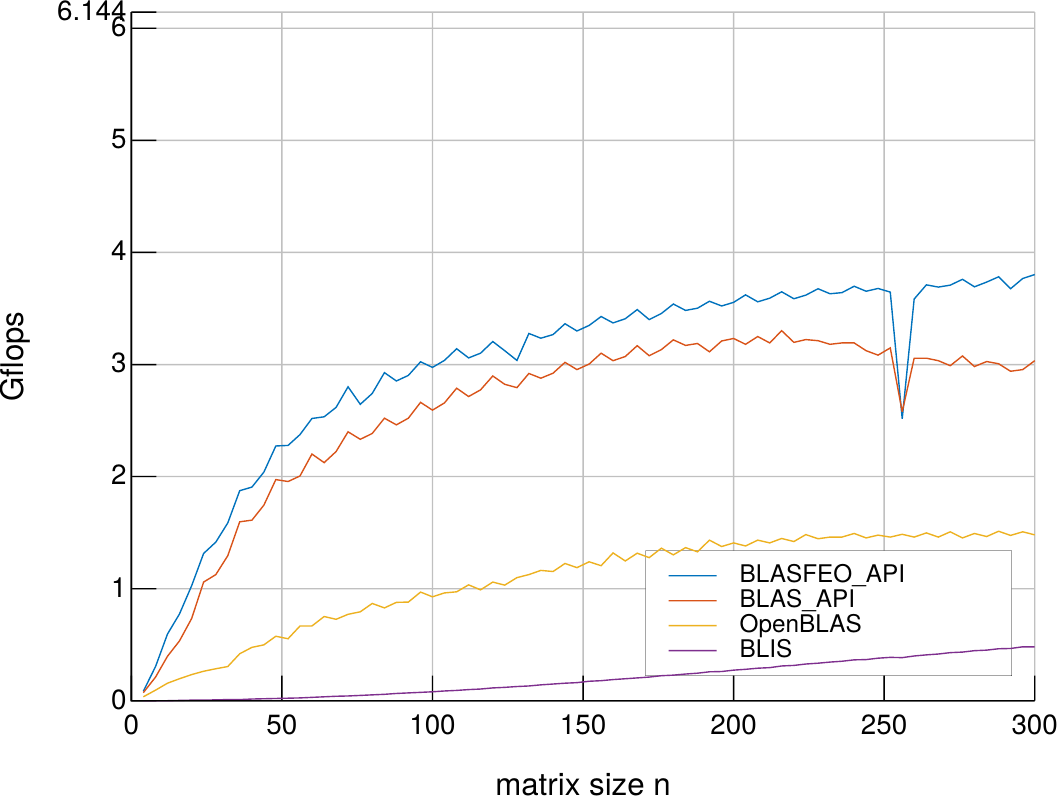} \label{fig:perf:cortexa53:dpotrf_l}} %\\
\subfloat[dgetrf]{\includegraphics[width=0.45\linewidth]{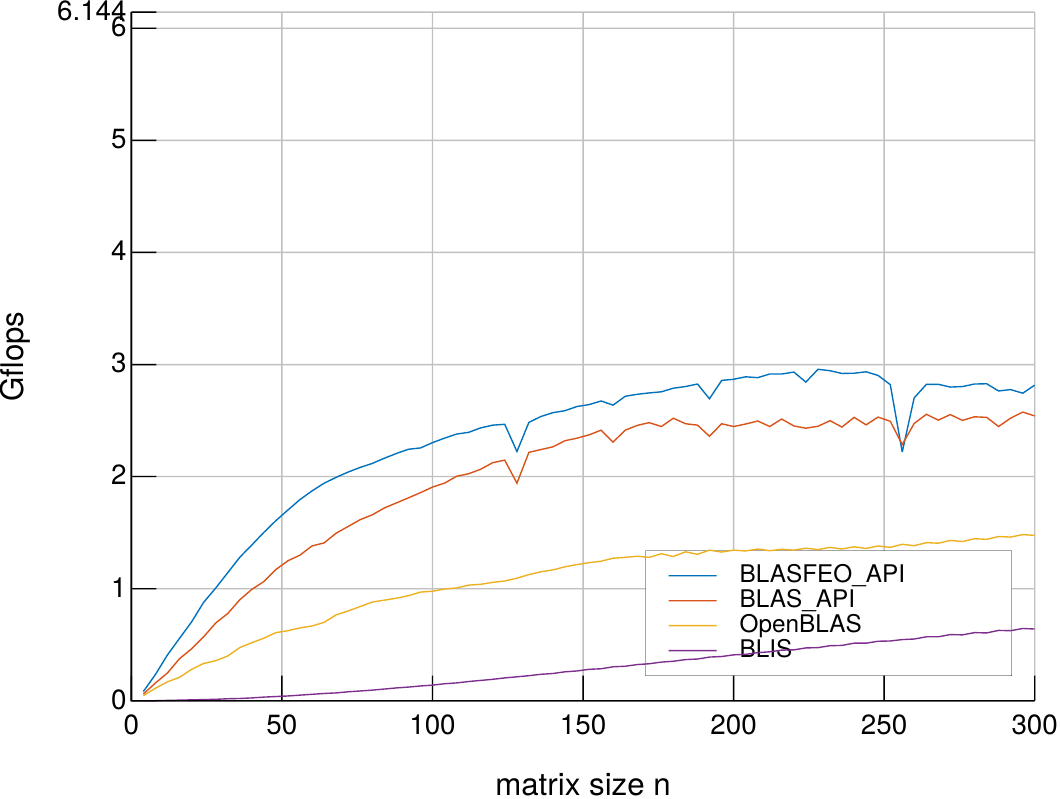} \label{fig:perf:cortexa53:dgetrf}} \\
\caption{Performance of some BLAS and LAPACK routines on one core of the ARM Cortex A53 @ 1.536 GHz.}
\label{fig:perf:cortexa53}
\end{figure}

Performance plots are in Figure~\ref{fig:perf:cortexa53}.
In case of BLIS, the code for the Cortex A53 target in version 0.5.0 terminates with a segfault\footnote{The issue is fixed in later version 0.6.0.}, so the code for the Cortex A57 had to be used instead.
Note that, in BLIS, the Cortex A53 and A57 targets share the same kernel code, and only differ by the choice of compilation flags.
This implies that there is not a large performance difference between the two targets.
The first thing worth noticing is that in general BLASFEO outperforms both OpenBLAS and BLIS by a wide margin due to its better-performing assembly kernels.
%On top of that, the BLASFEO framework provides similar speedups as in the case of Cortex A57.
The speedup is considerably larger than in the case of the Cortex A57, in which case the speedup is solely due to the low overhead of the BLASFEO framework.
Compared to the BLAS API of BLASFEO, the use of the BLASFEO API gives an additional 10-20\% speedup in most routines.

As in the case of the Cortex A57, the drop in performance for the BLASFEO API implementations of {\tt dgemm} is due to the fact that 3 square matrices of doubles exceed the L2 cache size for $n\geq147$ (again slightly lowered due to cache associativity and share of L2 cache between data and instruction).
The drop happens for smaller matrix sizes in the case of the BLAS API since this employs more memory internally.
The performance of the {\tt dgemm} kernel itself exceeds 70\% of peak FMA throughput, therefore getting very close to the 75\% maximally attainable by the scheme.

%%%%%%%%%%%%%%%%%%%%%%%%%%%%%%%%
\section{Application in scientific programming languages} \label{sec:appl}
%%%%%%%%%%%%%%%%%%%%%%%%%%%%%%%%

The BLAS API of BLASFEO currently supports a sub-set of the entire BLAS and LAPACK libraries, focusing on the routines with the largest impact on the intended application areas.
The main interest is in real double precision routines, particularly various types of matrix-matrix multiplications and solution of linear systems of equations.
Complex routines are currently not of interest.
Therefore, the BLAS API of BLASFEO does not aim at replacing existing BLAS implementations, but at complementing them, improving the performance of selected routines for small matrices.
This can be obtained even without any modification to the existing BLAS implementations.

In case of compiling a program from source, this can be achieved by simply choosing the linking order of libraries, placing the BLASFEO library before the other BLAS and LAPACK libraries.
In this way, the linker looks for the symbols first in BLASFEO and then it looks for the unresolved symbols in the BLAS and LAPACK libraries.

In case of an existing program making use of dynamic libraries (as e.g. programs installed with package managers in many Linux distributions, looking for libraries in the standard search directories), the Unix command {\tt LD\_PRELOAD} can be employed to resolve all symbols in the BLASFEO dynamic library first, and then to resolve any missing symbol in the default libraries.
This {\tt LD\_PRELOAD} trick is employed in the experiments with Octave and SciPy presented next.

%%%%%%%%%%%%%%%%
\subsection{Performance plots} \label{sec:appl:plot}
%%%%%%%%%%%%%%%%

\begin{figure}[!t]
\centering
\subfloat[dgemm\_nn]{\includegraphics[width=0.45\linewidth]{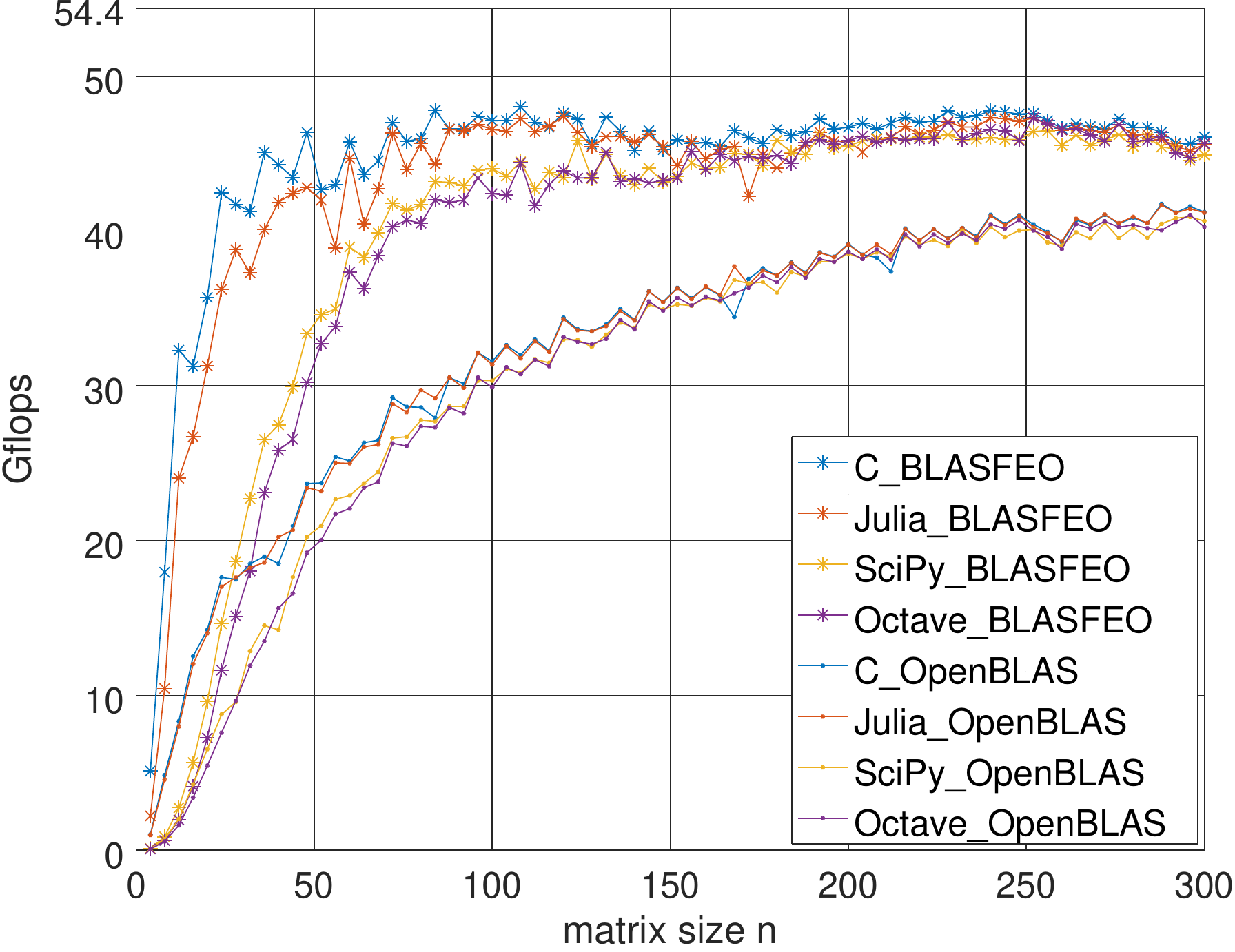} \label{fig:octave:perf:dgemm_nn}} %\\
\subfloat[dsyrk\_un]{\includegraphics[width=0.45\linewidth]{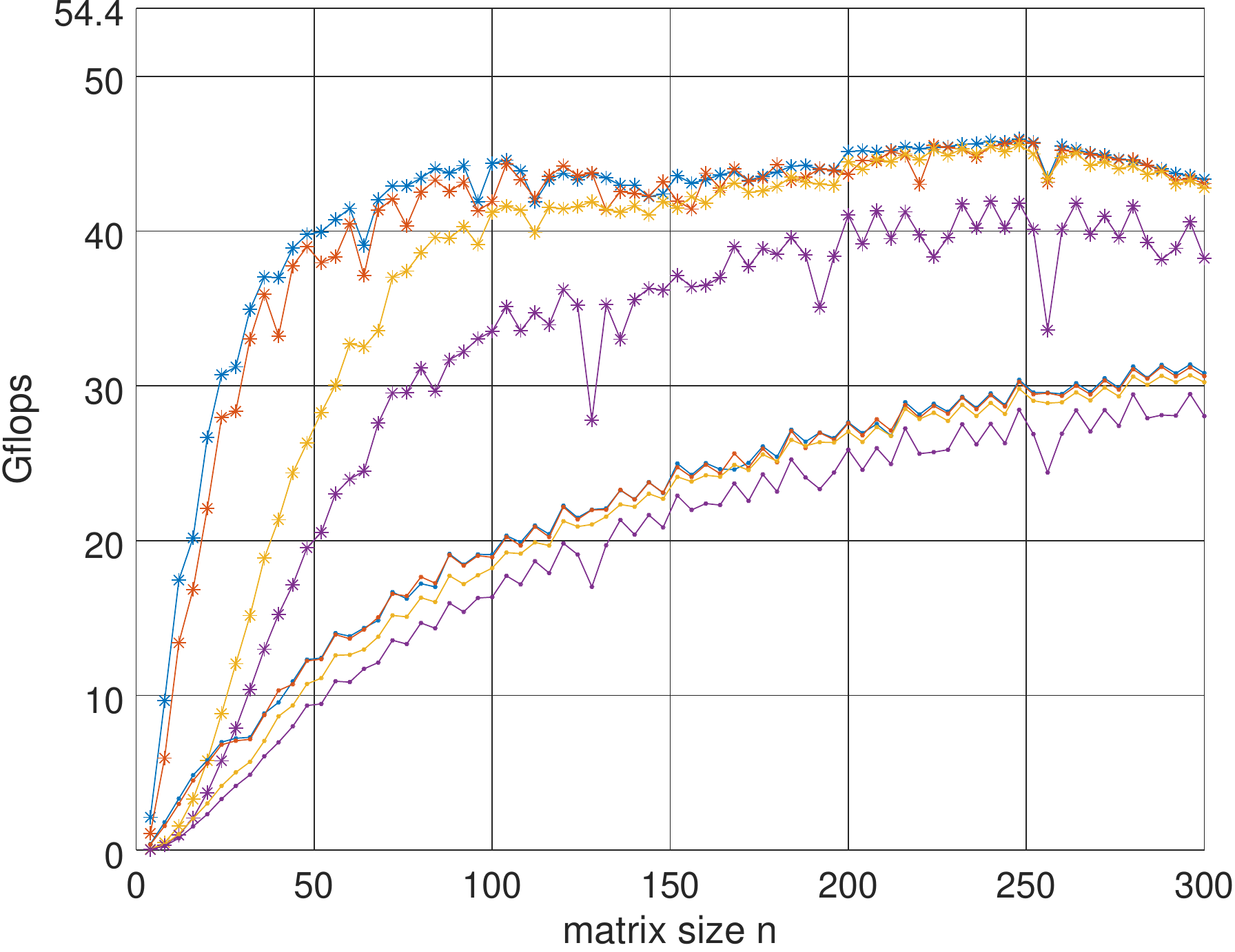} \label{fig:octave:perf:dsyrk_ut}} \\
\subfloat[dpotrf\_l]{\includegraphics[width=0.45\linewidth]{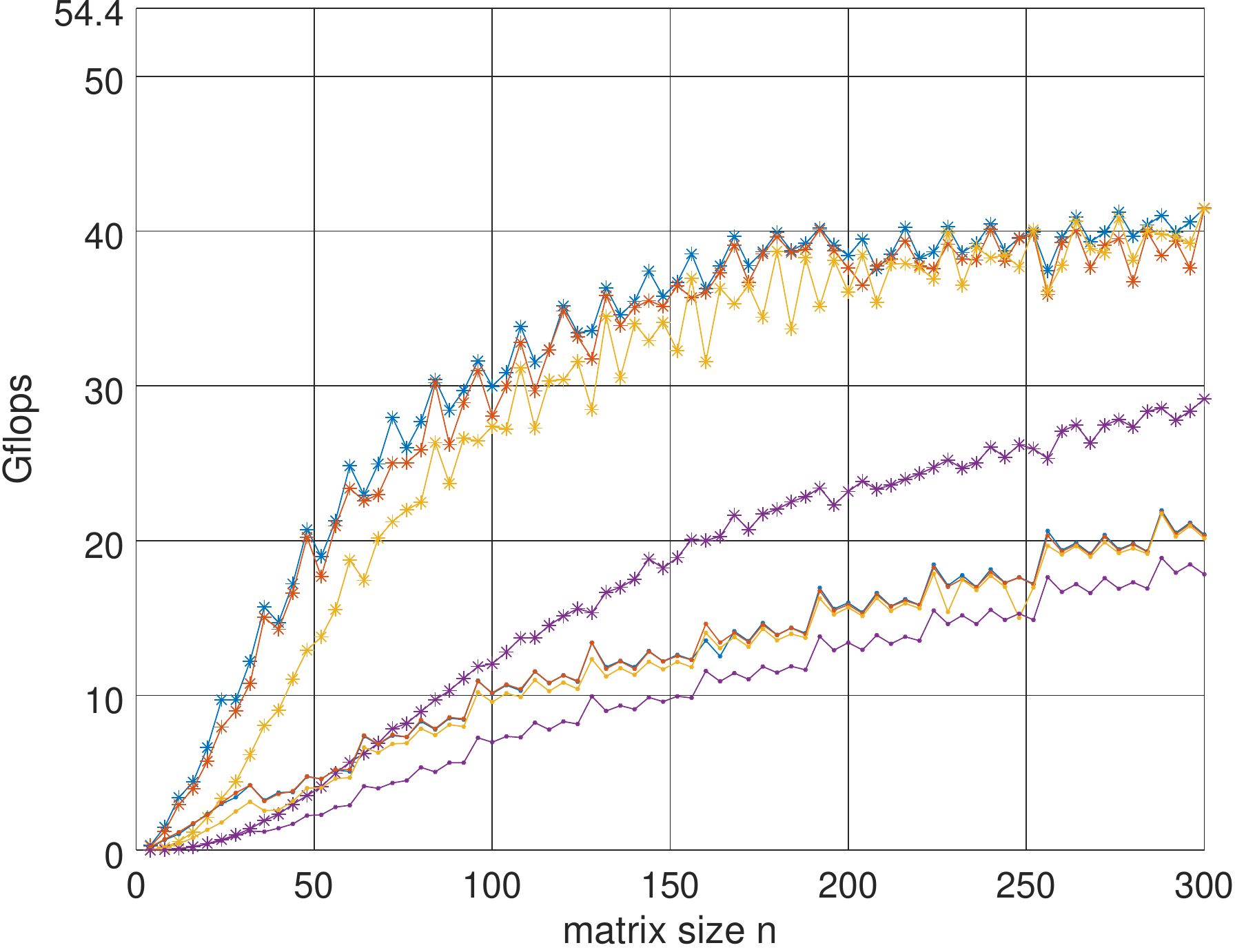} \label{fig:octave:perf:dpotrf_l}} %\\
\subfloat[dgetrf]{\includegraphics[width=0.45\linewidth]{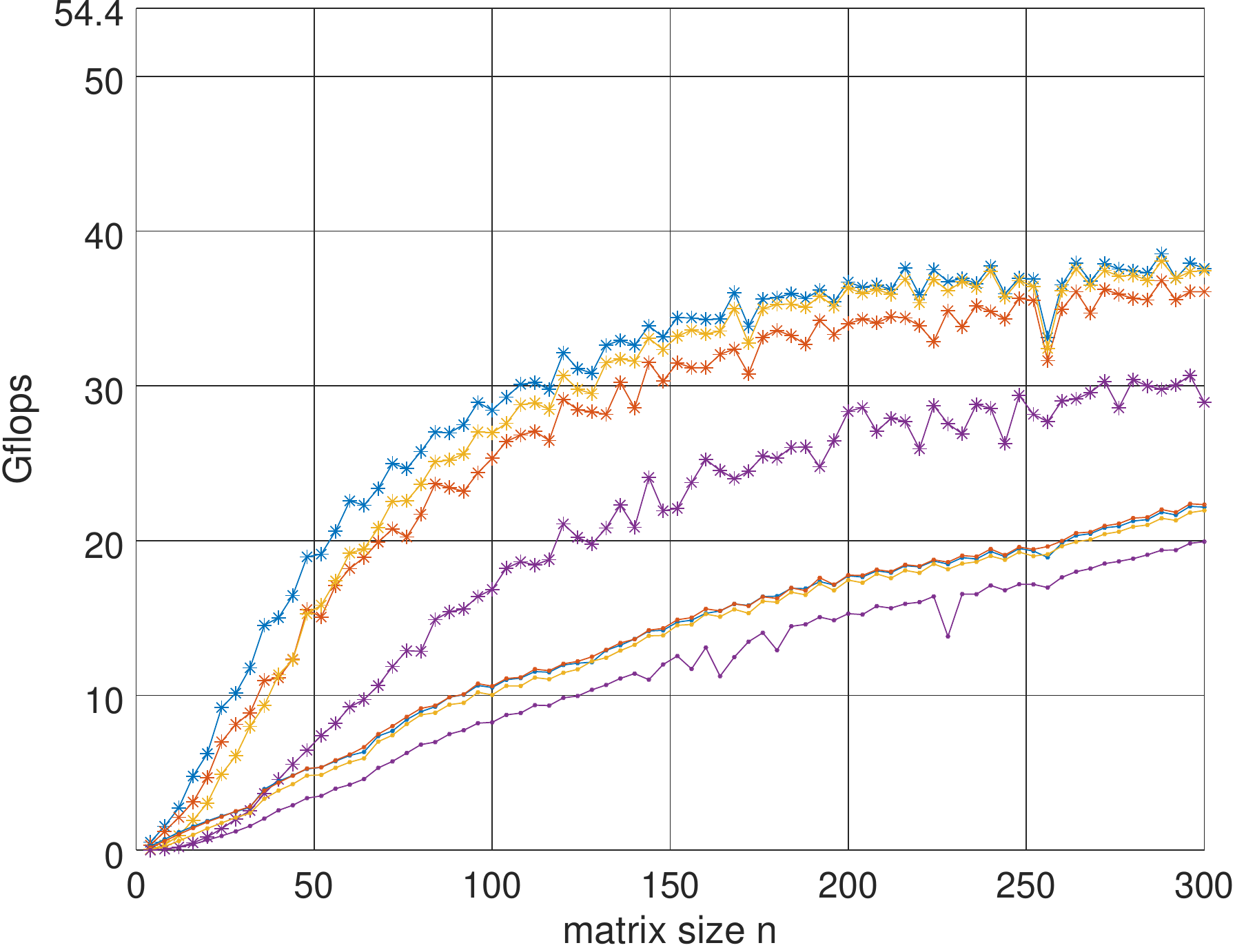} \label{fig:octave:perf:dgetrf}} \\
\caption{Performance of some BLAS or LAPACK routines in Octave 4.2.2, Python SciPy 1.2, Julia 1.0.1 and native calls in C, on Intel Haswell Core i7 4810MQ @ 3.4 GHz.
BLAS and LAPACK provided by single-threaded versions of either the BLAS API of BLASFEO or OpenBLAS 0.3.4.dev.}
\label{fig:octave:perf}
\end{figure}

This section compares the computational performance of some key linear algebra routines as employed in the scientific programming languages Octave, Python SciPy and Julia\footnote{For each language, when multiple commands call the same BLAS or LAPACK routine, the one with less overhead is employed.}, and with native calls to the BLAS or LAPACK routines from C code.
More detailed information about these scientific programming languages and their use of BLAS and LAPACK routines is reported in Section~\ref{sec:commands} in the appendix.
The performance of the BLAS API of BLASFEO is evaluated against OpenBLAS, as this is currently the best open-source implementation for the matrix sizes of interest, and generally the default BLAS and LAPACK version in open-source scientific software on Linux distributions.
The results are presented in Figure \ref{fig:octave:perf}.

%Figure \ref{fig:octave:perf} shows the computational performance of some native Octave commands and compares it with the performance of a call to the corresponding native BLAS or LAPACK routine, in the two cases of BLAS provided by BLASFEO and OpenBLAS.
The first fact worth noticing is that both Octave and (to a smaller extent) SciPy show some noticeable overhead. %, with Octave introducing more overhead over the native BLAS or LAPACK routines than SciPy. %, especially for specialized BLAS routines and LAPACK factorizations.
On the other hand, once the JIT compilation is terminated, the performance plots for Julia are almost indistinguishable from the ones for the native calls from C.
Since for the matrix sizes of interest BLASFEO routines are faster than the OpenBLAS counterparts, the impact of this overhead on performance is much more visible in the case of BLASFEO.
Nonetheless, the use of BLASFEO gives a significant speedup for all languages, especially for rather small matrices of size up to 100 where the speedup ranges from about 50\% in the case of {\tt dgemm}, up to doubling or more the performance in the case of the {\tt dpotrf} and {\tt dgetrf} factorizations.

In more detail, the overhead introduced by Octave is relatively small in the case of {\tt dgemm}, but it affects performance more in the case of {\tt dsyrk} (since the additional operation of copying the upper triangular part in the lower triangular part of the result matrix is performed) and in the case of the factorizations {\tt dpotrf} and {\tt dgetrf} (since additional operations are performed, like the copy in a new matrix before factorization and computation of the condition number after factorization).
In particular, all matrix copies are implemented with double nested loops in C++, and therefore they do not take advantage of optimized and vectorized level 1 BLAS routines.
Conversely, the low level SciPy wrappers to BLAS and LAPACK routines show considerably smaller overhead, which affects performance only for matrix sizes up to a few tens.
The Julia wrappers are JIT compiled and show almost unnoticeable overhead.

%Figure \ref{fig:octave:perf} shows the performance plot for selected BLAS and LAPACK routines, when called though native Octave commands.
%TODO

%%%%%%%%
\subsection{Riccati recursion} \label{sec:appl:ric}
%%%%%%%%

%As an example of the practical speedup given by the use of BLASFEO in Octave, this section presents the implementation of the Riccati recursion algorithm.
This section compares the running time of the Riccati recursion algorithm as implemented using BLAS and LAPACK routines through different programming languages, and it investigates to what extent the choice of the BLAS and LAPACK implementation affects performance.

The Riccati recursion algorithm is widely used in control applications, and for example it is encountered in the computation of the optimal state feedback in finite horizon linear quadratic regulator (LQR) or in the Kalman filter.
It can also be used to efficiently factorize the Karush-Kuhn-Tucker (KKT) matrix arising in constrained optimal control problems.
The backward Riccati recursion reads
\begin{equation*}
P_n = Q + A P_{n+1} A^T - (S + A P_{n+1} B^T) (R + B P_{n+1} B^T )^{-1} (S^T + B P_{n+1} A^T)
\end{equation*}
where the matrices $P_{n+1}$ and $\begin{bmatrix} R & S \\ S^T & Q \end{bmatrix}$ are assumed to be symmetric positive definite.
The matrices $A$, $Q$, $P$ have size $n_x\times n_x$, the matrices $B$ and $S^T$ have size $n_x\times n_u$ and the matrix $R$ has size $n_u\times n_u$, where $n_x$ is the number of states and $n_u$ is the number of controls in the dynamical system.
In finite horizon LQR or in the factorization of the KKT matrix in optimal control problems the recursion is repeated $N$ times, where $N$ is the control horizon length.
In the Kalman filter, one recursion step is performed at each sampling instant.

The Riccati recursion can be implemented efficiently as~\cite{Frison2014}
\begin{align*}
C \gets & \begin{bmatrix} B^T \\ A^T \end{bmatrix} \cdot \mathcal L_{n+1} \\
\begin{bmatrix} \Lambda_n & 0 \\ L_n & \mathcal L_n \end{bmatrix} \gets & \left(\begin{bmatrix} R & S \\ S^T & Q \end{bmatrix} + C \cdot C^T \right) ^{\tfrac 1 2}
\end{align*}
where $\mathcal L_{n}$ is the lower Cholesky factor of $P_{n}$ and the exponent $^{\tfrac 1 2}$ denotes the lower triangular Cholesky factorization.
The algorithm can be implemented using the {\tt dtrmm\_rlnn} and {\tt dsyrk\_ln} BLAS routines and the {\tt dpotrf\_l} LAPACK routine.
All these routines are available natively in C and through thin wrappers in Julia 1.0.1 and SciPy 1.2 (after the fix to {\tt dtrmm}, see Section~\ref{sec:commands:python} in the appendix).
The {\tt dtrmm} routine is unused in Octave 4.2.2.
If not fixed, the {\tt dtrmm} routine in SciPy 1.2 gives the wrong result and it needs to be substituted with {\tt dgemm} in the Riccati recursion implementation.

\begin{figure}[!t]
\centering
\subfloat[BLASFEO vs OpenBLAS: time]{\includegraphics[width=0.45\linewidth]{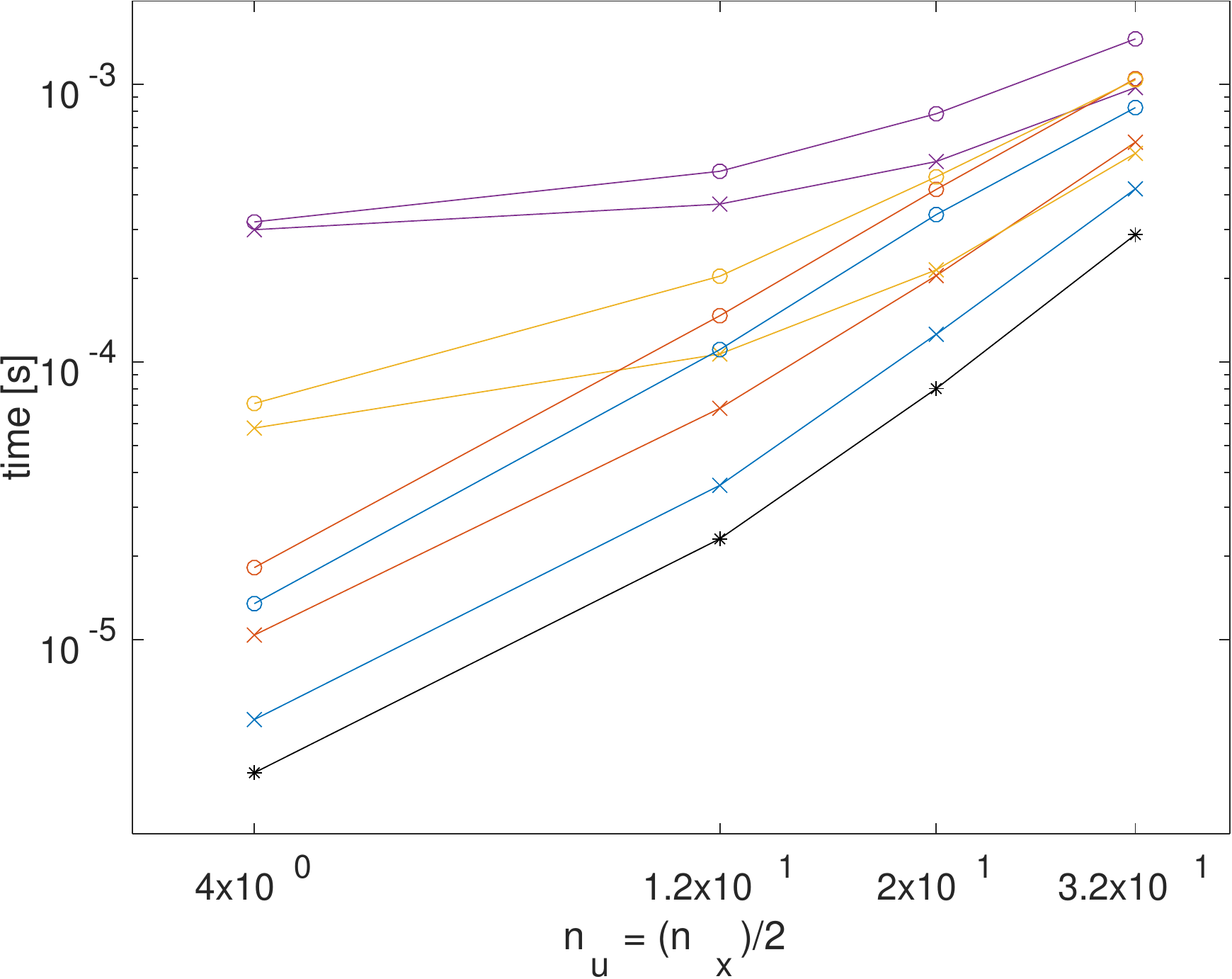} \label{fig:appl:ric:time}}
\subfloat[BLASFEO vs OpenBLAS: speedup]{\includegraphics[width=0.45\linewidth]{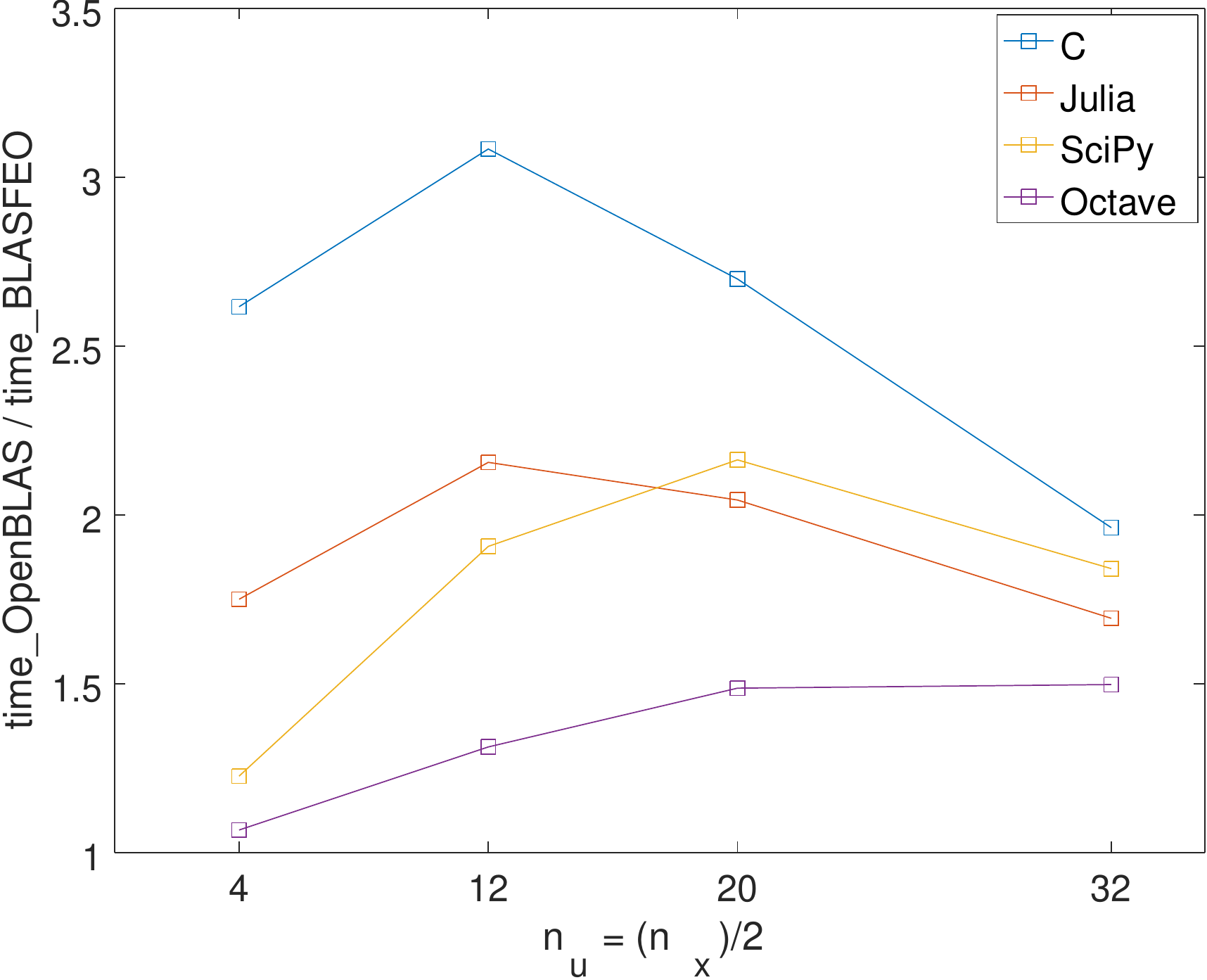} \label{fig:appl:ric:speedup}} \\
\subfloat[BLASFEO vs MKL: time]{\includegraphics[width=0.45\linewidth]{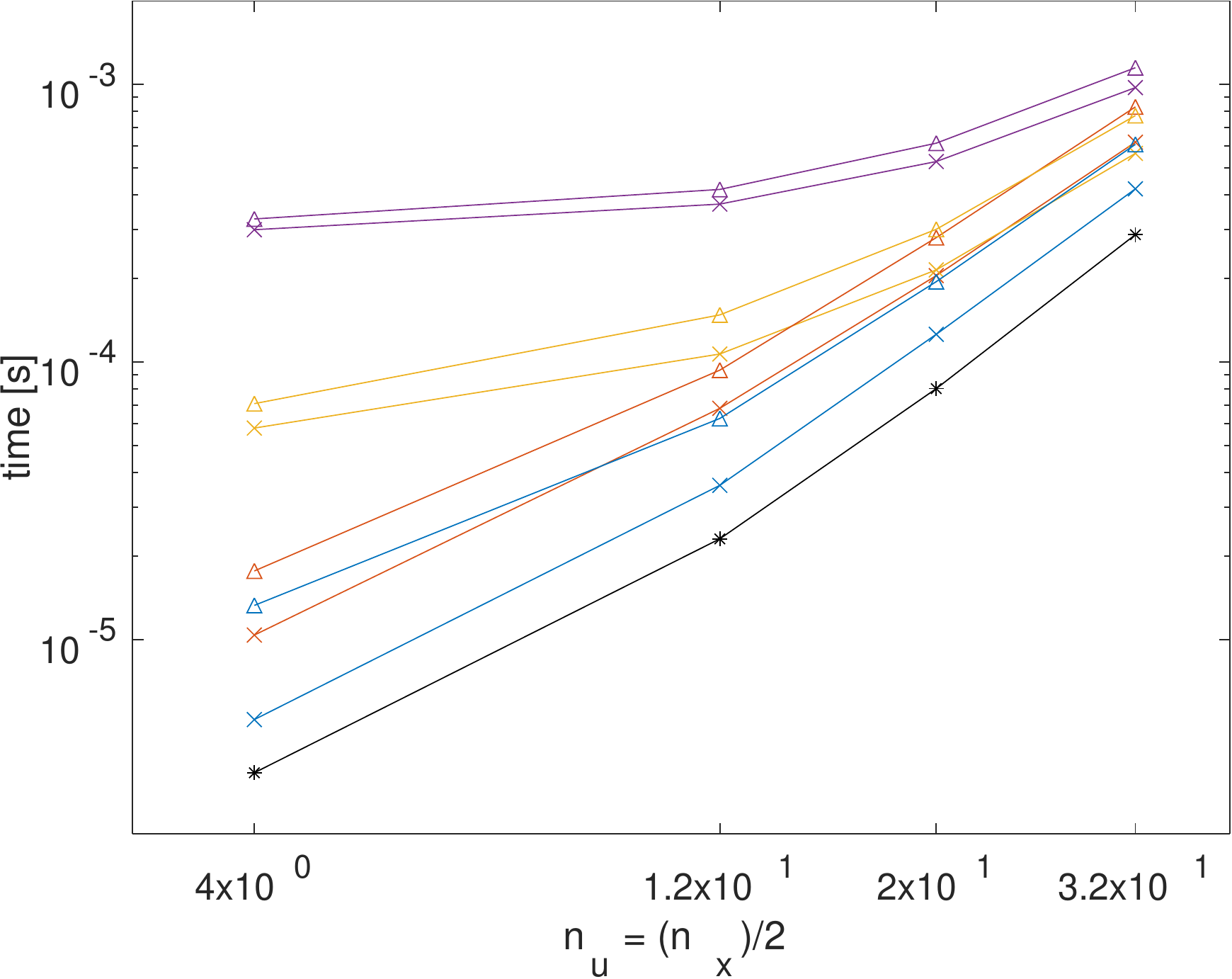} \label{fig:appl:ric:mkl_time}}
\subfloat[BLASFEO vs MKL: speedup]{\includegraphics[width=0.45\linewidth]{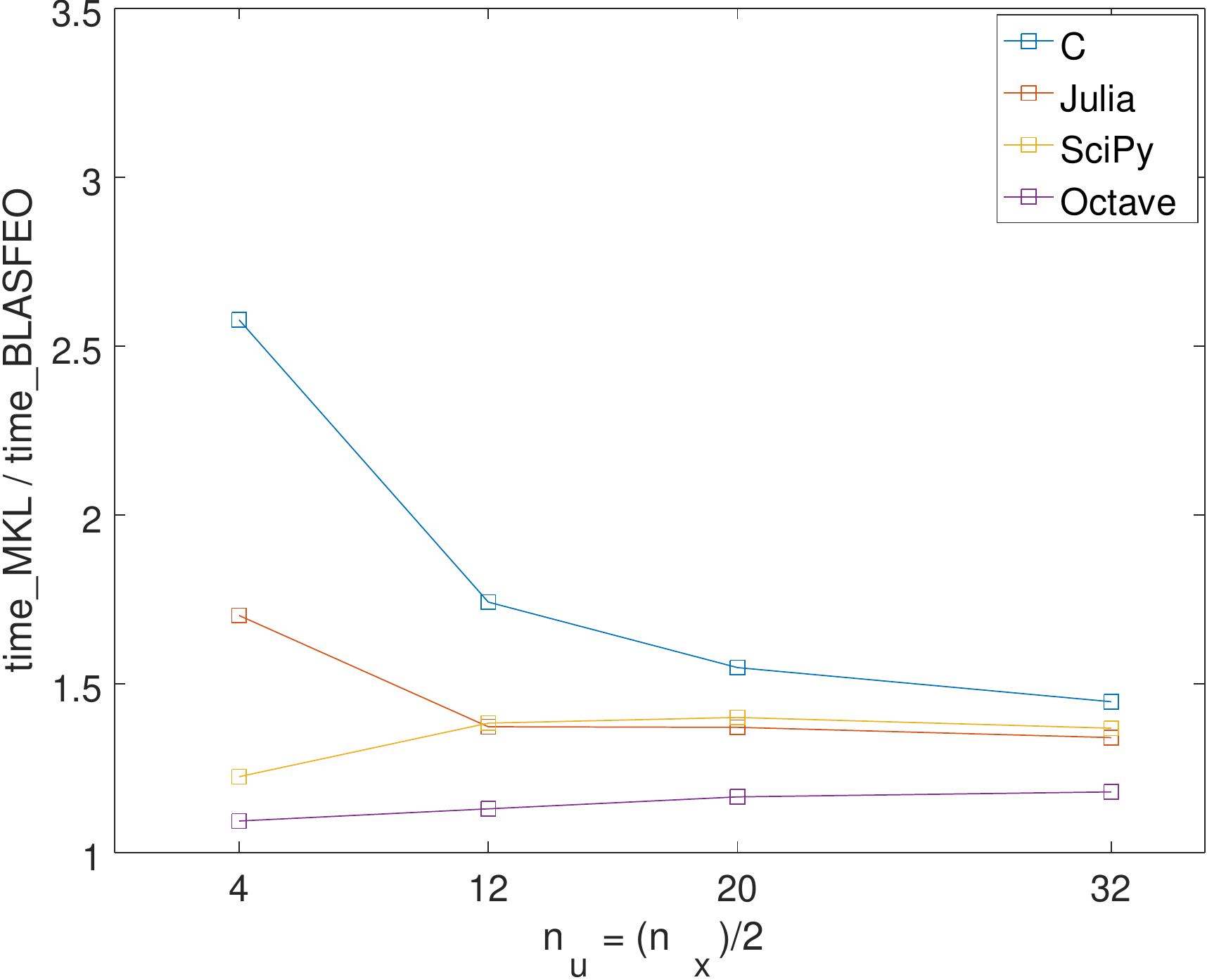} \label{fig:appl:ric:mkl_speedup}}
\caption{Graphical representation of the data in Table~\ref{tab:appl:ric}.
Computation time (in seconds) and speedup of Riccati recursion implemented natively in Octave 4.2.2 (plotted in purple), SciPy 1.2 (with fix to {\tt dtrmm}) (plotted in yellow), Julia 1.0.1 (plotted in red) and in C (plotted in blue), on Intel Haswell Core i7 4810MQ @ 3.4 GHz.
BLAS and LAPACK provided by single-threaded versions of either BLASFEO (plotted with the cross `-x-'), OpenBLAS 0.3.4-dev (plotted with the circle `-o-') or MKL 2019.1.144 (plotted with the triangle `-$\bigtriangleup$-').
The BLASFEO API is also added as a reference (plotted in black with the asterisk `-*-').
The control horizon length is fixed to $N=10$; the number of states $n_x$ and the number of controls $n_u$ vary.}
\label{fig:appl:ric}
\end{figure}

Table~\ref{tab:appl:ric} contains the solution time for the Riccati recursion implemented using native BLAS and LAPACK calls in C and through wrappers in Octave 4.2.2, SciPy 1.2 and Julia 1.0.1.
The BLAS and LAPACK routines are provided by either the BLAS API of BLASFEO, OpenBLAS 0.3.4.dev or MKL 2019.1.144.
The performance of the Riccati recursion implemented using the BLASFEO API is added as an additional comparison.
Figure~\ref{fig:appl:ric} presents graphical representations of the data in Table~\ref{tab:appl:ric}. % (except the unfixed SciPy).

In case of Octave and SciPy, the choice of the BLAS and LAPACK implementation barely affects performance for the smallest example (with matrix sizes in the order of the units), as the overhead of these interpreted scripting languages dominates the computation time.
For the other examples (with matrix sizes in the order of tens), the BLAS and LAPACK implementation has a significant effect on the performance, and the BLAS API of BLASFEO gives a speedup ranging between 1.5 and 2 times over OpenBLAS and between 1.2 and 1.4 times over MKL.
Julia employs JIT compilation and therefore, once the Riccati recursion module is compiled, the overhead is several times lower than interpreted scripting languages, and also in the smallest example the BLAS API of BLASFEO gives a speedup of about 1.7 times over both OpenBLAS and MKL.
The C implementation of the Riccati recursion has even lower overhead, and therefore the BLAS API of BLASFEO can show its full potential, with a speedup ranging between 2 and 3 times over OpenBLAS and between 1.5 and 2.5 times over MKL (which in the C implementation is used with the MKL\_DIRECT\_CALL\_SEQ option). %roughly 3 times for the smallest examples and 2 for the largest.

The BLASFEO API shows an additional speedup of 1.5 over the BLAS API of BLASFEO.
Note that this speedup is larger than the speedup of the single routines evaluated in the performance plots in Section~\ref{sec:perf:haswell}.
%The additional speedup comes from the fact that the BLASFEO API is non-destructive (there is an additional matrix argument reserved for the output, often avoiding the need for an explicit matrix copy), it provides specialized routines (e.g. fusing {\tt dsyrk} with a subsequent {\tt dpotrf} in a single routine) and by the fact that it employs none or very limited internal memory (the reduction in the memory footprint increases the amount of memory which is kept in cache in between iterations of the Riccati recursion).
The additional speedup comes from: the fact that the BLASFEO API is non-destructive (there is an additional matrix argument reserved for the output, often avoiding the need for an explicit matrix copy); the fact that the BLASFEO API provides specialized routines (e.g. fusing {\tt dsyrk} with a subsequent {\tt dpotrf} in a single routine); and the fact that the BLASFEO API employs none or very limited internal memory (the reduction in the memory footprint increases the amount of memory which is kept in cache in between iterations of the Riccati recursion).
This experimentally shows that the BLASFEO API can be a better fit for embedded applications than a similarly implemented BLAS API.

\begin{table}
\centering
\caption{Computation time (in seconds) of Riccati recursion implemented natively in Octave 4.2.2, SciPy 1.2 (with fix to {\tt dtrmm}), Julia 1.0.1 and in C, on Intel Haswell Core i7 4810MQ @ 3.4 GHz.
BLAS and LAPACK provided by single-threaded versions of either BLASFEO, OpenBLAS 0.3.4-dev or MKL 2019.1.144.
The BLASFEO API is also added as a reference.
The control horizon length is fixed to $N=10$; the number of states $n_x$ and the number of controls $n_u$ vary.}
\label{tab:appl:ric}
\begin{tabular}{c|c||c|c|c|c}
 & BLAS & $n_x=8$ & $n_x=24$ & $n_x=40$ & $n_x=64$ \\
API & implemen. & $n_u=4$ & $n_u=12$ & $n_u=20$ & $n_u=32$ \\
\hline
\hline
Octave 4.2.2 & OpenBLAS & $3.20\cdot 10^{-4}$ & $4.87\cdot 10^{-4}$ & $7.85\cdot 10^{-4}$ & $1.46\cdot 10^{-3}$ \\
Octave 4.2.2 & MKL & $3.28\cdot 10^{-4}$ & $4.19\cdot 10^{-4}$ & $6.15\cdot 10^{-4}$ & $1.15\cdot 10^{-3}$ \\
Octave 4.2.2 & BLASFEO & $3.00\cdot 10^{-4}$ & $3.71\cdot 10^{-4}$ & $5.28\cdot 10^{-4}$ & $9.75\cdot 10^{-4}$ \\
\hline
%Python SciPy 1.2 & OpenBLAS & $7.17\cdot 10^{-5}$ & $2.09\cdot 10^{-4}$ & $4.88\cdot 10^{-4}$ & $1.13\cdot 10^{-3}$ \\
%Python SciPy 1.2 & MKL & $7.67\cdot 10^{-5}$ & $1.43\cdot 10^{-4}$ & $3.33\cdot 10^{-4}$ & $8.55\cdot 10^{-4}$ \\
%Python SciPy 1.2 & BLASFEO & $5.96\cdot 10^{-5}$ & $1.11\cdot 10^{-4}$ & $2.40\cdot 10^{-4}$ & $6.47\cdot 10^{-4}$ \\
%\hline
%Python SciPy 1.2+fix & OpenBLAS & $7.10\cdot 10^{-5}$ & $2.04\cdot 10^{-4}$ & $4.65\cdot 10^{-4}$ & $1.04\cdot 10^{-3}$ \\
%Python SciPy 1.2+fix & MKL & $7.09\cdot 10^{-5}$ & $1.48\cdot 10^{-4}$ & $3.01\cdot 10^{-4}$ & $7.73\cdot 10^{-4}$ \\
%Python SciPy 1.2+fix & BLASFEO & $5.79\cdot 10^{-5}$ & $1.07\cdot 10^{-4}$ & $2.15\cdot 10^{-4}$ & $5.65\cdot 10^{-4}$ \\
Python SciPy 1.2 & OpenBLAS & $7.10\cdot 10^{-5}$ & $2.04\cdot 10^{-4}$ & $4.65\cdot 10^{-4}$ & $1.04\cdot 10^{-3}$ \\
Python SciPy 1.2 & MKL & $7.09\cdot 10^{-5}$ & $1.48\cdot 10^{-4}$ & $3.01\cdot 10^{-4}$ & $7.73\cdot 10^{-4}$ \\
Python SciPy 1.2 & BLASFEO & $5.79\cdot 10^{-5}$ & $1.07\cdot 10^{-4}$ & $2.15\cdot 10^{-4}$ & $5.65\cdot 10^{-4}$ \\
\hline
Julia 1.0.1 & OpenBLAS & $1.82\cdot 10^{-5}$ & $1.47\cdot 10^{-4}$ & $4.19\cdot 10^{-4}$ & $1.05\cdot 10^{-3}$ \\
Julia 1.0.1 & MKL & $1.77\cdot 10^{-5}$ & $9.36\cdot 10^{-5}$ & $2.81\cdot 10^{-4}$ & $8.31\cdot 10^{-4}$ \\
Julia 1.0.1 & BLASFEO & $1.04\cdot 10^{-5}$ & $6.82\cdot 10^{-5}$ & $2.05\cdot 10^{-4}$ & $6.20\cdot 10^{-4}$ \\
\hline
BLAS from C & OpenBLAS & $1.35\cdot 10^{-5}$ & $1.11\cdot 10^{-4}$ & $3.40\cdot 10^{-4}$ & $8.26\cdot 10^{-4}$ \\
BLAS from C & MKL & $1.33\cdot 10^{-5}$ & $6.27\cdot 10^{-5}$ & $1.95\cdot 10^{-4}$ & $6.09\cdot 10^{-4}$ \\
BLAS from C & BLASFEO & $5.11\cdot 10^{-6}$ & $3.60\cdot 10^{-5}$ & $1.25\cdot 10^{-4}$ & $4.19\cdot 10^{-4}$ \\
\hline
BLASFEO from C & BLASFEO & $3.24\cdot 10^{-6}$ & $2.29\cdot 10^{-5}$ & $7.95\cdot 10^{-5}$ & $2.87\cdot 10^{-4}$ \\
\end{tabular}
\end{table}

%%%%%%%%%%%%%%%%
%\subsection{Direct sparse solvers} \label{sec:appl:sparse}
%%%%%%%%%%%%%%%%

%%%%%%%%
%\subsubsection{MA57} \label{sec:appl:sparse:ma57}
%%%%%%%%

%TODO

%%%%%%%%%%%%%%%%%%%%%%%%%%%%%%%%
\section{Conclusion} \label{sec:concl}
%%%%%%%%%%%%%%%%%%%%%%%%%%%%%%%%

This paper describes the implementation of a standard BLAS API in the BLASFEO framework, which optimizes the performance for small matrices fitting in cache.
The proposed implementation scheme for a generic DLA routine switches between two algorithmic variants: the first one focuses on reducing overhead for small matrices (by limiting packing and avoiding dynamic memory allocation), and the other attains a performance level scaling better with the matrix size (by improving data layout and movement across the cache hierarchy). % and providing a performance level which scales better with the matrix size.
In special cases such as some {\tt dgemm} variants and when dealing with rectangular matrices, two supplementary algorithmic variants can provide additional speedups.
The flexibility of the assembly sub-routines framework is leveraged to conveniently implement kernels which operate on matrix arguments of different storage format.

Numerical experiments show the effectiveness of the approach.
In the BLASFEO framework, the routines of the BLAS API are on average 10-15\% slower than the BLASFEO API counterparts, but for the matrix sizes of interest they are nonetheless significantly faster than state-of-the-art open-source BLAS and LAPACK implementations, and in most cases they also outperform commercial implementations.
The performance advantage is large especially in the case of LAPACK routines such as factorizations, where the speedup factor is about 2 for matrices of size up to 100.

These results are in a way more significant than the BLASFEO API presented in~\cite{Frison2018}, since the implementation described in the current paper sticks to the standard BLAS API, and therefore it is directly comparable to the other BLAS and LAPACK implementations.

Future research will focus on adding routines operating on additional matrix formats (e.g. the general stride proposed in~\cite{Zee2015}) by leveraging the flexibility of the assembly sub-routines framework.
Furthermore, the implementation of multi-threaded versions of DLA routines in the BLASFEO framework will be investigated, maintaining the focus on performance for small matrices.

%%%%%%%%%%%%%%%%%%%%%%%%%%%%%%%%
\appendix
%%%%%%%%%%%%%%%%%%%%%%%%%%%%%%%%

%%%%%%%%%%%%%%%%%%%%%%%%%%%%%%%%
\section{Implementing BLASFEO for a new architecture} \label{sec:new_isa}
%%%%%%%%%%%%%%%%%%%%%%%%%%%%%%%%

This section discusses the effort required to implement BLASFEO for a new architecture, and shows the performance attained for different optimization levels for the Intel Haswell and ARM Cortex A57 assumed to be a new architectures in BLASFEO.

BLASFEO has been designed with the goal of enhancing performance for small matrices.
In this it differs from other BLAS implementations such as BLIS, which has been designed with the explicit goal of being a `framework for rapidly instantiating BLAS functionality'~\cite{Zee2015}.
As discussed in detail in the current paper as well as in the original BLASFEO paper \cite{Frison2018}, in order to attain its goal, BLASFEO employs a large number of assembly coded kernels, tailored to each linear algebra routine and optimized for each hardware target.
The use of the custom function calling convention described in Section~\ref{sec:impl:kernel} helps to manage this large amount of assembly code. %, by introducing modularity thanks to a systematic way of decomposing assembly routines into elementary sub-routines.
However, the question remains, of how large is the effort required to implement BLASFEO for a new architecture.

In BLASFEO, only the innermost loop is coded as an assembly kernel.
The framework around the kernels is coded in C, and it is tailored to a fixed panel size $p_s$.
All hardware targets characterized by the same $p_s$ share the same framework.
Furthermore, the `generic' target of BLASFEO is coded in C in such a way that all kernels for the level 3 BLAS and LAPACK routines of the BLASFEO API and algorithmic variant `B' of the BLAS API are built on top of the single {\tt kernel\_dgemm\_nt\_4x4\_lib4444} kernel (this is analogue to what happens in the corresponding assembly kernels for other targets, which are all built on top of the {\tt inner\_kernel\_dgemm\_nt\_XxX\_lib44} inner kernel).
Therefore, if an optimized implementation of this single kernel is available, all these routines automatically get optimized to some degree.

\begin{figure}[!t]
\centering
\subfloat[dgemm\_nn]{\includegraphics[width=0.45\linewidth]{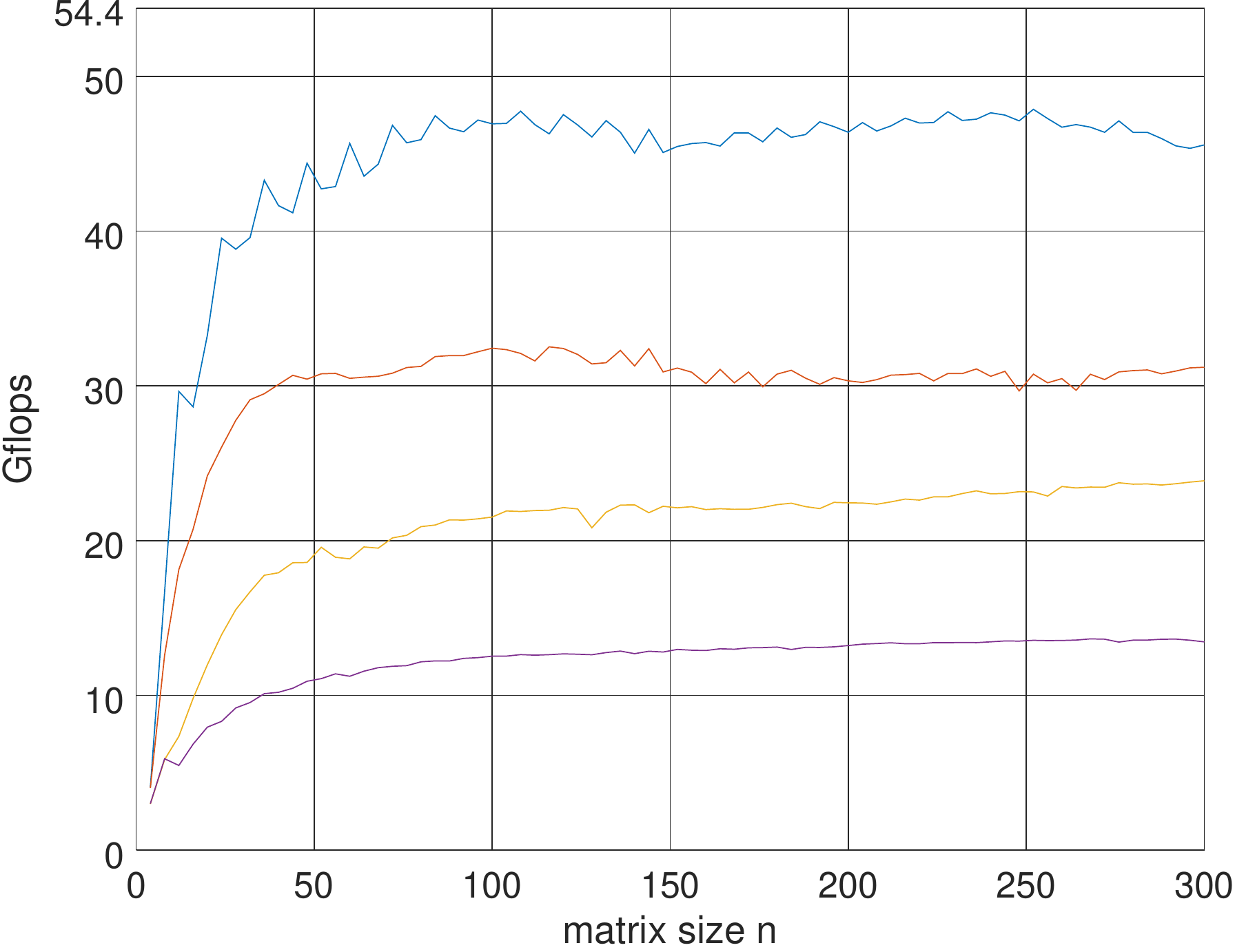} \label{fig:new_isa:haswell:dgemm_nn}} %\\
\subfloat[dpotrf\_l]{\includegraphics[width=0.45\linewidth]{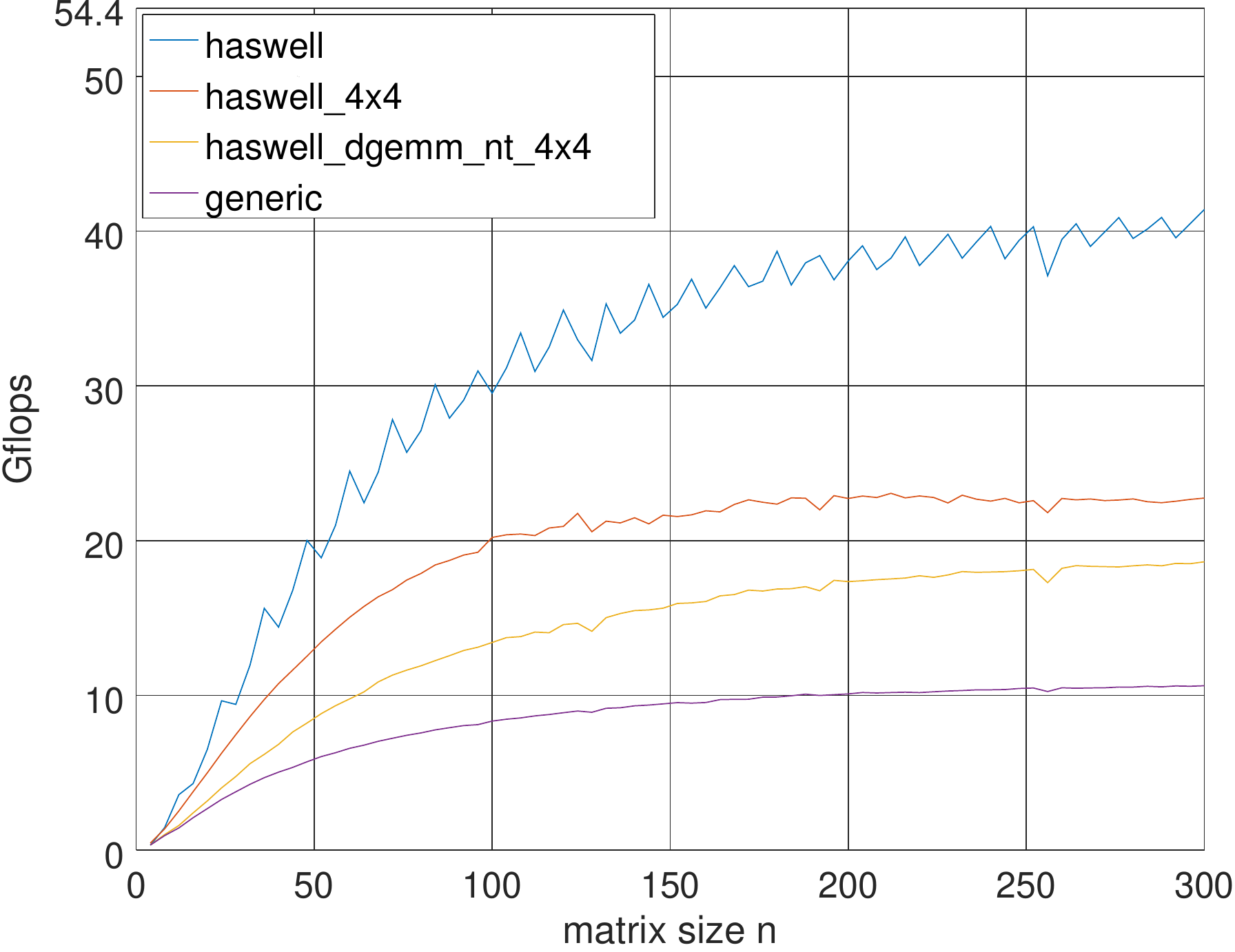} \label{fig:new_isa:haswell:dpotrf_l}} %\\
\caption{Comparison of different implementation efforts for the routines {\tt dgemm\_nn} and {\tt dpotrf\_l}: performance on Intel Haswell Core i7 4810MQ @ 3.4 GHz.
The optimization levels are: `haswell' target with optimal $12\times4$ kernel size ({\it haswell}); `haswell' target with $4\times4$ kernel size ({\it haswell\_4x4}); `generic` target with the only {\tt kernel\_dgemm\_nt\_4x4\_lib4444} optimized for the `haswell' target ({\it haswell\_dgemm\_nt\_4x4}); `generic' target ({\it generic}).}
\label{fig:new_isa:haswell}
\end{figure}

\begin{figure}[!t]
\centering
\subfloat[dgemm\_nn]{\includegraphics[width=0.45\linewidth]{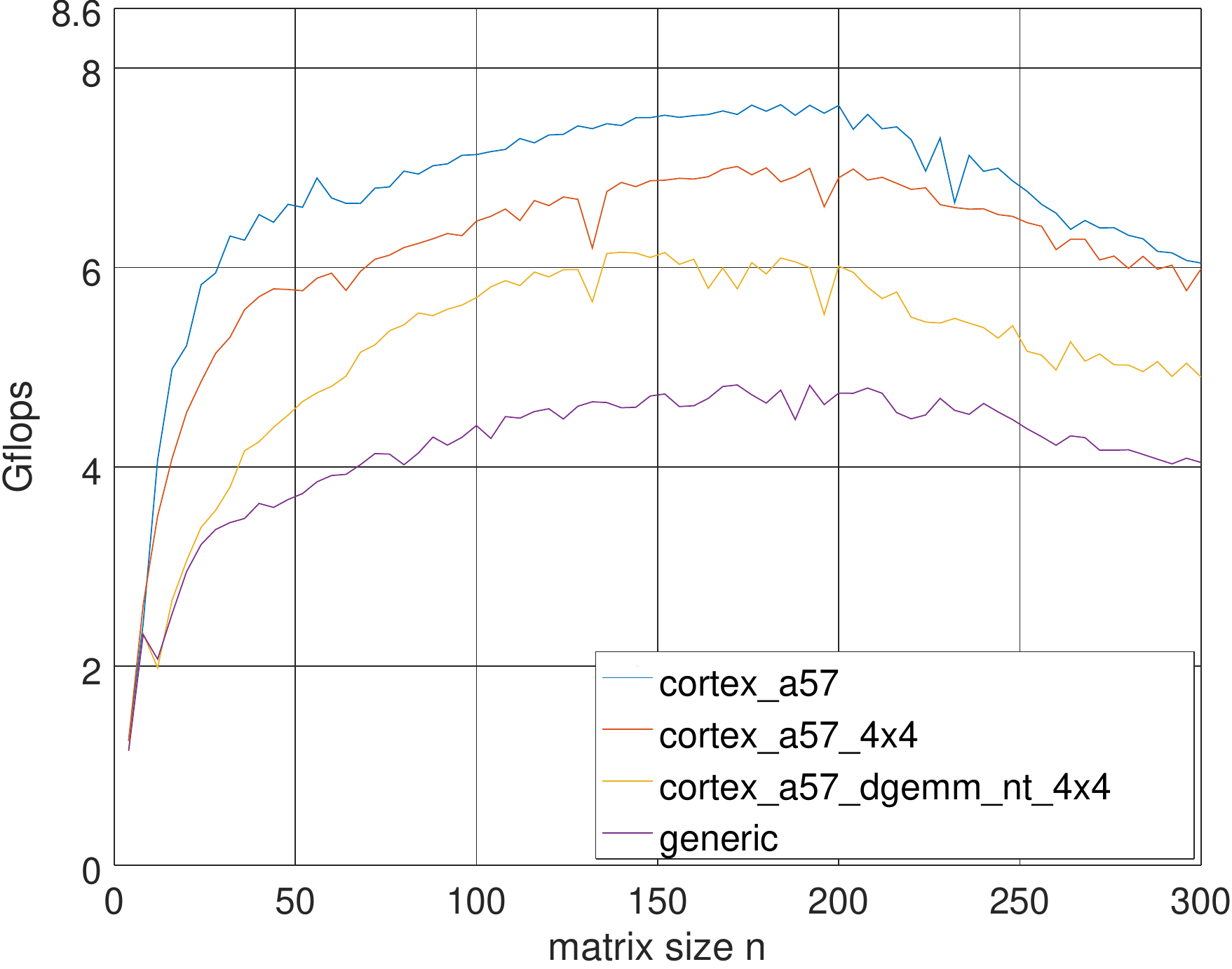} \label{fig:new_isa:cortexa_57:dgemm_nn}} %\\
\subfloat[dpotrf\_l]{\includegraphics[width=0.45\linewidth]{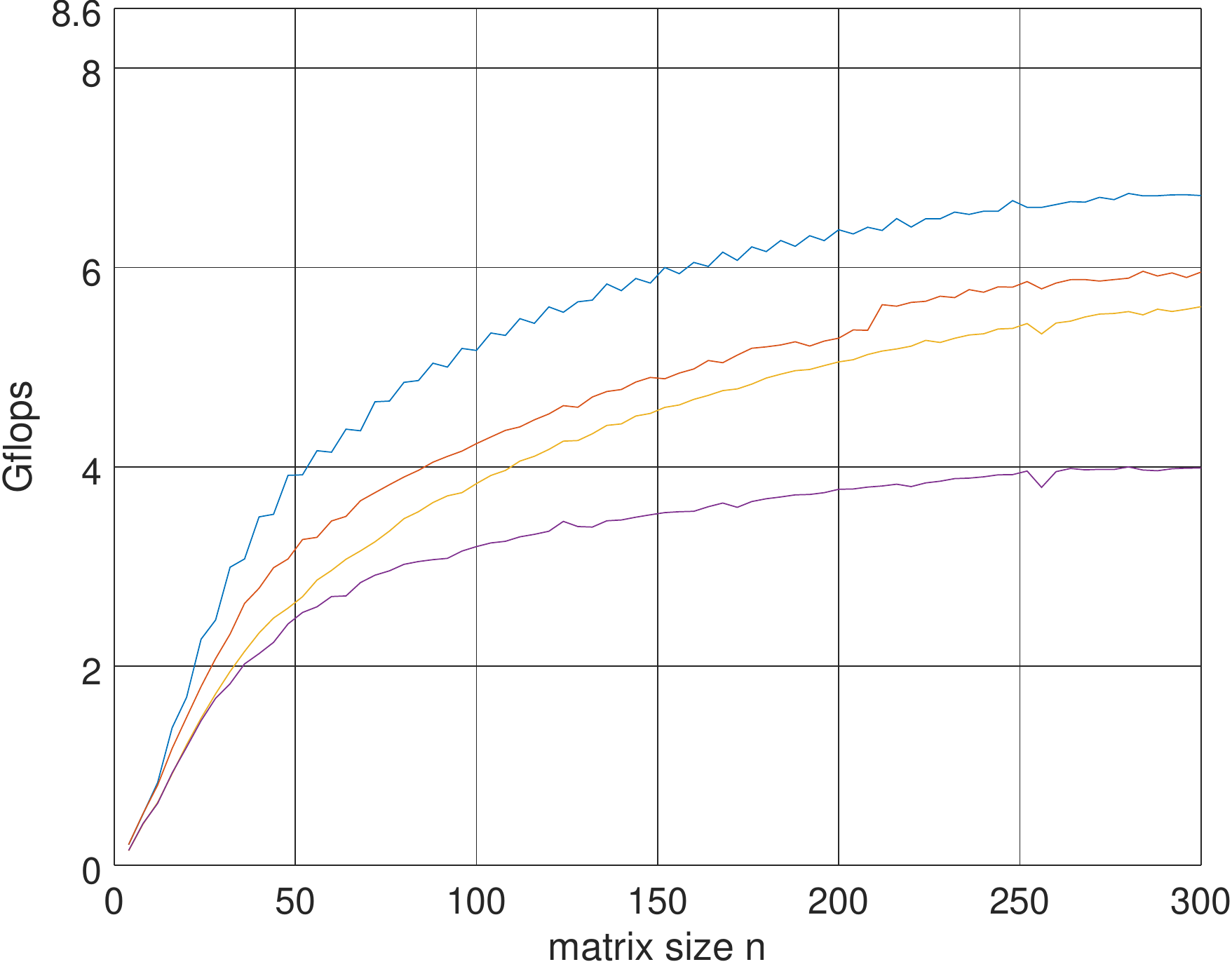} \label{fig:new_isa:cortexa_57:dpotrf_l}} %\\
\caption{Comparison of different implementation efforts for the routines {\tt dgemm\_nn} and {\tt dpotrf\_l}: performance on ARM Cortex A57 @ 2.15 GHz.
The optimization levels are: `cortex\_a57' target with optimal $8\times4$ kernel size ({\it cortex\_a57}); `cortex\_a57' target with $4\times4$ kernel size ({\it cortex\_a57\_4x4}); `generic' target with the only {\tt kernel\_dgemm\_nt\_4x4\_lib4444} optimized for the `cortex\_a57' target ({\it cortex\_a57\_dgemm\_nt\_4x4}); `generic' target ({\it generic}).}
\label{fig:new_isa:cortex_a57}
\end{figure}

Figures~\ref{fig:new_isa:haswell} and~\ref{fig:new_isa:cortex_a57} show the performance of different optimization efforts of the {\tt dgemm\_nn} and {\tt dpotrf\_l} routines on the Intel Haswell and ARM Cortex A57 (which, here, serve as new architectures for illustration purposes).
The `generic' target (purple line) performs rather poorly on Intel Haswell (since it cannot exploit the wide FP execution units of this microarchitecture), but it exceeds 50\% of the maximum performance on the lower power ARM Cortex A57.
The optimization of the single {\tt kernel\_dgemm\_nt\_4x4\_lib4444} kernel (yellow line) gives most of the performance improvement for the linear algebra routines based on {\tt 4x4} kernels, and as the matrix size increases, most of the computations are performed by this single kernel.
The optimization of all the other {\tt 4x4} kernels (red line) mainly improves performance for small matrices.
Finally, by optimizing all kernels for the optimal kernel size (blue line) the performance is further improved, especially in the case of Intel Haswell, which has an optimal kernel size of {\tt 12x4}.

These findings are representative of a more generic trend.
In case of low-power microarchitectures with narrow FP execution units, the `generic' target of BLASFEO generally provides a good basis, and the optimization of few key kernels gives most of the performance improvement.
In case of high-performance microarchitectures with wide FP execution units, the `generic' target gives rather poor performance, and it may not even use the optimal $p_s$ value.

In case of an architecture with a new value of $p_s$, a new version of the entire framework as well as of the assembly kernels needs to be implemented.
As an example, this is the case for the AVX512 ISA (which at the time of writing is not supported yet in BLASFEO) since doubling the FP registers width from 256 to 512 bits requires doubling the value of $p_s$ from 4 to 8 in double precision and from 8 to 16 in single precision.

%%%%%%%%%%%%%%%%%%%%%%%%%%%%%%%%
\section{Performance evaluation of additional parametric variants} \label{sec:more_variants}
%%%%%%%%%%%%%%%%%%%%%%%%%%%%%%%%

The previous sections of this paper evaluated the performance of the BLASFEO implementation of some BLAS and LAPACK routines against state-of-the-art open-source and proprietary implementations.
For compactness reasons, the comparison had to be limited to selected parametric variants of the routines like transposition, side, upper or lower triangular and so on.
These variants were chosen as representative of the performance of the routines in the general case.

\begin{figure}[!t]
\centering
\subfloat[dgemm]{\includegraphics[width=0.45\linewidth]{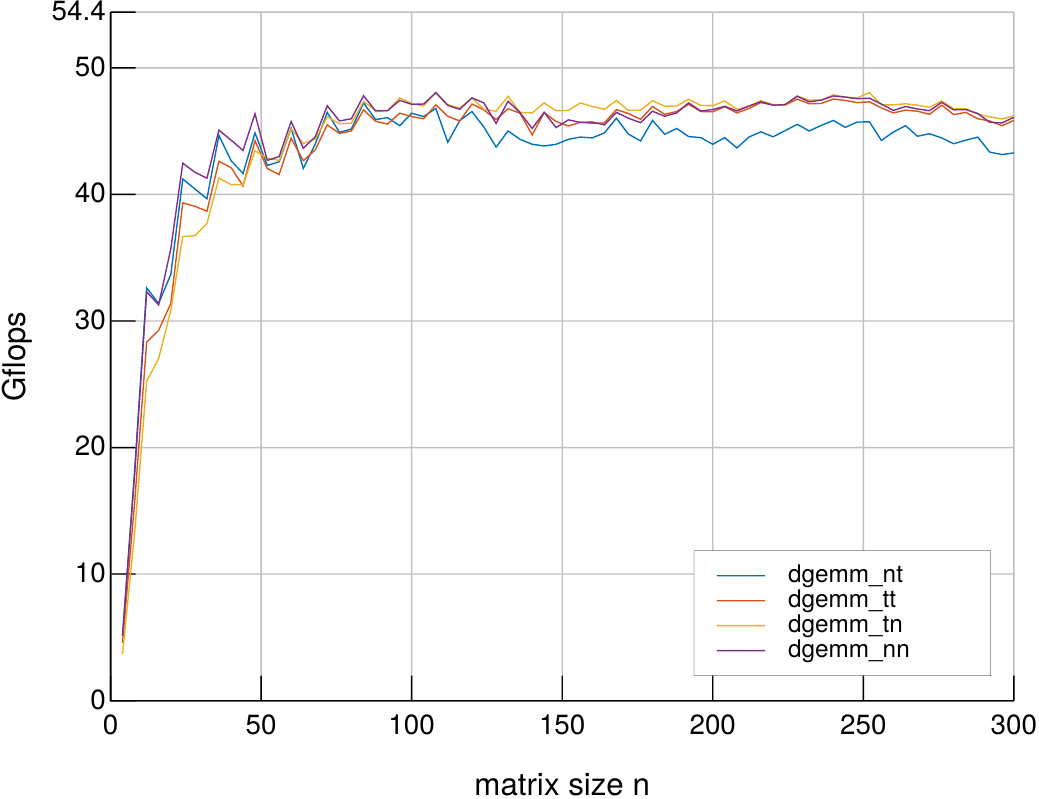} \label{fig:bench:haswell:dgemm}} %\\
\subfloat[dsyrk]{\includegraphics[width=0.45\linewidth]{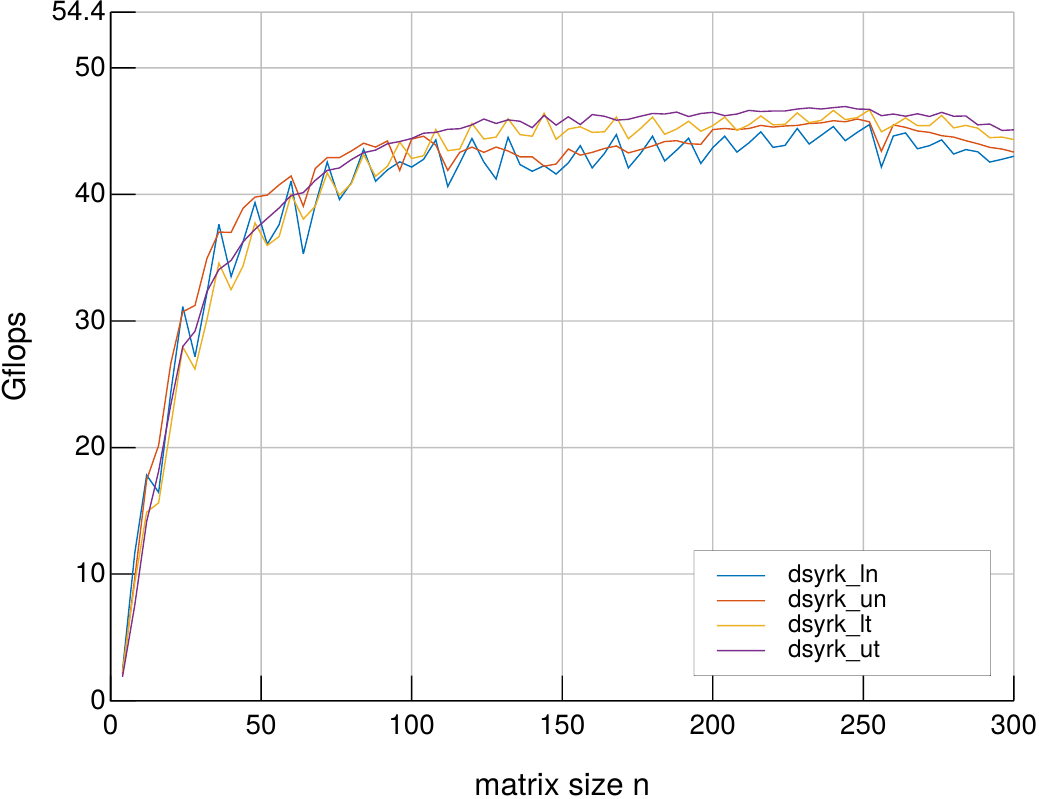} \label{fig:bench:haswell:dsyrk}} \\
\subfloat[dtrmm]{\includegraphics[width=0.45\linewidth]{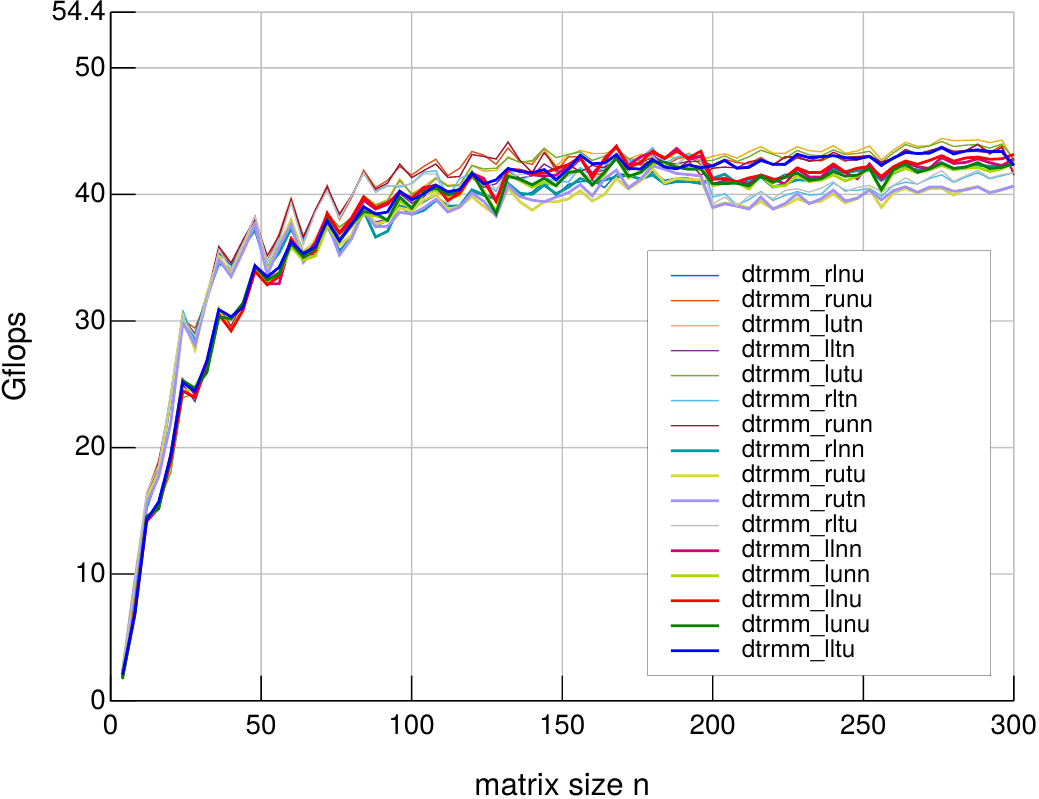} \label{fig:bench:haswell:dtrmm}} %\\
\subfloat[dtrsm]{\includegraphics[width=0.45\linewidth]{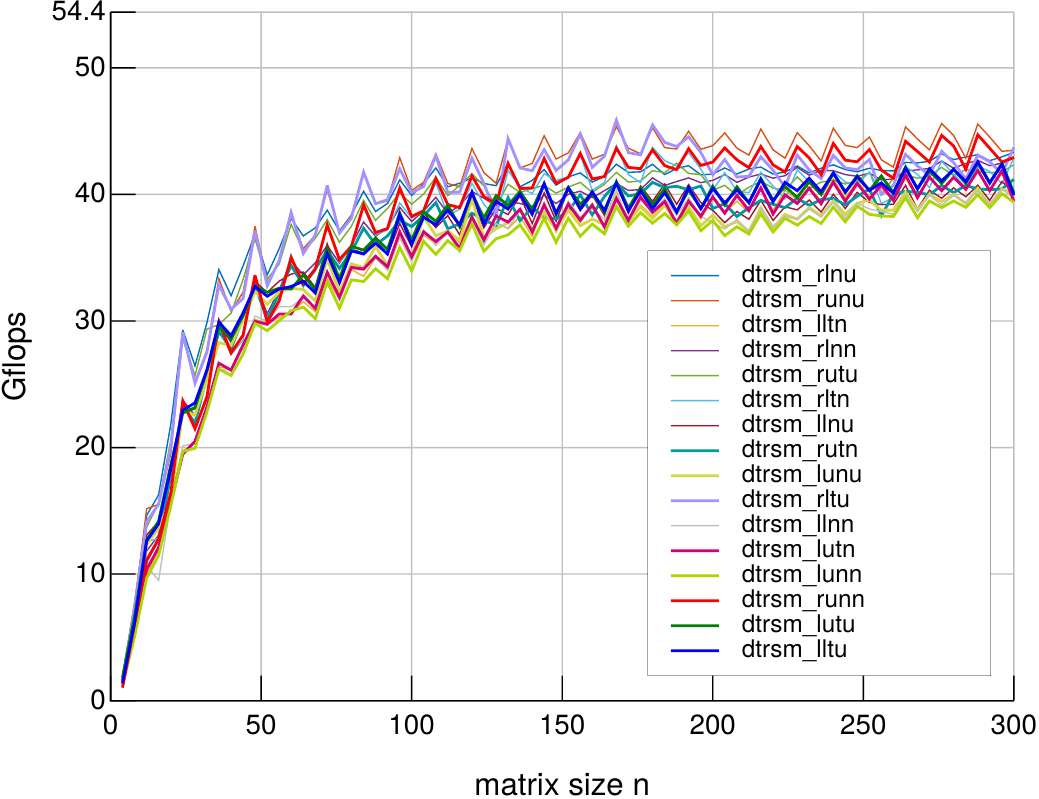} \label{fig:bench:haswell:dtrsm}} \\
\subfloat[dpotrf]{\includegraphics[width=0.45\linewidth]{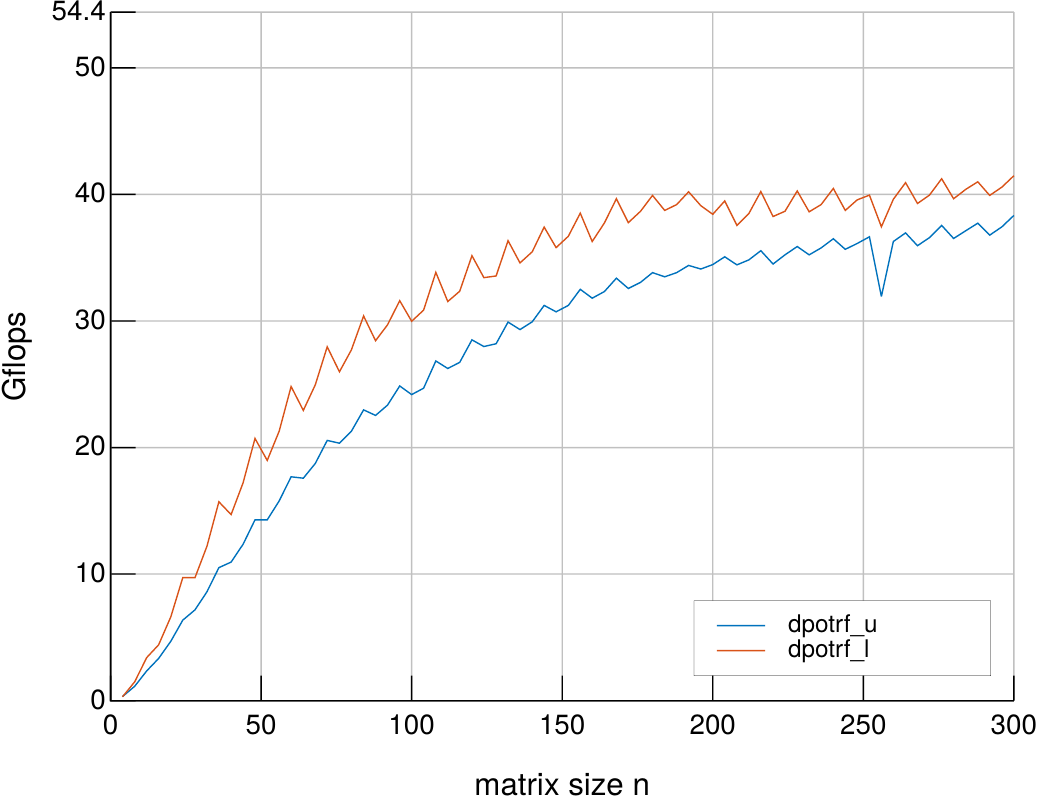} \label{fig:bench:haswell:dpotrf}} %\\
\subfloat[dgetrf]{\includegraphics[width=0.45\linewidth]{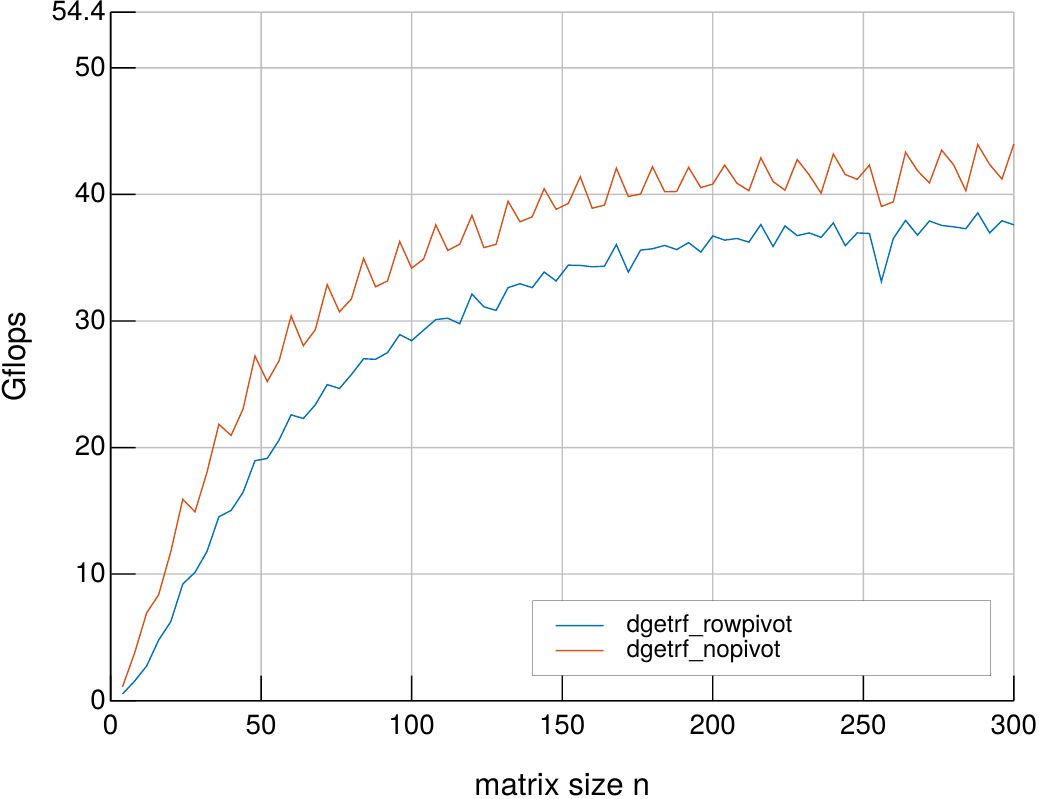} \label{fig:bench:haswell:dgetrf}} \\
\caption{Performance of all parametric variants of some BLAS and LAPACK routines on one core of the Intel Haswell Core i7 4810MQ @ 3.4 GHz.
All matrices are square of size $n$.}
\label{fig:bench:haswell}
\end{figure}

Conversely, this section aims at comparing the BLASFEO BLAS API implementation of all parametric variants of some BLAS and LAPACK routines against each others.
In Figure~\ref{fig:bench:haswell} the performance of the BLAS routines {\tt dgemm}, {\tt dsyrk}, {\tt dtrmm}, {\tt dtrsm} and the LAPACK routines {\tt dpotrf}, {\tt dgetrf} is evaluated for the case of square matrices on the Intel Haswell microarchitecture.
From the plots it is clear that the implementation approach presented in this paper performs consistently well on all parametric variants, with small performance variations due to the nature of the variants themselves.
No inherent performance drawback is found in the proposed implementation approach.

%%%%%%%%%%%%%%%%%%%%%%%%%%%%%%%%
\section{{\tt dgemm} performance evaluation on skinny matrices} \label{sec:skinny}
%%%%%%%%%%%%%%%%%%%%%%%%%%%%%%%%

Some considerations about the case of rectangular matrices are collected in Section~\ref{sec:alg:comp:skinny}.
However, in the previous sections of this paper, the performance evaluation of BLASFEO routines had to be limited to the case of square matrices.

\begin{figure}[!t]
\centering
\subfloat[dgemm\_nn, m=4, n=4, k=s]{\includegraphics[width=0.41\linewidth]{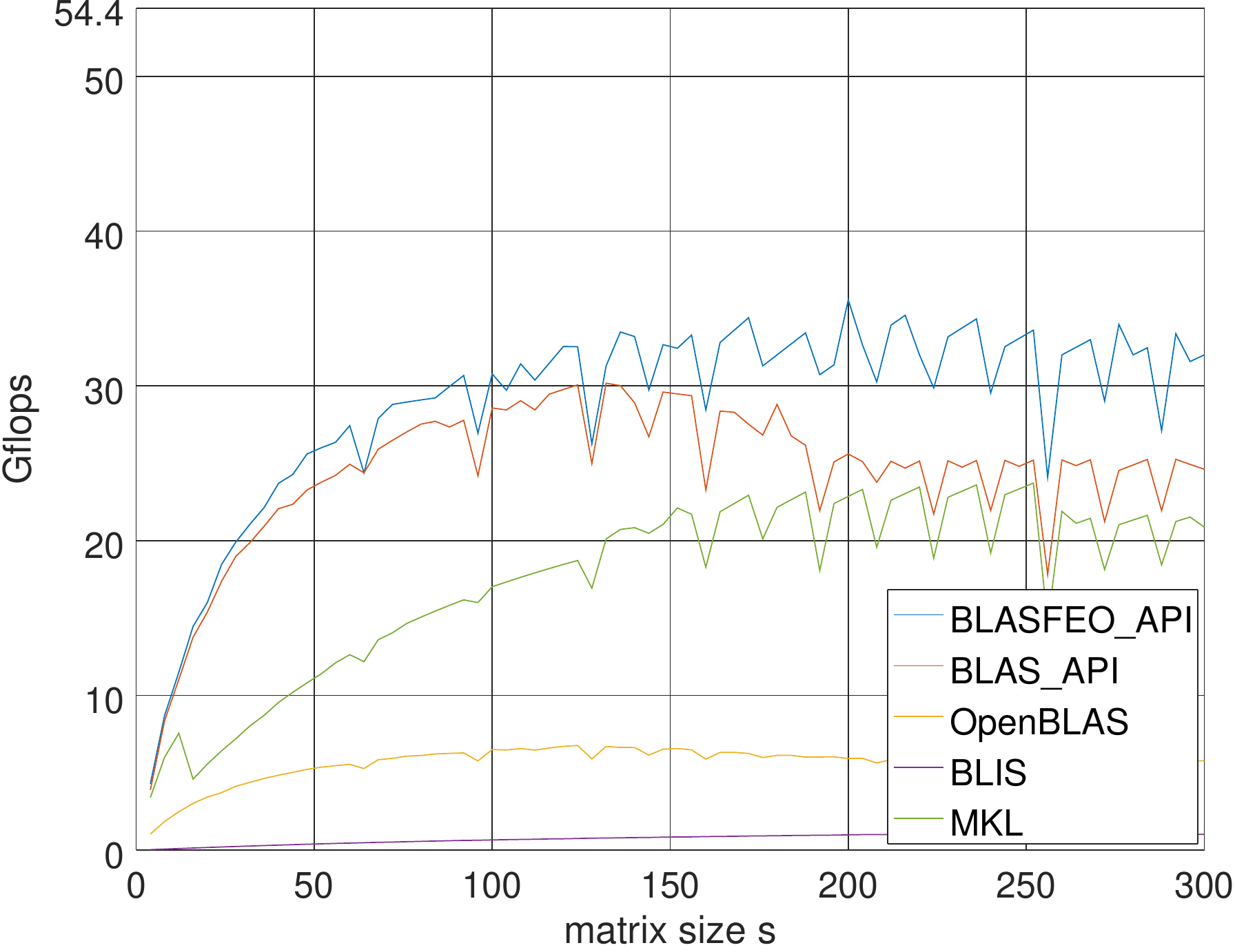} \label{fig:skinny:haswell:dgemm_nn_44k}} %\\
\subfloat[dgemm\_nn, m=4, n=s, k=s]{\includegraphics[width=0.41\linewidth]{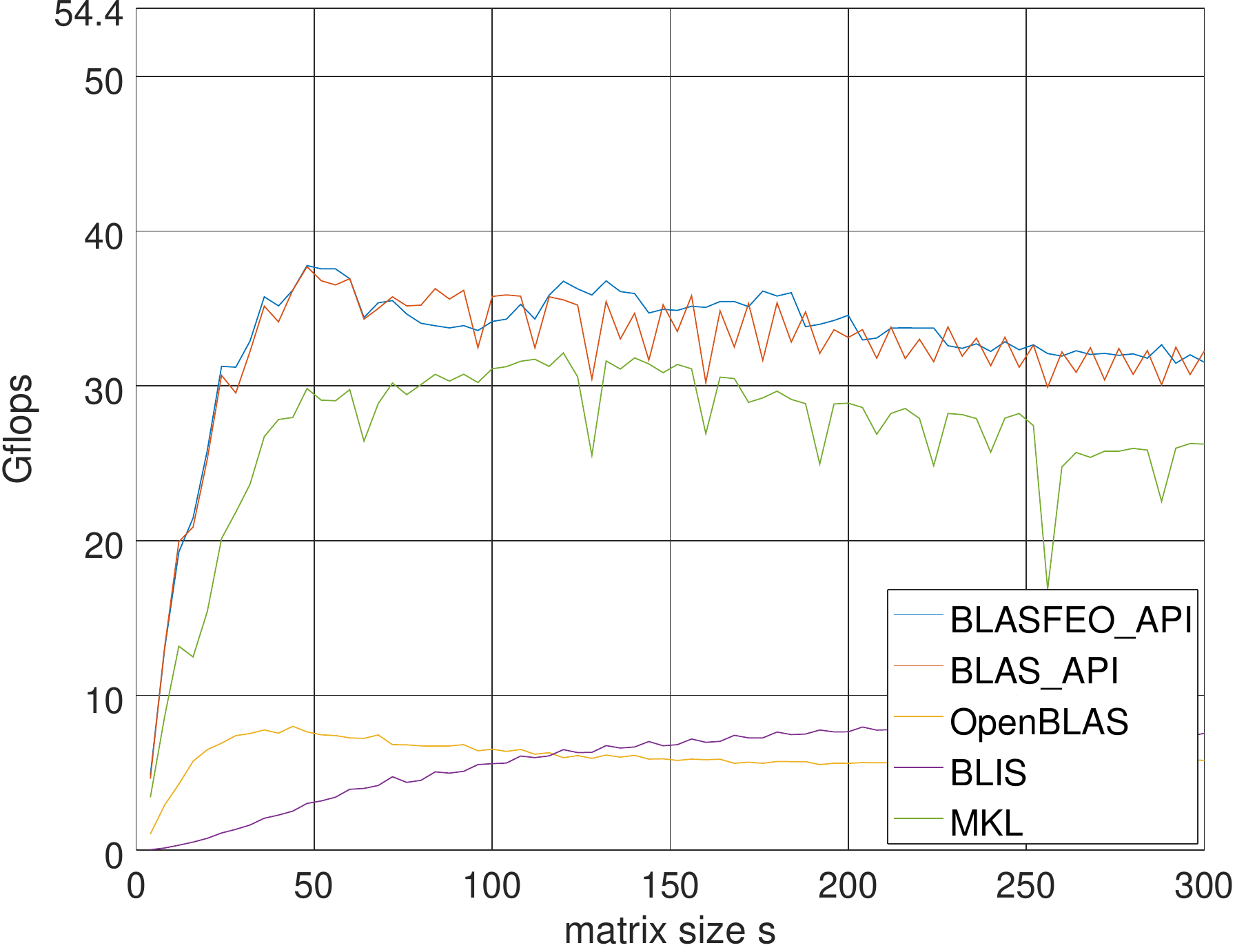} \label{fig:skinny:haswell:dgemm_nn_4nk}} \\
\subfloat[dgemm\_nt, m=4, n=4, k=s]{\includegraphics[width=0.41\linewidth]{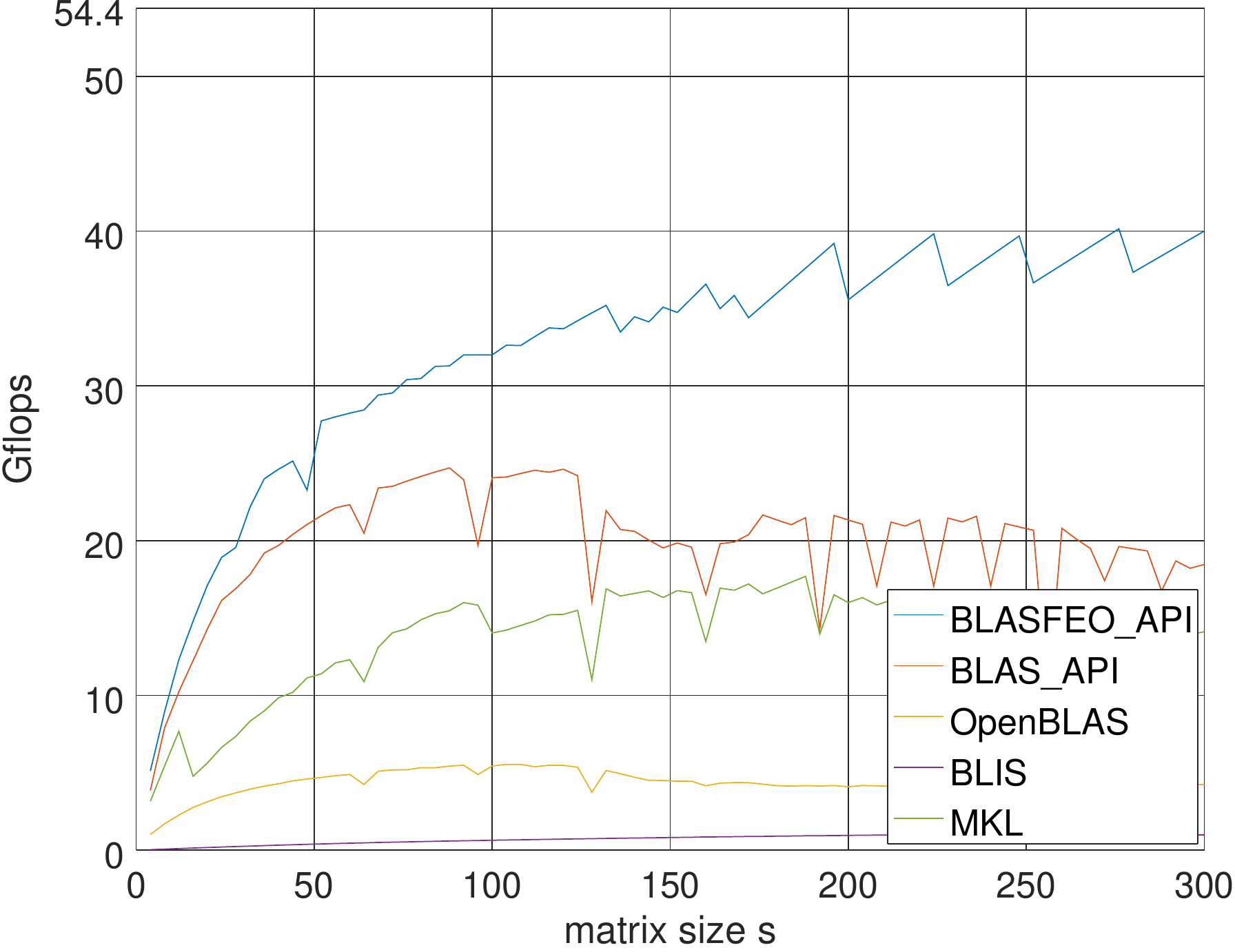} \label{fig:skinny:haswell:dgemm_nt_44k}} %\\
\subfloat[dgemm\_nt, m=4, n=s, k=s]{\includegraphics[width=0.41\linewidth]{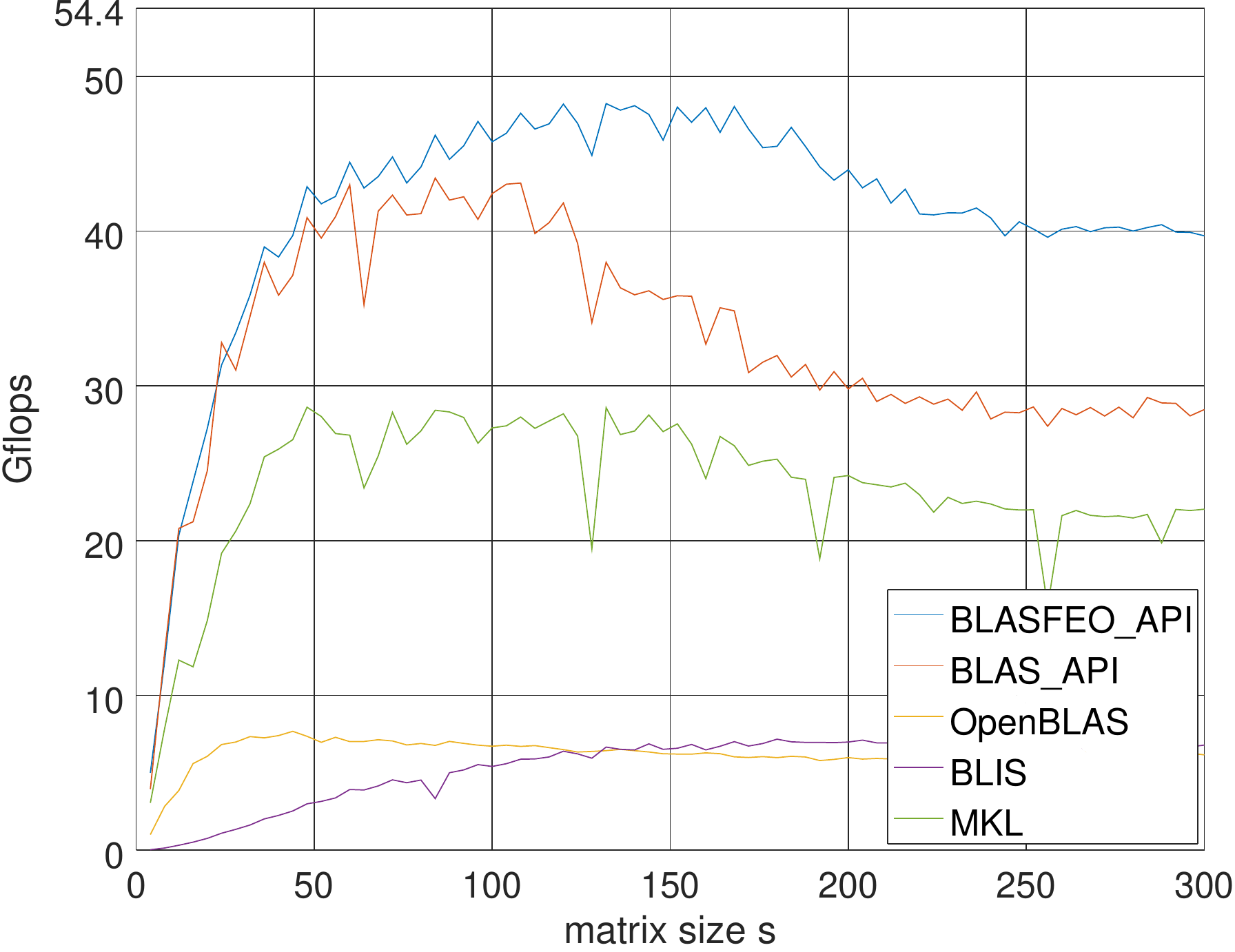} \label{fig:skinny:haswell:dgemm_nt_4nk}} \\
\subfloat[dgemm\_tn, m=4, n=4, k=s]{\includegraphics[width=0.41\linewidth]{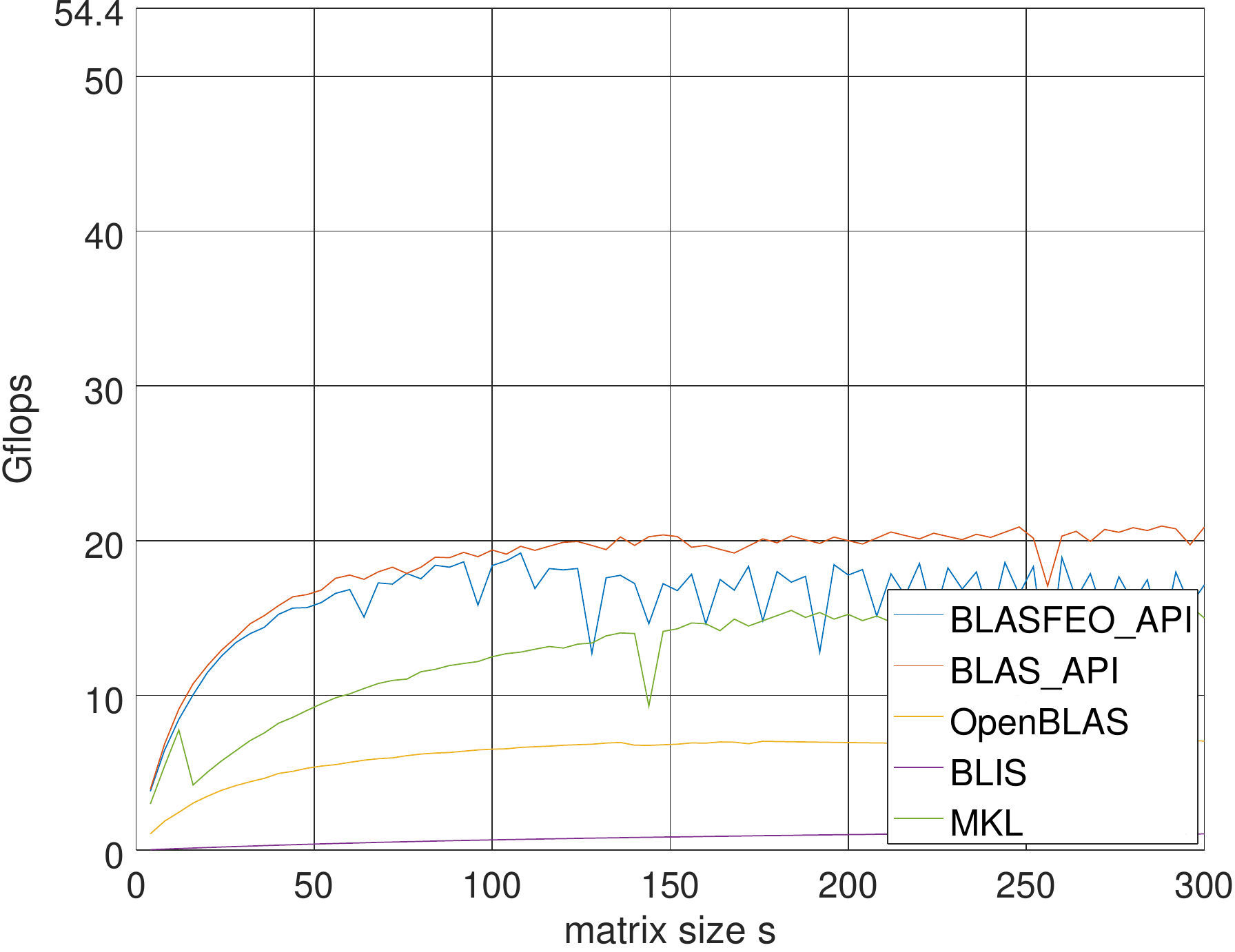} \label{fig:skinny:haswell:dgemm_tn_44k}} %\\
\subfloat[dgemm\_tn, m=4, n=s, k=s]{\includegraphics[width=0.41\linewidth]{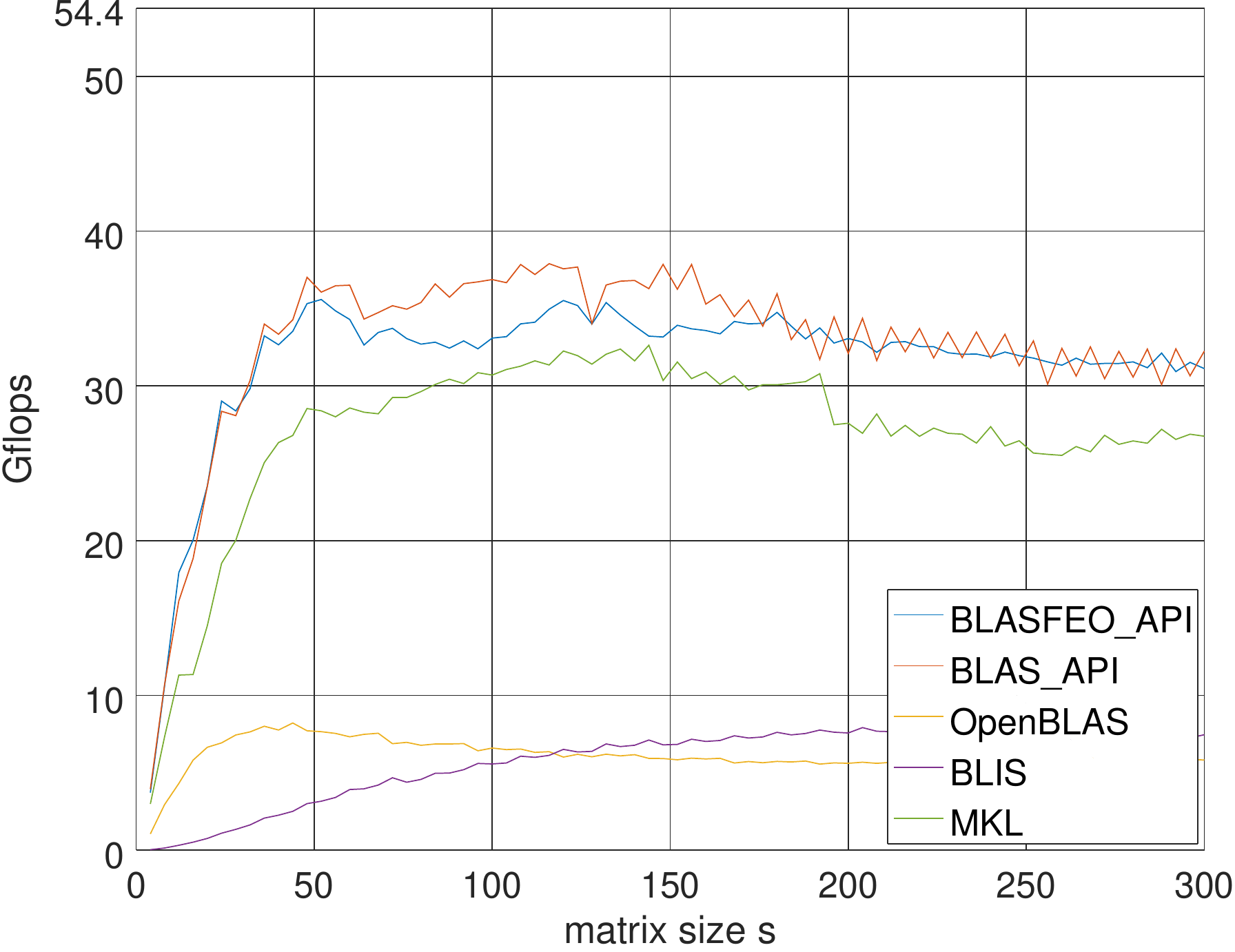} \label{fig:skinny:haswell:dgemm_tn_4nk}} \\
\subfloat[dgemm\_tt, m=4, n=4, k=s]{\includegraphics[width=0.41\linewidth]{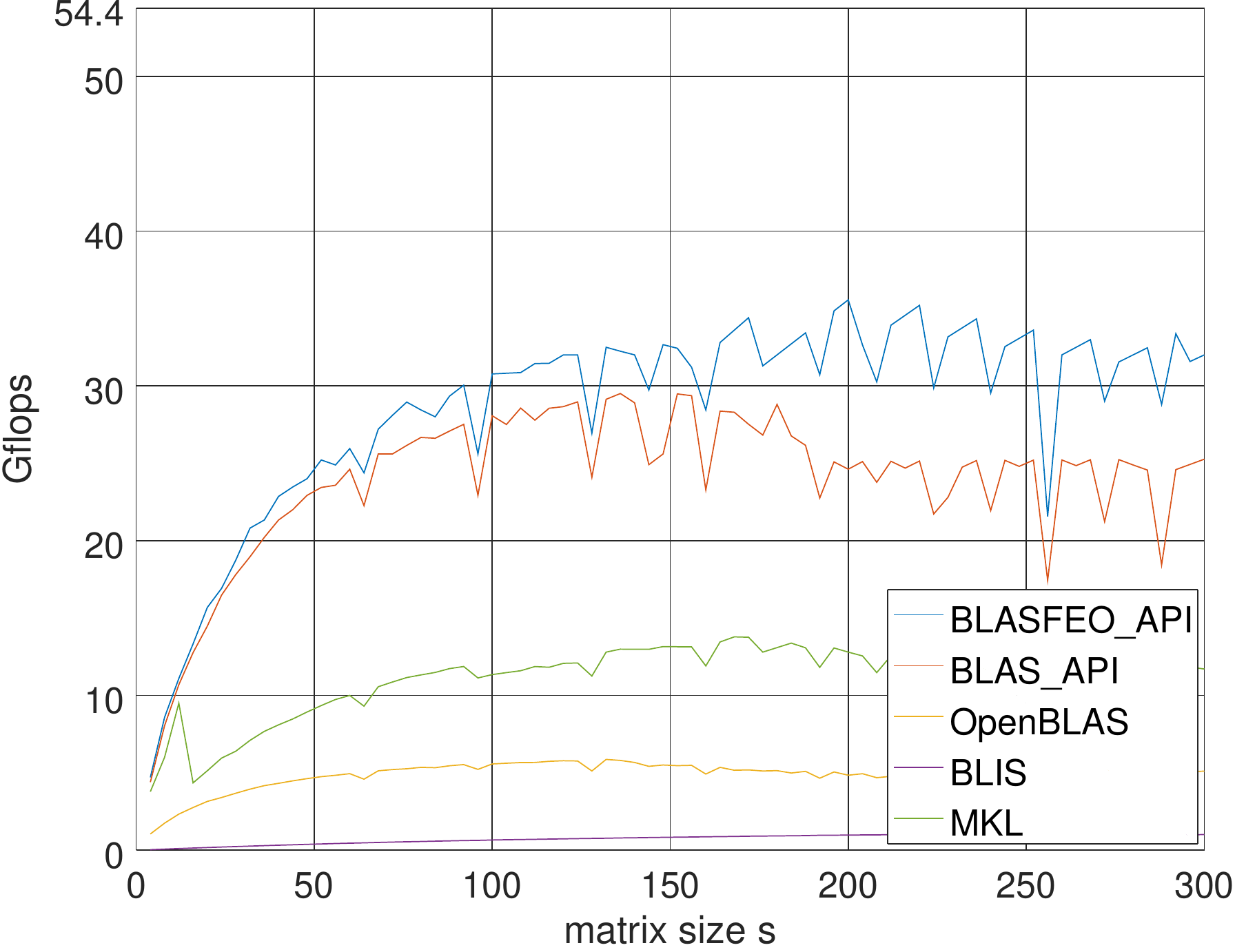} \label{fig:skinny:haswell:dgemm_tt_44k}} %\\
\subfloat[dgemm\_tt, m=4, n=s, k=s]{\includegraphics[width=0.41\linewidth]{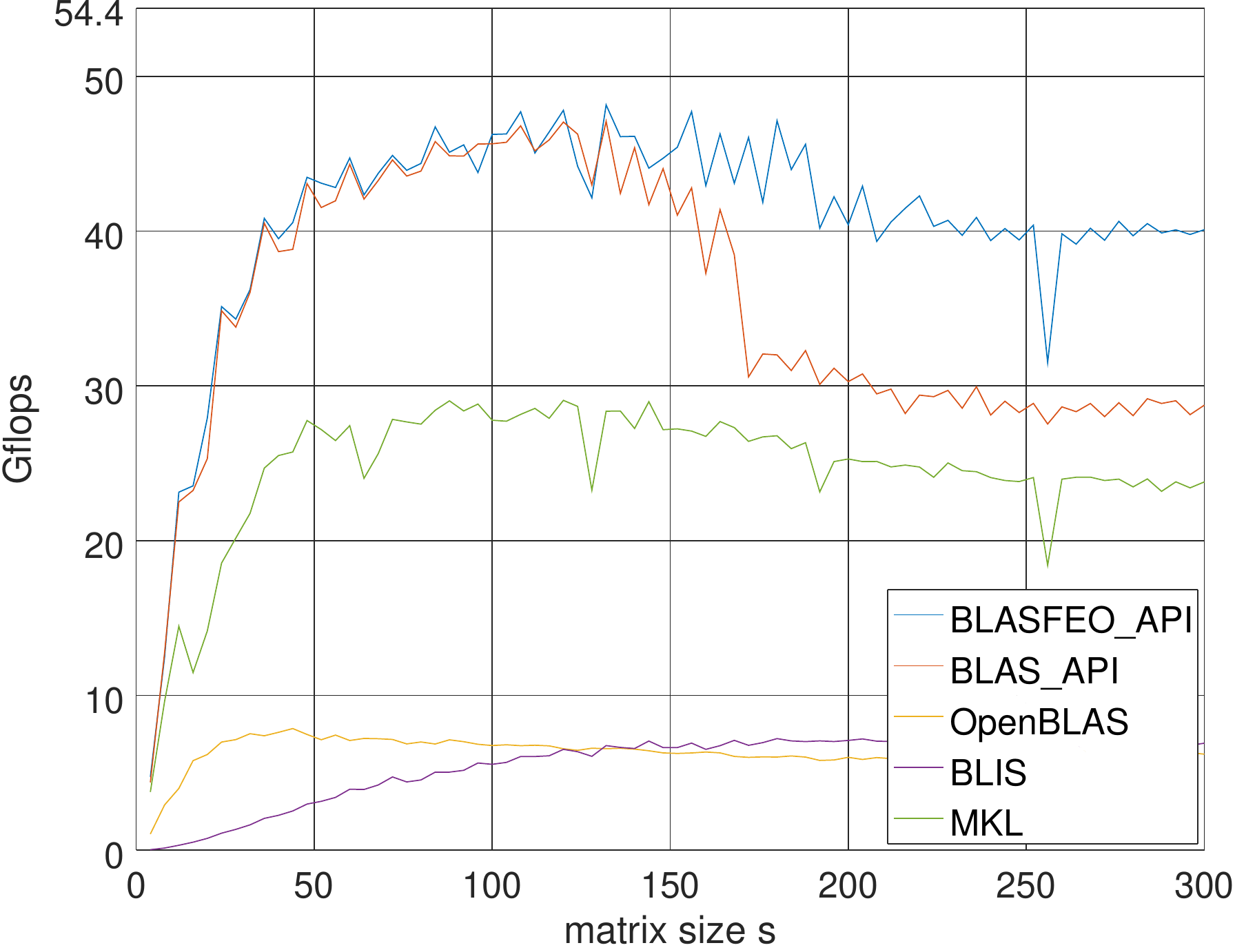} \label{fig:skinny:haswell:dgemm_tt_4nk}} \\
\caption{Performance of all parametric variants of {\tt dgemm} one core of the Intel Haswell Core i7 4810MQ @ 3.4 GHz.
Experiments for the cases $m=4$, $n=4$, $k$ between 4 and 300 (left column), and $m=4$, $n$ and $k$ between 4 and 300 (right column).}
\label{fig:skinny:fig1}
\end{figure}

\begin{figure}[!t]
\centering
\subfloat[dgemm\_nn, m=s, n=4, k=s]{\includegraphics[width=0.41\linewidth]{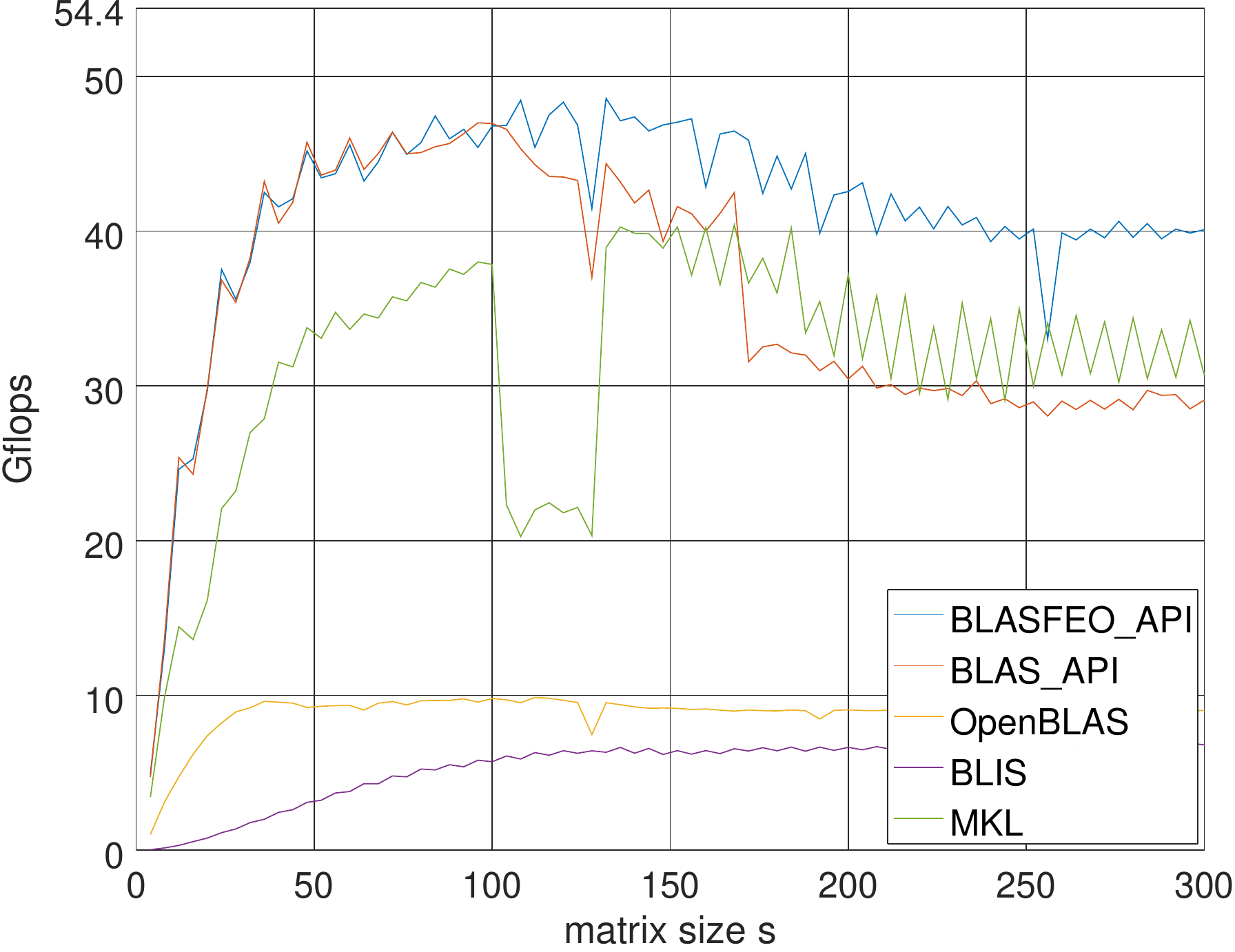} \label{fig:skinny:haswell:dgemm_nn_m4k}} %\\
\subfloat[dgemm\_nn, m=s, n=s, k=s]{\includegraphics[width=0.41\linewidth]{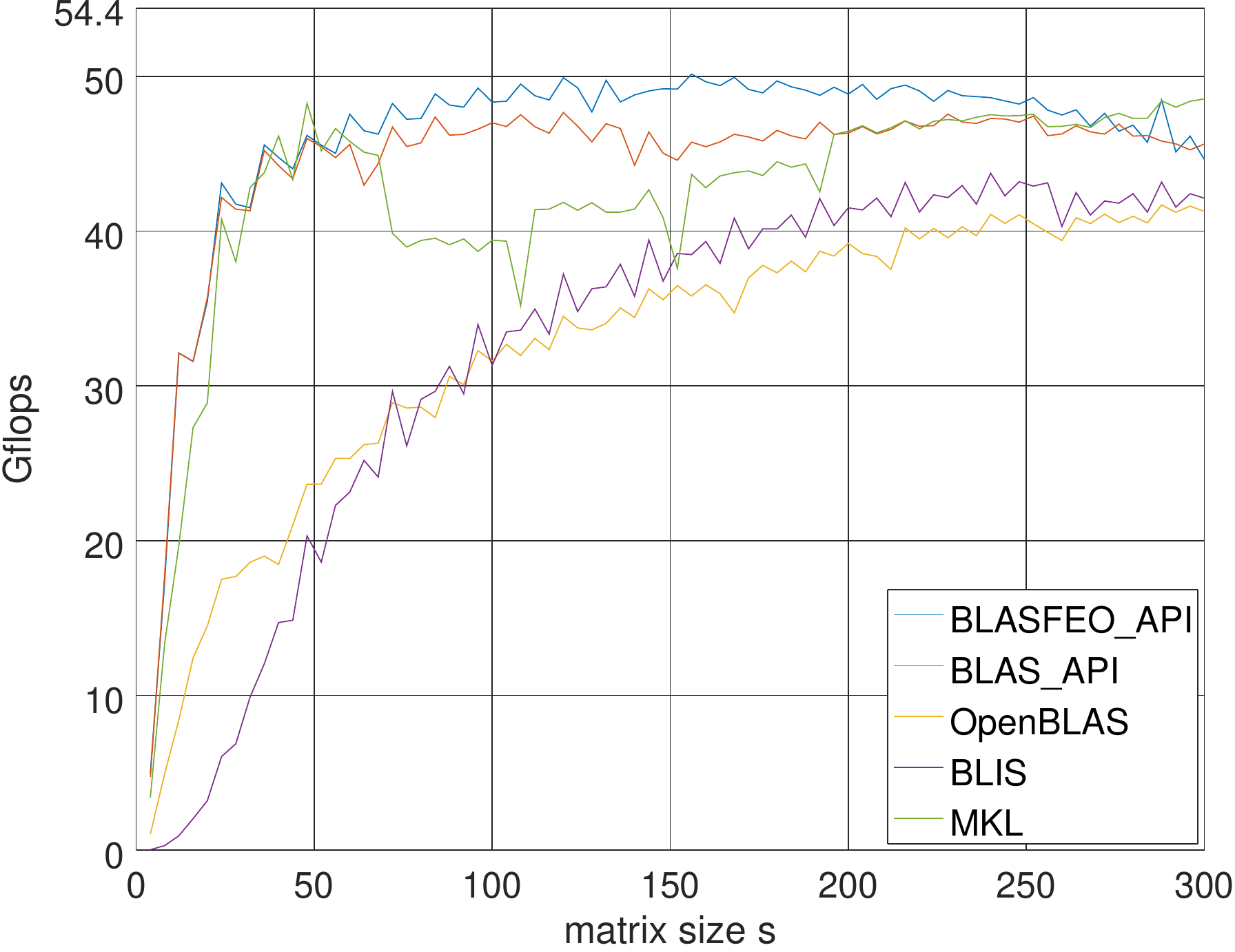} \label{fig:skinny:haswell:dgemm_nn_mnk}} \\
\subfloat[dgemm\_nt, m=s, n=4, k=s]{\includegraphics[width=0.41\linewidth]{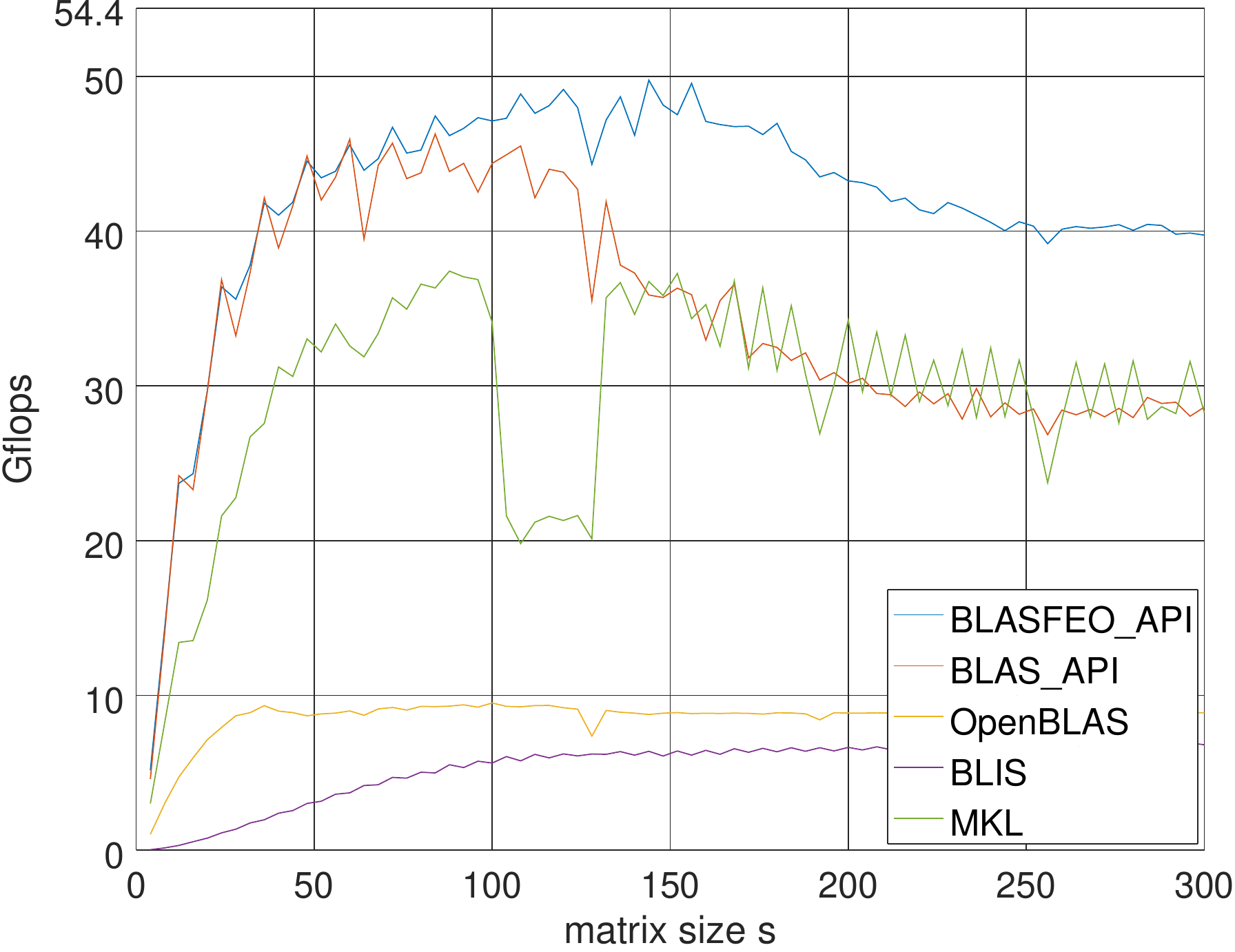} \label{fig:skinny:haswell:dgemm_nt_m4k}} %\\
\subfloat[dgemm\_nt, m=s, n=s, k=s]{\includegraphics[width=0.41\linewidth]{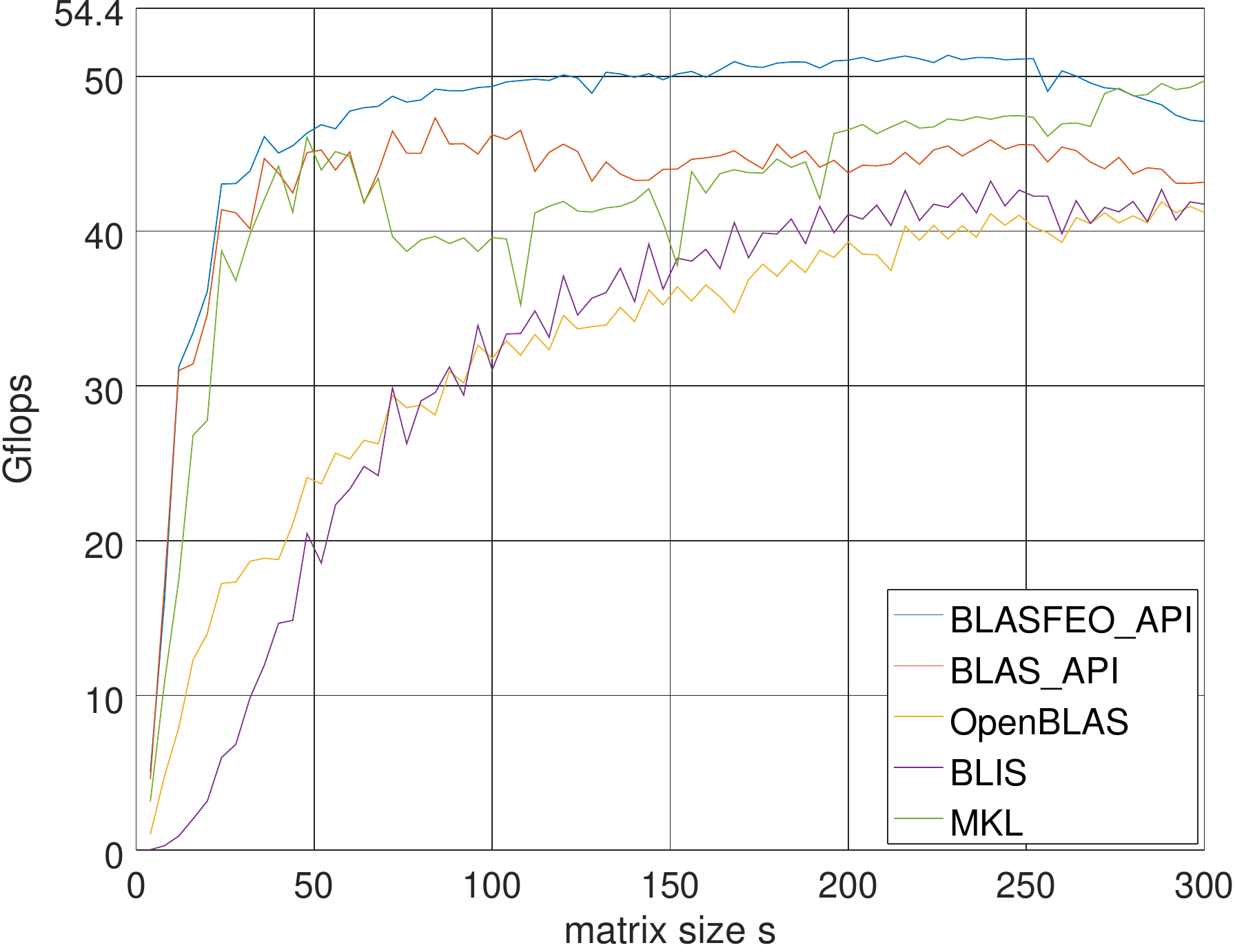} \label{fig:skinny:haswell:dgemm_nt_mnk}} \\
\subfloat[dgemm\_tn, m=s, n=4, k=s]{\includegraphics[width=0.41\linewidth]{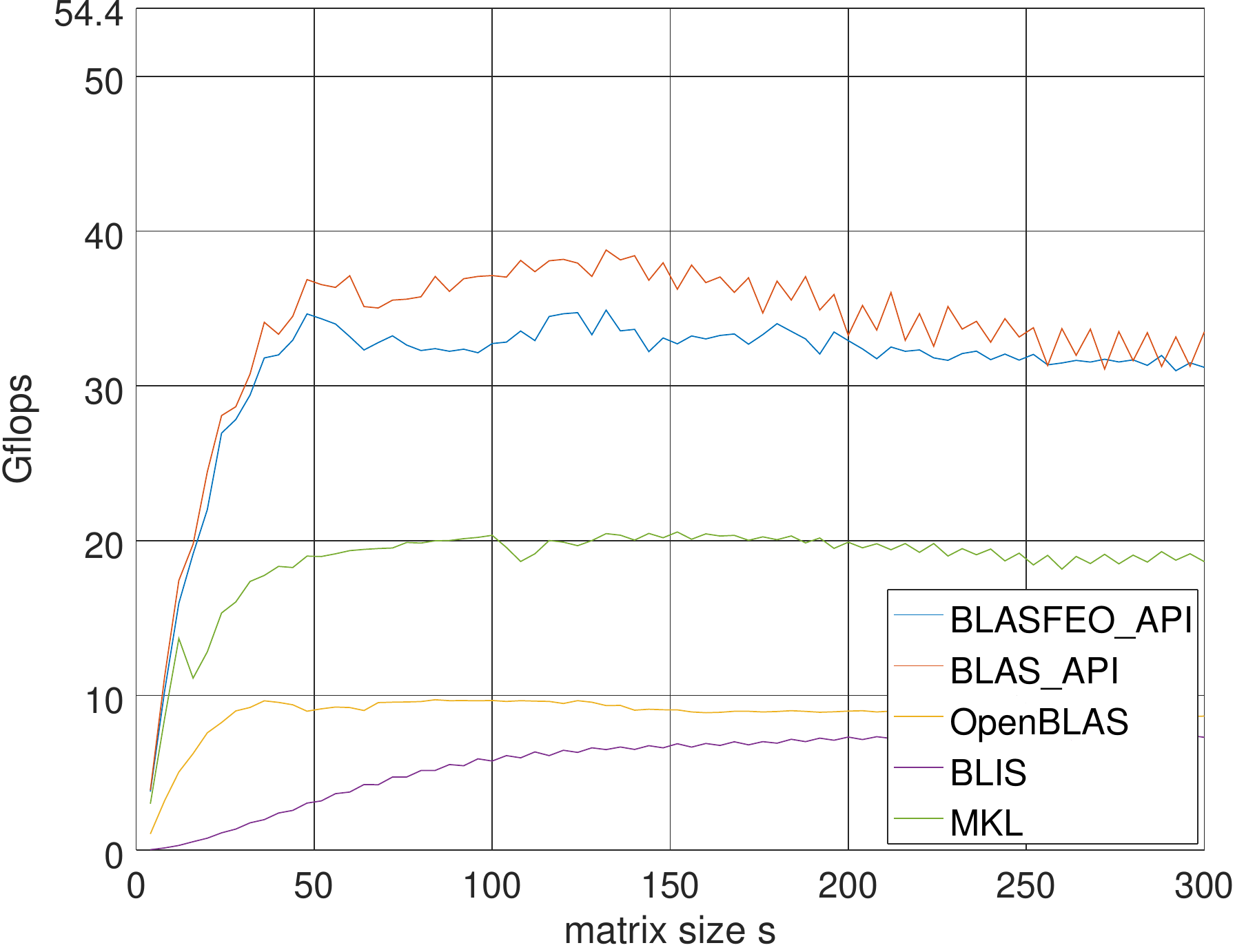} \label{fig:skinny:haswell:dgemm_tn_m4k}} %\\
\subfloat[dgemm\_tn, m=s, n=s, k=s]{\includegraphics[width=0.41\linewidth]{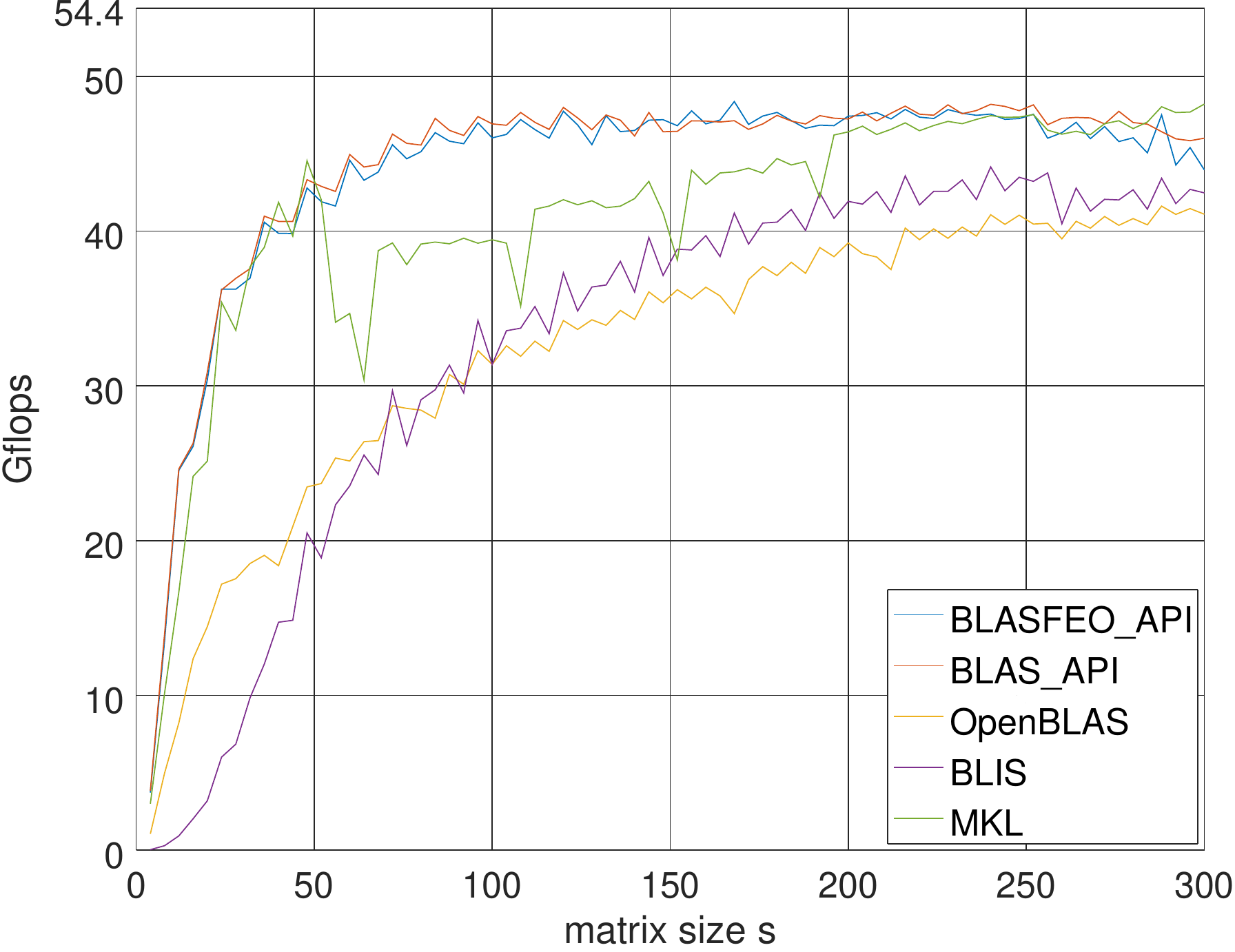} \label{fig:skinny:haswell:dgemm_tn_mnk}} \\
\subfloat[dgemm\_tt, m=s, n=4, k=s]{\includegraphics[width=0.41\linewidth]{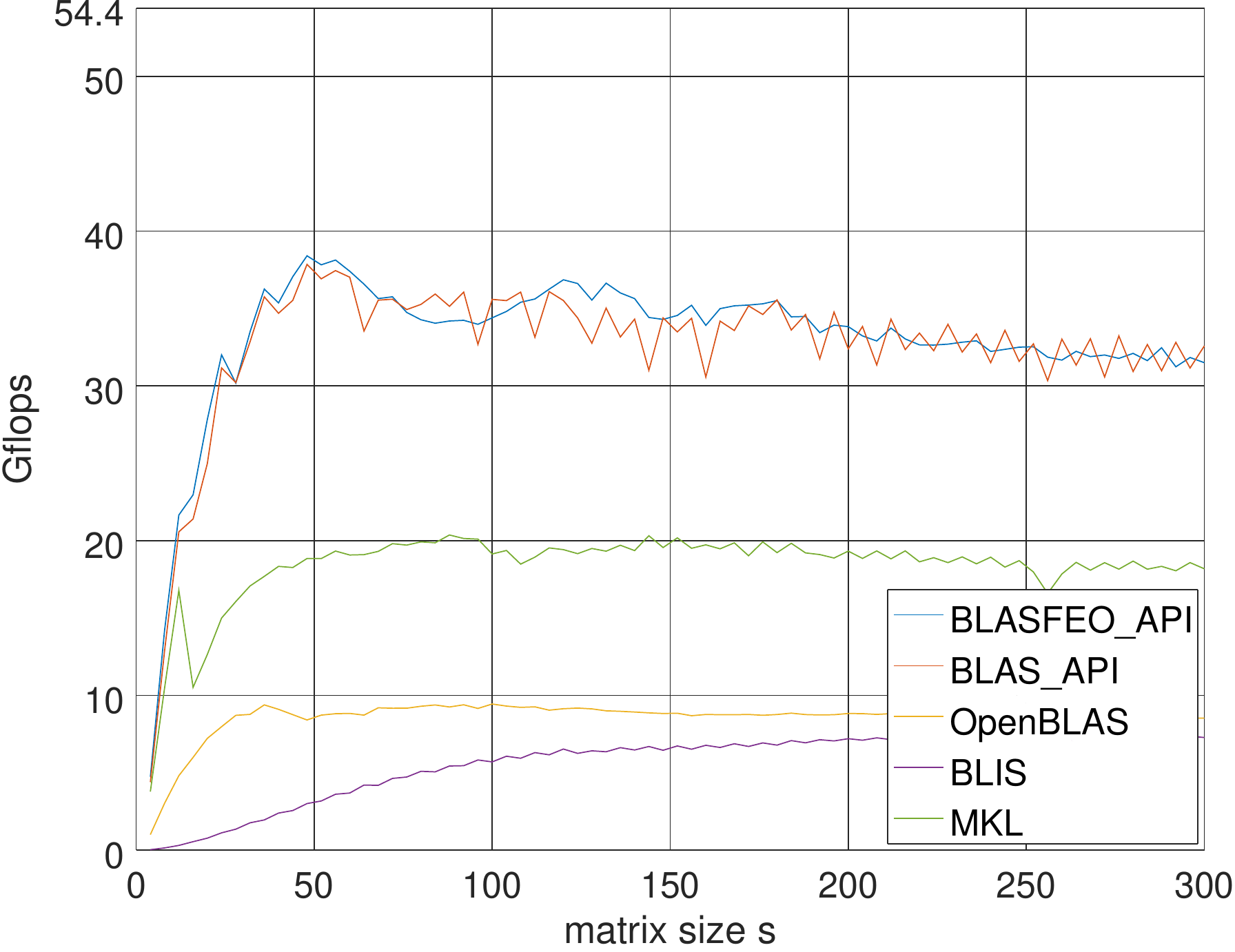} \label{fig:skinny:haswell:dgemm_tt_m4k}} %\\
\subfloat[dgemm\_tt, m=s, n=s, k=s]{\includegraphics[width=0.41\linewidth]{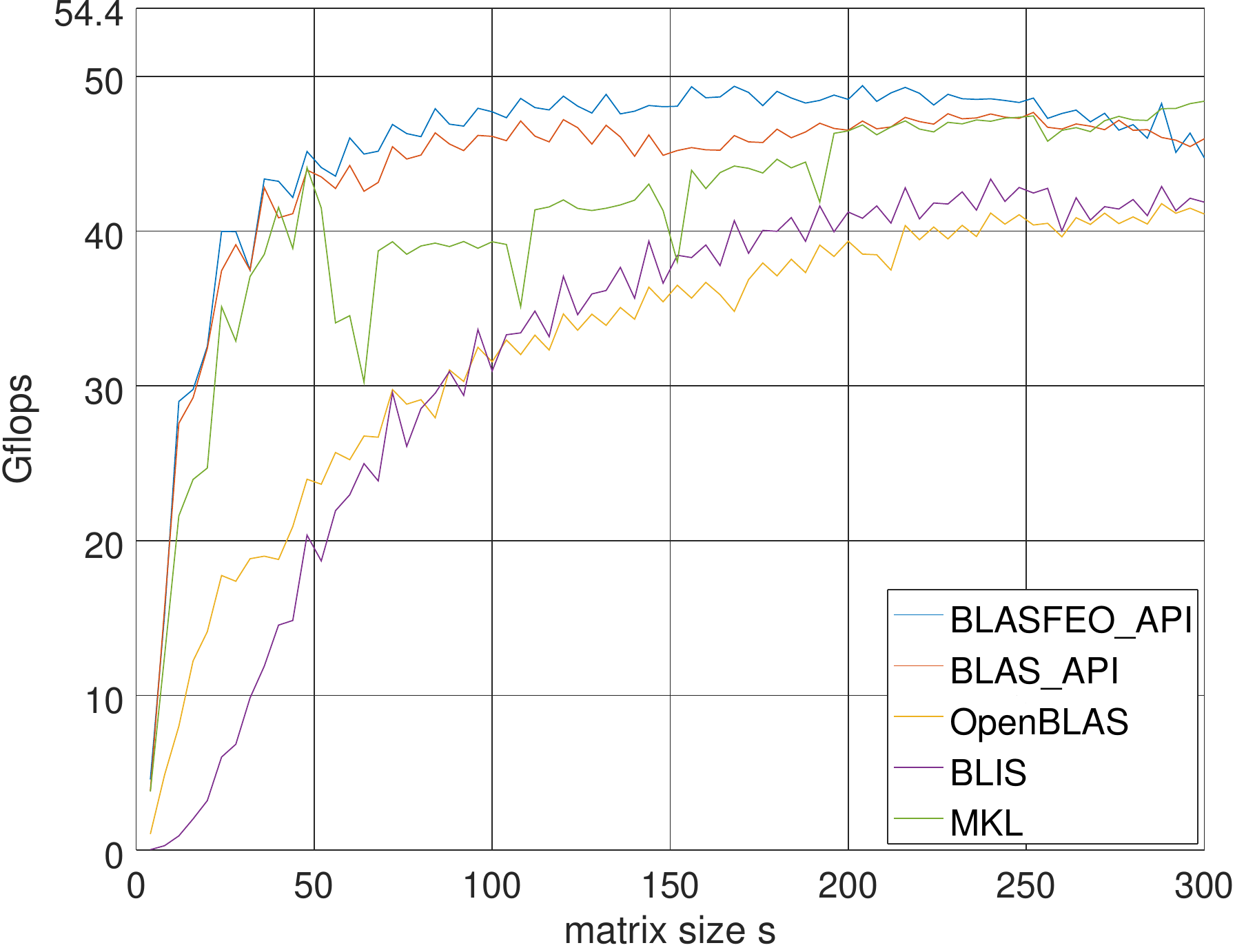} \label{fig:skinny:haswell:dgemm_tt_mnk}} \\
\caption{Performance of all parametric variants of {\tt dgemm} one core of the Intel Haswell Core i7 4810MQ @ 3.4 GHz.
Experiments for the cases $n=4$, $m$ and $k$ between 4 and 300 (left column), and $m$, $n$ and $k$ between 4 and 300 (right column).}
\label{fig:skinny:fig2}
\end{figure}

This section compares the performance of all four {\tt dgemm} variants in the case of both square and skinny (i.e. where one dimension is much smaller than the other) matrices on the Intel Haswell microarchitecture.
In particular, for these benchmarks the small dimension is chosen to be fixed and equal to 4.
The performance is evaluated in the four cases
\begin{itemize}
\item $m=4$, $n=4$, $k$ varied between 4 and 300 (Figure~\ref{fig:skinny:fig1}, left column).
\item $m=4$, $n$ and $k$ varied between 4 and 300 (Figure~\ref{fig:skinny:fig1}, right column).
\item $n=4$, $m$ and $k$ varied between 4 and 300 (Figure~\ref{fig:skinny:fig2}, left column).
\item $m$, $n$ and $k$ varied between 4 and 300 (Figure~\ref{fig:skinny:fig2}, right column).
\end{itemize}
The {\tt dgemm} variants `NN', `NT' and `TT' are implemented using algorithmic variants `B' (for large matrices), `C' and `Ct' (for medium matrices, when $m<m$ and $m\geq n$ respectively) and `D' (for small matrices).
The {\tt dgemm} variant `TN' is implemented using algorithmic variants `B' (for large matrices), and `C' and `Ct' (for medium and small matrices), since a native algorithmic variant `D` would not perform better than the other algorithmic variants in any case.

From the plots it is clear that for most of the considered cases the BLAS API of BLASFEO outperforms the other BLAS implementations, except a few cases where MKL is slightly faster.
%At the moment the switching between different algorithmic variants is computed using a very simple and conservative strategy, and in some cases performance can still improve by developing a more performing switching strategy.
Therefore, the proposed approach shows a good and consistent performance also in the case of rectangular matrices, and no inherent performance drawback is found.

In these experiments the BLASFEO API of BLASFEO is added as a reference.
In the BLASFEO API, all input and output matrix arguments are stored in the panel-major format, so {\tt dgemm} variants `NN', `NT' and `TT' are all implemented using only the equivalent of algorithmic variant `D' (i.e. no packing is performed).
The performance is good for any matrix size, as there is no overhead from packing and the panel-major matrix format enhances data caching and movement.
Therefore, it can be seen as an upper bound on the performance of the BLAS API, a bound which is approached when the packing overhead is minimized and the data memory layout allows the computational kernels to remain well fed.

Conversely, since algorithmic variant `D' would perform poorly, the {\tt dgemm} variant `TN' is implemented using the equivalent of algorithmic variants `C' or `Ct' and packing is employed to explicitly transpose either the matrix $A$ or $B$ and reduce to variants `NN' or `TT' respectively.
As a consequence, the performance is analogue to the performance of the BLAS API, and for every matrix size the best algorithmic variant has to be employed to minimize packing overhead.

%%%%%%%%%%%%%%%%%%%%%%%%%%%%%%%%
\section{Common commands in scientific programming languages and their implementation using BLAS and LAPACK} \label{sec:commands}
%%%%%%%%%%%%%%%%%%%%%%%%%%%%%%%%

This section contains tables of native commands in Octave, Python NumPy/SciPy and Julia, and their implementation using calls to BLAS and LAPACK routines.

Note that the {\tt char} options in the original Fortran interface are added to the routine name, after an underscore, in the same order as they appear in the routine signature.
For example, {\tt dsyrk\_un} is the {\tt dsyrk} routine called with options {\tt uplo=`u'} and {\tt trans=`n'}.

%%%%%%%%%%%%%%%%
\subsection{Octave} \label{sec:commands:octave}
%%%%%%%%%%%%%%%%

GNU Octave is open-source software featuring a high-level programming language, primarily intended for numerical computing.
Its language is mostly compatible with --- and therefore the leading free alternative to --- MATLAB.

Octave is written in C++ and it uses the C++ standard library.
It can be extended with oct-files (using the native Octave API) or mex-files (for MATLAB compatibility), and it can use dynamically loaded libraries.
Octave uses an interpreter to execute the Octave scripting language.
The syntax, as with MATLAB, is matrix-based.
Octave employs calls to BLAS and LAPACK routines to implement the most common matrix and vector operations in its native language.
Therefore, these operations are particularly efficient in Octave, especially if optimized BLAS and LAPACK implementations are employed.

%On a Linux distribution, Octave is generally configured to use OpenBLAS, which is currently the most performing open-source alternative for the matrix sizes of interest.
%Therefore, in this section the BLAS API of BLASFEO is compared to OpenBLAS (compiled in single thread mode, as this reduces overhead for small matrices).

The matrix multiplication in Octave is implemented with different versions of the BLAS routines {\tt dgemm} and {\tt dsyrk}.
In cases where one of the two matrix factors is triangular, the general matrix-matrix multiplication routine {\tt dgemm} is employed instead of the specialized routine {\tt dtrmm}.
The back- and forward-slash operators are implemented with a range of LAPACK routines depending on the matrix argument type, like {\tt dpotrf}, {\tt dpotrs}, {\tt dgetrf}, {\tt dgetrs}, {\tt dtrtrs}.
The logic to choose between these routines adds considerable overhead to the routine calls, especially in the case of very small matrices.
Direct calls to the BLAS routine {\tt dtrsm} are not employed in the solution of triangular linear systems, and the LAPACK routine {\tt dtrtrs} is employed in all cases, even if this implies the cost of an additional matrix transposition if the triangular matrix appears on the right-hand side.
%Table~\ref{tab:octave} in Section~\ref{sec:commands:octave} in the appendix contains a list of Octave native commands, and the version and number of flops (assuming square $n\times n$ matrices) of the corresponding BLAS or LAPACK routine.

Table~\ref{tab:octave} contains a list of some of the most common Octave commands and their implementation using BLAS and LAPACK routines.
In Octave, matrices are stored in column-major, and therefore they can be passed directly to BLAS and LAPACK routines without need for transposition.
The findings apply to the Octave version 4.2.2.

\begin{table}[t!]
\centering
\caption{Version of BLAS and LAPACK routines employed in Octave native language.
{\tt A}, {\tt B} and {\tt C} denote generic matrices, {\tt Q} denotes a symmetric positive definite matrix. %, {\tt L} and {\tt U} denote lower and upper triangular matrices.
All flops counts refer to $n\times n$ matrices.}
\label{tab:octave}
\begin{tabular}{l||l|c}
Octave command & BLAS and LAPACK routine & flops \\
\hline
\hline
{\tt C = A * B} & {\tt dgemm\_nn} & $2 \cdot n^3$ \\
{\tt C = A * B'} & {\tt dgemm\_nt} & $2 \cdot n^3$ \\
{\tt C = A' * B} & {\tt dgemm\_tn} & $2 \cdot n^3$ \\
{\tt C = A' * B'} & {\tt dgemm\_tt} & $2 \cdot n^3$ \\
\hline
{\tt C = A * A'} & {\tt dsyrk\_un} & $1 \cdot n^3$ \\
{\tt C = A' * A} & {\tt dsyrk\_ut} & $1 \cdot n^3$ \\
\hline
{\tt C = chol(Q, 'lower')} & {\tt dpotrf\_l} & $\sfrac 1 3 \cdot n^3$ \\
{\tt C = chol(Q, 'upper')} & {\tt dpotrf\_u} & $\sfrac 1 3 \cdot n^3$ \\
%{\tt C = A / Q} & {\tt dpotrf\_l} + {\tt dpotrs\_l} & $\sfrac 7 3 \cdot  n^3$ \\
%{\tt C = lu(A)} & {\tt dgetrf} & $\sfrac 2 3 \cdot  n^3$ \\
%{\tt C = A / B} & {\tt dgetrf} + {\tt dgetrs\_n} & $\sfrac 8 3 \cdot  n^3$ \\
%{\tt C = A / L} & {\tt dtrtrs\_ltn} & $1 \cdot  n^3$ \\
%{\tt C = A / L'} & {\tt dtrtrs\_utn} & $1 \cdot  n^3$ \\
%{\tt C = A / U} & {\tt dtrtrs\_utn} & $1 \cdot  n^3$ \\
%{\tt C = A / U'} & {\tt dtrtrs\_ltn} & $1 \cdot  n^3$ \\
%{\tt C = L $\backslash$ A} & {\tt dtrtrs\_lnn} & $1 \cdot  n^3$ \\
%{\tt C = L' $\backslash$ A} & {\tt dtrtrs\_ltn} & $1 \cdot  n^3$ \\
%{\tt C = U $\backslash$ A} & {\tt dtrtrs\_unn} & $1 \cdot  n^3$ \\
%{\tt C = U' $\backslash$ A} & {\tt dtrtrs\_utn} & $1 \cdot  n^3$ \\
\end{tabular}
\end{table}

%%%%%%%%%%%%%%%%
\subsection{NumPy/SciPy} \label{sec:commands:python}
%%%%%%%%%%%%%%%%

Python is an interpreted high-level language with a dynamic type system and automatic memory management.
Due to its large standard library and availability of interpreters for many operating systems, it has become a popular choice in many fields of application.
In particular, thanks to the presence of an extensive mathematics library and the third-party library NumPy, it is often used as a scientific scripting language.
Although speed is not the main focus of the language, as opposed to readability and simplicity~\cite{Peters2004} of use, time-critical functionalities can be moved to extension modules written in C or delegated to external high-performance code, which can be accessed through foreign function libraries.

NumPy~\cite{Oliphant2006} is an open-source library providing the Python language with multidimensional array objects and a large collection of high-level mathematical functions.
In particular, it provides an interface for linear algebra operations on matrices and vectors, and in this it is analogous to other numerical programming languages like Octave.
These operations are ultimately implemented using calls to BLAS and LAPACK routines, even if NumPy provides a more convenient interface, which hides most low-level details of BLAS and LAPACK APIs and performs additional sanity checks.
This is analogous to the Octave implementation, and it adds a similar level of overhead.
NumPy provides functionality to operate on matrices on both column- and row-major formats, even if the default storage scheme is row-major.
Therefore, BLAS is effectively accessed through the CBLAS interface, while LAPACK routines are called directly, possibly after transposition between row- and column-major.

SciPy is an open-source Python library for scientific and technical computing; it is part of the NumPy stack and builds on the NumPy array object.
In particular, SciPy provides a rich library for linear algebra operations, and a set of functions that is both wider range and lower level compared to NumPy.
For some linear algebra routines multiple implementations exist, with the lower level being a thin wrapper to the corresponding BLAS or LAPACK routine.
These low level routines allow one to perform specialized operations (like exploiting the fact that a matrix is triangular) and do not perform any additional sanity checks, resulting in much lower overhead.
SciPy always directly calls the BLAS and LAPACK APIs.
If the NumPy matrices are stored in row-major order, they are explicitly transposed into dynamically allocated memory buffers, which are then passed to the BLAS and LAPACK routines.
Therefore, in order to obtain full performance from the SciPy routines, all matrices must be stored in column-major order.

The Cholesky factorization is used as an example.
This factorization is implemented in the NumPy routine {\tt numpy.linalg.cholesky}, which computes the lower triangular factor of a positive definite matrix.
The SciPy routine {\tt scipy.linalg.cholesky} additionally provides options to compute the upper/lower triangular factor, to overwrite the input matrix and to enable/disable checks for finiteness of the elements of the input matrix.
The lowest level interface is provided by the SciPy routine {\tt scipy.linalg.lapack.dpotrf}, which roughly provides the same functionality as the corresponding LAPACK routine.

As NumPy provides similar levels of functionality and overhead as Octave, only SciPy is considered in the performance evaluations performed in this paper.
In particular, the version 1.0 of SciPy introduces new low level functions providing wrappers to the BLAS routines {\tt dtrsm} and {\tt dtrmm}.
Unfortunately, up to the current version 1.2 a bug affects the functionality of {\tt dtrmm} as needed in the implementation of the Riccati recursion in Section~\ref{sec:appl:ric}.
The replacement of this routine with {\tt dgemm} does not allow one to fully exploit the structure of the Riccati recursion.
Since SciPy is open-source software, we were able to examine the source code and implement a fix for the bug.
This fixed version of SciPy 1.2 is also considered in the tests in Section~\ref{sec:appl:ric}.

Table~\ref{tab:scipy} contains a list of SciPy/NumPy commands and their respective calls to BLAS and LAPACK routines.
In particular, different ways of calling the BLAS routines {\tt dgemm} and {\tt dsyrk} and the LAPACK routine {\tt dpotrf} are considered.
The findings apply to the NumPy version 1.11 and SciPy version 1.2.

\begin{table}[t!]
\centering
\caption{Version of BLAS and LAPACK routines employed in NumPy.
{\tt A}, {\tt B} and {\tt C} denote generic matrices, {\tt Q} denotes a symmetric positive definite matrix.
All flops counts refer to $n\times n$ matrices.
Notice that calls to CBLAS in NumPy are carried out with {\tt Order = CblasRowMajor}. % by default.
} 
\label{tab:scipy}
\begin{tabular}{l||l|c}
NumPy / SciPy command & BLAS and LAPACK routine & flops \\
\hline
\hline
{\tt C = np.dot(A, B)} 		& {\tt cblas $\rightarrow$ dgemm\_nn}& $2  \cdot n^3$ \\
{\tt C = np.dot(A, B.T)} 	& {\tt cblas $\rightarrow$ dgemm\_tn} & $2 \cdot n^3$ \\
{\tt C = np.dot(A.T, B)} 	& {\tt cblas $\rightarrow$ dgemm\_nt} & $2 \cdot n^3$ \\
{\tt C = np.dot(A.T, B.T)} 	& {\tt cblas $\rightarrow$ dgemm\_tt} & $2 \cdot n^3$ \\
{\tt sp.linalg.blas.dgemm(1.0, A, B,} & {\tt dgemm\_nn}& $2 \cdot n^3$\\
{\tt $\quad$c=C, trans\_a=False, trans\_b=False)} & & \\
{\tt sp.linalg.blas.dgemm(1.0, A, B,} & {\tt dgemm\_nt}& $2 \cdot n^3$\\
{\tt $\quad$c=C, trans\_a=False, trans\_b=True)} & & \\
{\tt sp.linalg.blas.dgemm(1.0, A, B,} & {\tt dgemm\_tn}& $2 \cdot n^3$\\
{\tt $\quad$c=C, trans\_a=True, trans\_b=False)} & & \\
{\tt sp.linalg.blas.dgemm(1.0, A, B,} & {\tt dgemm\_tt}& $2 \cdot n^3$\\
{\tt $\quad$c=C, trans\_a=True, trans\_b=True)} & & \\
\hline
{\tt C = np.dot(A, A.T)} 	& {\tt cblas $\rightarrow$ dsyrk\_lt} & $1 \cdot n^3$ \\
{\tt C = np.dot(A.T, A)} 	& {\tt cblas $\rightarrow$ dsyrk\_ln} & $1 \cdot n^3$ \\
{\tt sp.linalg.blas.dsyrk(1.0, A, trans=0, lower=0} &{\tt dsyrk\_un} & $1 \cdot n^3$ \\
{\tt $\quad$ c=C, overwrite\_c=True)} & & \\
{\tt sp.linalg.blas.dsyrk(1.0, A, trans=1, lower=0} &{\tt dsyrk\_ut} & $1 \cdot n^3$ \\
{\tt $\quad$ c=C, overwrite\_c=True)} & & \\
{\tt sp.linalg.blas.dsyrk(1.0, A, trans=0, lower=1} &{\tt dsyrk\_ln} & $1 \cdot n^3$ \\
{\tt $\quad$ c=C, overwrite\_c=True)} & & \\
{\tt sp.linalg.blas.dsyrk(1.0, A, trans=1, lower=1} &{\tt dsyrk\_lt} & $1 \cdot n^3$ \\
{\tt $\quad$ c=C, overwrite\_c=True)} & & \\
\hline
%{\tt C = np.linalg.cholesky(Q)} 	& {\tt dcopy, dpotrf\_l} & $\sfrac 1 3 \cdot n^3$ \\
{\tt C = np.linalg.cholesky(Q)} 	& {\tt dpotrf\_l} & $\sfrac 1 3 \cdot n^3$ \\
{\tt C = sp.linalg.cholesky(Q, lower=True,} & {\tt dpotrf\_l} & $\sfrac 1 3 \cdot n^3$\\
{\tt \quad overwrite\_a=False, check\_finite=False)} 	&  \\
{\tt C = sp.linalg.cholesky(Q, lower=False,} & {\tt dpotrf\_u} & $\sfrac 1 3 \cdot n^3$\\
{\tt \quad overwrite\_a=False, check\_finite=False)} 	&  \\
%{\tt (C, info) = sp.linalg.cho\_factor(Q, lower=True,} & {\tt dpotrf\_l} & $\sfrac 1 3 \cdot n^3$ \\ 
%{\tt \quad overwrite\_a=False, check\_finite=False)} & & \\ 
{\tt sp.linalg.lapack.dpotrf(Q, lower=1,} & {\tt dpotrf\_l} & $\sfrac 1 3 \cdot n^3$ \\
{\tt \quad clean=0, overwrite\_a=True)} & & \\
{\tt sp.linalg.lapack.dpotrf(Q, lower=0,} & {\tt dpotrf\_u} & $\sfrac 1 3 \cdot n^3$ \\
{\tt \quad clean=0, overwrite\_a=True)} & & \\
\end{tabular}
\end{table}

%%%%%%%%%%%%%%%%
\subsection{Julia} \label{sec:commands:julia}
%%%%%%%%%%%%%%%%

Julia is an open-source high-level general-purpose programming language.
It is dynamically typed and in that regard it resembles scripting languages such as Octave and Python.
However, Julia makes use of JIT compilation, and its compiler is based on LLVM and therefore can generate optimized native code for multiple architectures.
Once compiled, the speed of Julia code approaches the speed of statically typed and compiled languages such as C and C++.
This combination of a dynamically typed language and high performance is an explicit design choice of Julia.

Regarding numeric computations, the Julia syntax has similarities with the Matlab and Octave syntax.
Besides support for multi-dimensional arrays, Julia provides native implementations for many linear algebra operations in the {\tt LinearAlgebra} module.
Basic matrix operations are implemented with calls to BLAS and LAPACK routines; the default implementation is OpenBLAS.
Calls to these routine do not need any additional transposition since the native matrix format in Julia is column-major.

Similarly to Python with NumPy and SciPy, in Julia there are several commands to perform the same linear algebra operations, trading off performance and easiness of use.
For example, the general matrix-matrix multiplication can be performed using the operator {\tt `*'} (which returns the result in a newly allocated matrix), or calling the function {\tt mul!} (which requires a pre-allocated output matrix) of the {\tt LinearAlgebra} module.
Furthermore, wrappers to BLAS and LAPACK routines are part of the {\tt LinearAlgebra.BLAS} and {\tt LinearAlgebra.LAPACK} modules.
The functions in these modules provide the same functionality as the BLAS and LAPACK routines.
As an example, the {\tt gemm!} function accepts also the scalar {\tt alpha} and {\tt beta} arguments.
More importantly, specialized routines like {\tt syrk}, {\tt symm} or {\tt trmm} can be called directly through these wrappers.

When operating with sub-matrices, it is important to notice that in Julia slicing an array creates a copy of the selected sub-array.
This has important consequences on performance, since unnecessary copying can be expensive.
Furthermore, this can lead to undesired behaviors: if a sliced matrix is passed as the return argument of an in-place linear algebra function, the result of the operations gets lost as it is not copied back in the original matrix.
The Julia command {\tt view} should be used instead to get an in-place reference to a sub-matrix, which can be used to modify the original matrix.
Notice that in this paper static arrays (from the module {\tt StaticArrays}) are not considered, since basic linear algebra operations on them are implemented with JIT compiled native Julia routines and not with calls to BLAS and LAPACK routines.

Table~\ref{tab:julia} contains a list of some common Julia commands for matrix operations, and their implementation using BLAS and LAPACK routines.
%In Julia, matrices are stored in column-major, and therefore they can be passed directly to BLAS and LAPACK routines without need for transposition.
The findings apply to the Julia version 1.0.1.

\begin{table}[t!]
\centering
\caption{Version of BLAS and LAPACK routines employed in Julia native language.
{\tt A}, {\tt B} and {\tt C} denote generic matrices, {\tt Q} denotes a symmetric positive definite matrix. %, {\tt L} and {\tt U} denote lower and upper triangular matrices.
All flops counts refer to $n\times n$ matrices.}
\label{tab:julia}
\begin{tabular}{l||l|c}
Julia command & BLAS and LAPACK routine & flops \\
\hline
\hline
{\tt C = A * B} & {\tt dgemm\_nn} & $2 \cdot n^3$ \\
{\tt C = A * tranpose(B)} & {\tt dgemm\_nt} & $2 \cdot n^3$ \\
{\tt C = transpose(A) * B} & {\tt dgemm\_tn} & $2 \cdot n^3$ \\
{\tt C = transpose(A) * tranpose(B)} & {\tt dgemm\_tt} & $2 \cdot n^3$ \\
{\tt mul!(C, A, B)} & {\tt dgemm\_nn} & $2 \cdot n^3$ \\
{\tt mul!(C, A, transpose(B))} & {\tt dgemm\_nt} & $2 \cdot n^3$ \\
{\tt mul!(C, transpose(A), B)} & {\tt dgemm\_tn} & $2 \cdot n^3$ \\
{\tt mul!(C, transpose(A), transpose(B))} & {\tt dgemm\_tt} & $2 \cdot n^3$ \\
{\tt BLAS.gemm!('N', 'N', alpha, A, B, beta, C)} & {\tt dgemm\_nn} & $2 \cdot n^3$ \\
{\tt BLAS.gemm!('N', 'T', alpha, A, B, beta, C)} & {\tt dgemm\_nt} & $2 \cdot n^3$ \\
{\tt BLAS.gemm!('T', 'N', alpha, A, B, beta, C)} & {\tt dgemm\_tn} & $2 \cdot n^3$ \\
{\tt BLAS.gemm!('T', 'T', alpha, A, B, beta, C)} & {\tt dgemm\_tt} & $2 \cdot n^3$ \\
\hline
{\tt C = A * transpose(A)} & {\tt dsyrk\_un} & $1 \cdot n^3$ \\
{\tt C = transpose(A) * A} & {\tt dsyrk\_ut} & $1 \cdot n^3$ \\
{\tt mul!(C, A, transpose(A))} & {\tt dsyrk\_un} & $1 \cdot n^3$ \\
{\tt mul!(C, transpose(A), A)} & {\tt dsyrk\_ut} & $1 \cdot n^3$ \\
{\tt BLAS.syrk!('L', 'N', alpha, A, beta, C)} & {\tt dsyrk\_ln} & $1 \cdot n^3$ \\
{\tt BLAS.syrk!('L', 'T', alpha, A, beta, C)} & {\tt dsyrk\_lt} & $1 \cdot n^3$ \\
{\tt BLAS.syrk!('U', 'N', alpha, A, beta, C)} & {\tt dsyrk\_un} & $1 \cdot n^3$ \\
{\tt BLAS.syrk!('U', 'T', alpha, A, beta, C)} & {\tt dsyrk\_ut} & $1 \cdot n^3$ \\
\hline
{\tt cholesky!(Q)} & {\tt dpotrf\_u} & $\sfrac 1 3  \cdot n^3$ \\
{\tt LAPACK.potrf!('L', Q)} & {\tt dpotrf\_l} & $\sfrac 1 3 \cdot n^3$ \\
{\tt LAPACK.potrf!('U', Q)} & {\tt dpotrf\_u} & $\sfrac 1 3 \cdot n^3$ \\
\end{tabular}
\end{table}

%%%%%%%%%%%%%%%%%%%%%%%%%%%%%%%%

%\bibliographystyle{plain}
%\bibliography{syscop}

%\begin{comment}

%\end{comment}